\documentclass[structabstract]{aa}
\usepackage{graphicx}
\usepackage{array}
\usepackage{txfonts}
\usepackage{natbib}
\usepackage{longtable}
\usepackage{float}
\usepackage{dcolumn}
\newcolumntype{k}[1]{D{.}{.}{#1}}
\newcolumntype{d}{D{.}{.}{-1}}
\newcolumntype{,}{D{,}{,}{-1}}

\begin{document}

\title{Laboratory Characterization and Astrophysical Detection of
Vibrationally Excited States of Vinyl Cyanide \object{in Orion-KL}.\\ 
\thanks{Appendix A (online Tables and Figures) and Appendix B (column density table of ethyl cyanide)
are only available in electronic form via http://www.edpscience.org; Tables A.6-A.14 are only available
in electronic form at the CDS via http://cdsweb.u-strasbg.fr/cgi-bin/qcat?J/A+A/
}\fnmsep\thanks{This work was based on observations carried out with the IRAM 30-meter telescope.
IRAM is supported by INSU/CNRS (France), MPG (Germany), and IGN (Spain).}}

\author{A. L\'opez\inst{1}
 \and B. Tercero\inst{1}
 \and Z. Kisiel\inst{2}
 \and A. M. Daly\inst{3,4}
 \and C. Berm\'udez\inst{4}
 \and H. Calcutt\inst{5}
 \and N. Marcelino\inst{6}
 \and S. Viti\inst{5}
 \and B.J. Drouin\inst{3}
 \and I. R. Medvedev\inst{7}
 \and C. F. Neese\inst{8}
 \and L.~Pszcz\'o\l kowski\inst{2}
 \and J. L. Alonso\inst{4}
  \and J. Cernicharo\inst{1}
  }
   \institute{Centro de Astrobiolog\'ia (CSIC-INTA). Departamento de
              Astrof\'isica Molecular. Ctra. de Ajalvir Km 4, 28850
              Torrej\'on de Ardoz, Madrid, Spain.
    \and Institute of Physics, Polish Academy of Sciences, Al. Lotnik\'ow 32/46, 02-668 Warszawa, Poland.
    \and Jet Propulsion Laboratory, California Institute of Technology, 4800 Oak Grove Dr., Pasadena, CA 91109, USA.
    \and Grupo de Espectroscopia Molecular (GEM), Unidad Asociada CSIC, Edificio Quifima, Laboratorios de Espectroscopia y Bioespectroscopia, 
Parque Científico UVa, Universidad de Valladolid, 47011, Valladolid, Spain
    \and Department of Physics and Astronomy, University College London, Gower Street, London WC1E 6B, UK.
    \and NRAO, 520 Edgemont Road, Charlottesville, VA 22902, USA.
    \and Wright State University, 3640 Colonel Glenn Hwy, Dayton, OH 45435 USA 
    \and Ohio State University, 191 W. Wooddruff Ave., Columbus, OH 43210 USA
    \\
\email{lopezja@cab.inta-csic.es; terceromb@cab.inta-csic.es; kisiel@ifpan.edu.pl; Adam.M.Daly@jpl.nasa.gov
;cbermu@qf.uva.es; hcalcutt@star.ucl.ac.uk; nmarceli@nrao.edu; sv@star.ucl.ac.uk; brian.j.drouin@jpl.nasa.gov; ivan.medvedev@wright.edu; cfneese@mps.ohio-state.edu;lbee@ifpan.edu.pl; jlalonso@qf.uva.es; jcernicharo@cab.inta-csic.es}
}
%}
%}
%         \and
%             University of Alexandria, Department of Geography, ...\\
%             \email{c.ptolemy@hipparch.uheaven.space}
%             \thanks{The university of heaven temporarily does not
%                     accept e-mails}

   \date{Received     ; accepted     }

% \abstract{}{}{}{}{}
% 5 {} token are mandatory
  \abstract
%  {}
{Laboratory characterization (18-1893 GHz) and astronomical
detection (IRAM-30m: 80-280 GHz, Orion-KL) of CH$_2$CHCN (vinyl cyanide) in its
ground and vibrationally excited states.
}
{Improving the understanding of rotational spectra of vibrationally excited vinyl cyanide with
new laboratory data and analysis. The laboratory results allow searching for these excited state
transitions in the Orion-KL line survey. Furthermore, rotational lines of
CH$_2$CHCN contribute to the understanding of the physical and chemical properties of the
cloud.
}
{Laboratory measurements of CH$_2$CHCN made on several different frequency-modulated 
spectrometers were combined into a single broadband 50-1900 GHz spectrum and its assignment was
confirmed by Stark modulation spectra recorded in the 18-40 GHz region 
and by ab-initio anharmonic force field calculations. For analyzing the emission lines of vinyl cyanide
detected in Orion-KL we used the excitation and radiative transfer code (MADEX)
at LTE conditions.
}
{Detailed characterisation of laboratory spectra of CH$_2$CHCN in 9 different excited vibrational states 
($\varv_{11}$=1, $\varv_{15}$=1, $\varv_{11}$=2, $\varv_{10}$=1$\Leftrightarrow$($\varv_{11}$=1,$\varv_{15}$=1), 
$\varv_{11}$=3/$\varv_{15}$=2/$\varv_{14}$=1, ($\varv_{11}$=1,$\varv_{10}$=1)$\Leftrightarrow$($\varv_{11}$=2,$\varv_{15}$=1),
$\varv_{9}$=1, ($\varv_{11}$=1,$\varv_{15}$=2)$\Leftrightarrow$($\varv_{10}$=1,$\varv_{15}$=1)$\Leftrightarrow$
($\varv_{11}$=1,$\varv_{14}$=1), and $\varv_{11}$=4)
and detection of transitions in the $\varv_{11}$=2 and $\varv_{11}$=3 states for the first
time in Orion-KL, and of those in the $\varv_{10}$=1$\Leftrightarrow$($\varv_{11}$=1,$\varv_{15}$=1) dyad of states 
for the first time in space. The rotational transitions of the ground state of this molecule emerge
from four cloud components of hot core nature which trace the physical and
chemical conditions of high mass star forming regions in the Orion-KL Nebula.
The lowest energy vibrationally excited states of vinyl cyanide such as
$\varv_{11}$=1 (at 328.5 K),  $\varv_{15}$=1 (at 478.6 K), $\varv_{11}$=2 (at 657.8 K),
the $\varv_{10}$=1$\Leftrightarrow$($\varv_{11}$=1,$\varv_{15}$=1) dyad (at 806.4/809.9 K), 
and $\varv_{11}$=3 (at 987.9 K) are 
populated under warm and dense conditions, so they probe the hottest parts of
the Orion-KL source. The
vibrational temperatures derived for the $\varv_{11}$=1, $\varv_{11}$=2, and $\varv_{15}$=1 states are 252$\pm$76\,K,
242$\pm$121\,K, and 227$\pm$68\,K, respectively; all of them close to the mean kinetic
temperature of the hot core component (210\,K). The total column density of CH$_2$CHCN in
the ground state is (3.0$\pm$0.9)$\times$10$^{15}$ cm$^{-2}$. We report the detection
of methyl isocyanide (CH$_3$NC) for the first time in Orion-KL and a tentative 
detection of vinyl isocyanide (CH$_2$CHNC) 
and give column density ratios between the cyanide and isocyanide isomers obtaining
a $N$(CH$_3$NC)/$N$(CH$_3$CN) ratio of 0.002.
}
{Laboratory characterisation of many previously unassigned vibrationally excited states of vinyl cyanide at
microwave to THz frequencies allowed us to detect these molecular species
in Orion-KL. Column density and rotational and
vibrational temperatures for CH$_2$CHCN in their ground and excited states, as
well as for the isotopologues, have been constrained by means of a sample of more than 1000
lines in this survey.
}


\titlerunning{Vibrationally excited Vinyl Cyanide in Orion-KL. }
\authorrunning{A. L\'opez et al.}

\maketitle
%
%________________________________________________________________

\section{Introduction}
\label{sect_int}
The rotational spectrum of vinyl cyanide (CH$_2$CHCN) was studied in 1954 by
Wilcox and collaborators, and somewhat later by \citeauthor{cos59} (\citeyear{cos59})
who also investigated the singly-substituted $^{13}$C species as well as the $^{15}$N and the
CH$_2$CDCN species. This molecule was detected for the first time in the
interstellar medium (ISM) in 1973 toward the Sagittarius B2 (Sgr B2) molecular
cloud (\citeauthor{gar75} \citeyear{gar75}). Since then, CH$_2$CHCN has been
detected toward different sources such as Orion (\citeauthor{sch97}
\citeyear{sch97}), the dark cloud TMC-1 (\citeauthor{mat83} \citeyear{mat83}),
the circumstellar envelope of the late-type star IRC+10216 (\citeauthor{agu08}
\citeyear{agu08}), and the Titan atmosphere (\citeauthor{cap81}
\citeyear{cap81}). CH$_2$CHCN is one of the molecules whose high abundance and
significant dipole moment allow radioastronomical detection even of its
rare isotopologue species. Hence, vinyl cyanide makes an important contribution to
the millimeter and submillimeter spectral emissions covered by high sensitivity
facilities such as ALMA and the Herschel Space Telescope. However, there has not
yet been a comprehensive study of its low-lying vibrational excited states.

Vinyl cyanide is a planar molecule
(six internuclear distances and five independent bond angles) and is a slightly asymmetric prolate
rotor with two non-zero electric dipole moment components, which leads to a rich rotational spectrum.
The first detailed discussion of the vinyl cyanide 
microwave spectrum was in 1973 by \citeauthor{ger73}.  Subsequent studies of
the rotational spectrum of vinyl cyanide resulted in the determination of its
electrical dipole moment components by \citeauthor{sto85} (\citeyear{sto85}); these values
were later improved by \citeauthor{kra11} (\citeyear{kra11}) who reported the
values $\mu_a$=3.821(3)\,D, $\mu_b$=0.687(8)\,D, and $\mu_{\rm TOT}$=3.882(3)\,D.
Additional studies upgraded the molecular structure as
\citeauthor{dem94} (\citeyear{dem94}), \citeauthor{col97} (\citeyear{col97}),  and
\citeauthor{krasnicki11} (\citeyear{krasnicki11}) 
derived successively more refined structural parameters from the rotational constants. The
$^{14}$N nuclear quadrupole hyperfine structure has been studied by
\citeauthor{col97} (\citeyear{col97}), \citeauthor{sto85} (\citeyear{sto85}),
and \citeauthor{bas96} (\citeyear{bas96}). 

\citeauthor{kis09} (\citeyear{kis09}) updated the rotational constants by simultaneously 
fitting the rotational lines of CH$_2$CHCN in its ground and the lowest excited state $\varv_{11}$=1, and also 
fitting the ground states of the $^{13}$C and the $^{15}$N isotopologues. 
More detailed analysis of the isotopologue spectra was later reported by 
\citeauthor{krasnicki11} (\citeyear{krasnicki11}).
The ground state rotational $a$-type
and $b$-type transitions of the parent vinyl cyanide have been assigned up to
$J$=129 with measurements in the laboratory reaching 1.67 THz
(\citeauthor{kis09} \citeyear{kis09}). They showed the influence of
temperature on the partition function and consequently on the spectrum of vinyl
cyanide. Fig.1 of \citeauthor{kis09} (\citeyear{kis09}) identifies this effect and
the dominance of the millimeter and submillimeter region by the $^a$$R$-type
transitions. However, at high frequencies (THz region) the $b$-type $R$-branch
rotational transitions are one order of magnitude more intense than those of
$a$-type due to smaller values of the rotational quantum numbers $J$.

The rotational transitions of CH$_2$CHCN in several of the
lowest vibrational excited states, $\varv_{11}$=1,2,3 and $\varv_{15}$=1, were
assigned by \citeauthor{caz88} (\citeyear{caz88}), and the measurements were
extended by \citeauthor{dem94} (\citeyear{dem94}) ($\varv_{11}$=1 and the ground
state). The data for $\varv_{11}$=3 was more limited hindering the determination
of all sextic or even quartic constants. Recently, the analysis of
broadband rotational spectra of vinyl cyanide revealed that there are perturbations 
between all pairs 
of adjacent vibrational states extending upwards from the ground state (g.s.), see Fig.2 of
\citeauthor{kis09} (\citeyear{kis09}). \citeauthor{kis12} (\citeyear{kis12})
covered a broader frequency region (90-1900 GHz), identifying and fitting the perturbations in
frequencies of rotational transitions due to $a$-, $b$- or $c$-axis Coriolis-type or Fermi
type interactions between the four lowest states of vinyl cyanide (g.s., $\varv_{11}$=1, 
$\varv_{15}$=1, and $\varv_{11}$=2).  The need for perturbation treatment of 
the $\varv_{10}$/$\varv_{11}\varv_{15}$
dyad at about 560 cm$^{-1}$ and the 3$\varv_{11}$/2$\varv_{15}$/$\varv_{14}$
triad of states at about 680 cm$^{-1}$ was also identified, and initial results for 
the dyad were reported in
\citeauthor{kisosu11} (\citeyear{kisosu11}). 
Thus a meticulous analysis aiming towards an eventual 
global fit of transitions in all states of vinyl cyanide is necessary. The
low resolution, gas-phase infrared spectrum of vinyl cyanide and its vibrational normal modes were studied
by \citeauthor{hal48} (\citeyear{hal48}) and by \citeauthor{khl99} (\citeyear{khl99}).  Partial rotational resolution
of the vibration-rotation spectrum of the two lowest wavenumber modes was also reported in the far-infrared study
by \citeauthor{col73} (\citeyear{col73}).

The first detection in the ISM of vinyl cyanide was in 1973 by means of the 
2$_{11}$-2$_{12}$ line in emission in Sgr B2 and was confirmed in 1975 by \citet{gar75},
suggesting the presence of the simplest olefin in the ISM, CH$_2$=CH$_2$
(ethylene) based on the evidence of the reactive vinyl radical.
\citeauthor{betz81} (\citeyear{betz81}) observed for the first time the 
non-polar organic molecule CH$_2$=CH$_2$ toward the red giant C-rich star IRC+10216,
specifically the $\varv_7$ band in the rotation-vibration spectral region (28
THz). Owing to the symmetry of ethylene the dipole rotational transitions are
forbidden, and \citeauthor{occ13} (\citeyear{occ13}) estimated a column density
of 1.26$\times$10$^{14}$ cm$^{-2}$ in standard hot cores for this molecule based
on the abundance of its derivative molecule, i.e. hydrocarbon methylacetylene
(CH$_3$CCH).

The dense and hot molecular clouds such as Orion and Sgr B2 give rise to
emission lines of vibrationally excited states of vinyl cyanide. Rotational transitions in
the two lowest
frequency modes $\varv_{11}$ and $\varv_{15}$ were detected in Orion by
\citeauthor{sch97} (\citeyear{sch97}) (as tentative detection of 3 and 2 lines,
respectively) and in Sgr B2 by \citeauthor{num99} (\citeyear{num99}) (64 and 45
identified lines, respectively). The latter authors also made the tentative
detection transitions in the $2\varv_{11}$ mode (5 lines). Recently, \citeauthor{bel13}
(\citeyear{bel13}) detected six vibrational states in a line survey of Sgr B2(N)
($\varv_{11}$=1,2,3, $\varv_{15}$=1,2, $\varv_{11}$=$\varv_{15}$=1) among which they
detected the higher-lying vibrational states for the first time in space.

On the other hand, the ground states of rare isotopologues have been well characterized in the
laboratory (\citeauthor{col97} \citeyear{col97}, \citeauthor{mul08} \citeyear{mul08},
\citeauthor{kis09} \citeyear{kis09}, \citeauthor{krasnicki11} \citeyear{krasnicki11}). 
All monosubstituted species containing $^{13}$C,
$^{15}$N, and D, as well as those of all $^{13}$C-monosubstituted species of
H$_2$C=CDCN, of both cis- and trans- conformers of HDC=CHCN, HDC=CDCN, and
D$_2$C=CDCN have been characterized. The double $^{13}$C and $^{13}$C$^{15}$N species have also 
been assigned by \citeauthor{krasnicki11} (\citeyear{krasnicki11}).
The detection of $^{13}$C species of vinyl cyanide in the ISM was
carried out toward Sgr B2 by \citeauthor{mul08} (\citeyear{mul08}) with 26
detected features.

The millimeter line survey of Orion-KL carried out with the IRAM 30-m telescope
by Tercero and collaborators (\citeauthor{Tercero10} \citeyear{Tercero10};
\citeauthor{Tercero11} \citeyear{Tercero11}; \citeauthor{the12}
\citeyear{the12}) presented initially 8000 unidentified lines.
Many of these features (near 4000) have been subsequently identified as lines
arising from isotopologues and vibrationally excited states of abundant species
such as ethyl cyanide and methyl formate thanks to a close collaboration with
different spectroscopic laboratories (\citeauthor{dem07} \citeyear{dem07};
\citeauthor{mar09} \citeyear{mar09}; \citeauthor{car09} \citeyear{car09};
\citeauthor{mar10} \citeyear{mar10}; \citeauthor{Tercero12}
\citeyear{Tercero12}; \citeauthor{mot12} \citeyear{mot12}; \citeauthor{adm13}
\citeyear{adm13}, \citeauthor{cou13} \citeyear{cou13}, \citeauthor{hay14}
\citeyear{hay14}). In this work we followed the procedure of our previous
papers, searching for all isotopologues and vibrationally excited states of
vinyl cyanide in this line survey. These identifications are essential to probe
new molecular species which contribute to reduce the number of U-lines and helps
to reduce the line confusion in the spectra. At this point we were ready to
begin the search for new molecular species in this cloud providing
clues to the formation of complex organic molecules on the grain surfaces and/or
in the gas phase (see the discovery of methyl acetate and $gauche$ ethyl formate
in \citeauthor{Ter13} \citeyear{Ter13}, the detection of the ammonium ion in
\citeauthor{Cer13} \citeyear{Cer13}, and the first detection of ethyl mercaptan in 
\citeauthor{kol14} \citeyear{kol14}).

%===============================================================================
%  FIGURE: energy levels
%===============================================================================
\begin{figure}
\centering
\resizebox{0.9\hsize}{!}{\includegraphics[angle=0,width=0.4\textwidth]{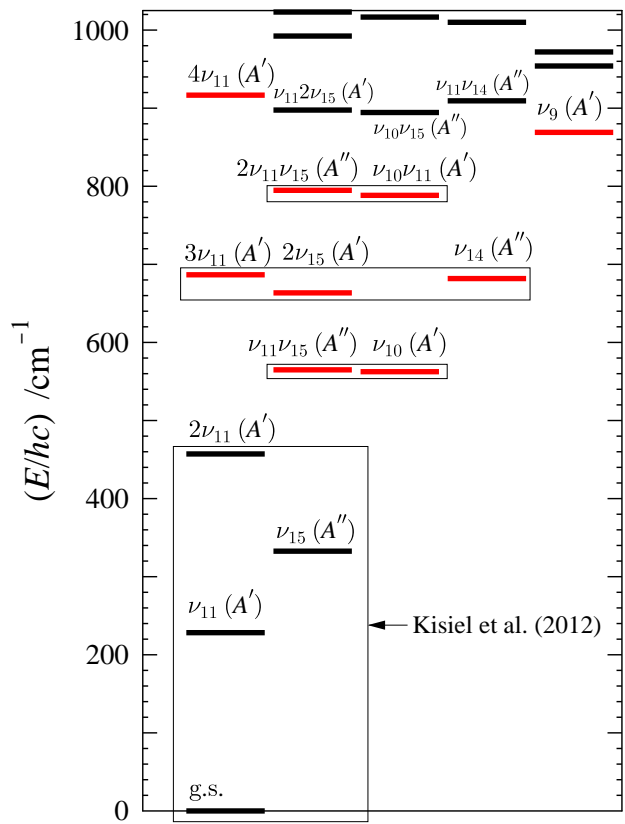}}
\caption{All vibrational levels of vinyl cyanide up to 1000 cm$^{-1}$.
The levels in red are those for which rotational transitions have
been analysed in this work.  The boxes identify sets of levels treated by means of 
coupled fits accounting for interstate perturbations.}
\label{fig_vincne}
\end{figure}
%===============================================================================

We report extensive characterization of 9 different 
excited vibrational states of vinyl cyanide (see Fig.\,\ref{fig_vincne}) positioned in energy 
immediately above $\varv_{11}=2$, which, up to this 
point, has been the highest vibrational state subjected to a 
detailed study (\citeauthor{kis12}, \citeyear{kis12}).
The assignment is confirmed by using the Stark modulation spectrometer of the spectroscopic laboratory
(GEM) of the University of Valladolid and ab initio calculations.
The new
laboratory assignments of $\varv_{11}$=2, $\varv_{11}$=3, and
$\varv_{10}$=1$\Leftrightarrow$($\varv_{11}$=1,$\varv_{15}$=1) vibrational modes of vinyl cyanide were used
successfully to identify these three states in Orion-KL, the latter for the
first time in the ISM. We also detected the $\varv_{11}$=1 and $\varv_{15}$=1
excited states in Orion-KL, as well as the ground state, the $^{13}$C
isotopologues (see Sect.\ref{sect_det}).

Because isomerism is a key issue for a more accurate understanding of
the formation of interstellar molecules, we report observations of some related
isocyanide isomers. 
\citeauthor{bol70} (\citeyear{bol70}) carried out the first laboratory
study of the pure rotation (10-40 GHz) spectrum of vinyl isocyanide
and also studied its 200-4400 cm$^{-1}$ vibrational spectrum.
Laboratory measurements were subsequently extended up to 175 GHz by
\citeauthor{yam75} (\citeyear{yam75}) and the
hyperfine structure of some cm-wave lines was measured by
\citeauthor{bes82} (\citeyear{bes82}).
In Section \ref{isoCN}, we searched for
all isocyanides corresponding to the detected cyanides in Orion-KL: methyl
cyanide (\citeauthor{tom14} \citeyear{tom14}), ethyl cyanide (\citeauthor{adm13}
\citeyear{adm13}), cyanoacetylene (\citeauthor{esp13b} \citeyear{esp13b}),
cyanamide, and vinyl cyanide.  In this study, we have tentatively detected vinyl isocyanide (CH$_2$CHNC)
in Orion-KL (see Section \ref{isoCN}). In addition, we
observed methyl isocyanide (CH$_3$NC) for the first time in Orion-KL 
$-$firstly observed by \citeauthor{cer88}
(\citeyear{cer88}) in the Sgr B2(OH) source$-$, and we provide a tentative
detection of ethyl isocyanide and isomers HCCNC and HNCCC of isocyanoacetylene.
After the detection of cyanamide (NH$_2$CN) by \citeauthor{tur75}
(\citeyear{tur75}) in Sgr B2, we report the tentative detection of this molecule
in Orion, as well as a tentative detection for isocyanamide.

Finally, in Sect. \ref{disc} and \ref{summ} we discuss and summarize all results.

\section{Experimental}
\label{sect_exp}

The present spectroscopic analysis is based largely on the broadband rotational 
spectrum of vinyl cyanide compiled from segments recorded in several different 
laboratories.  
That spectrum provided a total of 1170 GHz of coverage
and its makeup was detailed in Table\,1 of 
\citeauthor{kis12} (\citeyear{kis12}). In the present work the previous spectrum has been complemented by 
two additional segments: 50-90 GHz and 140-170 GHz, recorded at GEM by 
using cascaded multiplication of microwave synthesizer output.  
The addition of these segments provides
practically uninterrupted laboratory coverage of the room-temperature rotational 

spectrum of vinyl cyanide over the 50-640
 GHz region, which is key to the analysis 
of vibrational state transitions.

\begin{figure*}[!ht] 
\centering
\includegraphics[angle=270,width=1.0\textwidth]{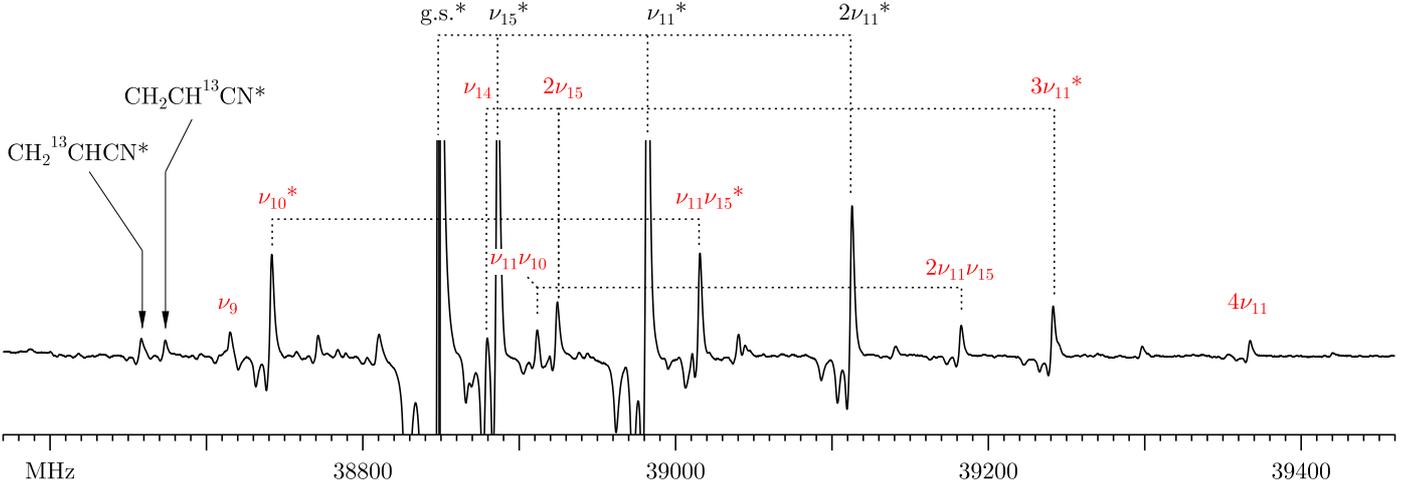}
\caption{The room-temperature laboratory spectrum of vinyl cyanide in the region 
of the $4_{13}-3_{12}$ rotational transition recorded with a Stark modulation 
spectrometer.  All marked lines are for the $4_{13}-3_{12}$ transition in a given vibrational
or isotopic species and display a characteristic pattern of negative lobes due to the non-zero 
field cycle of Stark modulation.
Dotted lines connect vibrational states analysed as 
perturbing polyads, red denotes vibrational states analysed in the present work, and asterisks 
identify states detected presently in Orion-KL.  It can be seen that laboratory analysis 
is now available for excited vibrational state transitions that are comparable in room-temperature intensity 
to those for $^{13}$C isotopologues in terrestrial natural abundance.} 
\label{fig_starkspec}
\end{figure*}

Another laboratory technique brought in by GEM is Stark spectroscopy at 
cm-wave frequencies. The Stark-modulation technique has the useful property of preferentially 
recording a given low-$J$ rotational transition by a suitable choice of the 
modulation voltage.  This is particularly the case for the lowest-$J$, $K_a$=1 
transitions.  Due to asymmetry splitting these transitions are significantly shifted in 
frequency relative to other transitions for the same $J$ value.
An example spectrum of this type is shown in 
Fig.~\ref{fig_starkspec} where all, but some of the weakest lines, correspond to the 
$4_{13}-3_{12}$ transition in either a vibrational state of the parent vinyl 
cyanide or in the ground state of an isotopic species. Such spectra are particularly 
useful for initial assignment since vibrationally induced frequency differences from the 
ground state are near additive. Relative intensities of transitions also give an immediate measure
of relative population of assigned vibrational states and isotopic species.

The analysis of the spectra was carried out with the AABS graphical  
package for {\it Assignment and Analysis of Broadband Spectra} 
(\citeauthor{kis05}, \citeyear{kis05}, \citeyear{kis12})
freely available on the PROSPE 
database (\citeauthor{kis01}, \citeyear{kis01})\footnote{http://info.ifpan.edu.pl/{\~{ }}kisiel/prospe.htm}.  
The AABS package was complemented
by the SPFIT/SPCAT program package (\citeauthor{pic91}, \citeyear{pic91})\footnote{http://spec.jpl.nasa.gov}
used for setting up the Hamiltonian, fitting and prediction.

Supporting ab initio calculations were carried out with 
GAUSSIAN 09\footnote{Frisch, M. J.; Trucks, G. W.; Schlegel, et al.,
Gaussian 09, Revision B.01; Gaussian: Wallingford, CT, 2010.}
 and 
CFOUR\footnote{Stanton, J. F., Gauss, J.; Harding, M. E. et al.,
CFOUR, a quantum chemical quantum package with
integrated packages MOLECULE (Almlöf, J.; Taylor, P. R.) and ECP
routines (Mitin A. V.; Wüllen, van C.), http://www.cfour.de}
packages.  The key parameters for vibrational assignment are vibrational changes in rotational
constants, which require relatively lengthy anharmonic force field calculations.  
Two strategies were used for this purpose: a relatively long basis set combined with a basic electron 
correlation correction (MP2/6-311++G(d,p)) and a more thorough correlation correction
with a relatively simple basis set (CCSD(T)/6-31G(d,p)).  The final results minimally favored the second 
approach but, in practice, both were found to be equally suitable.

%===============================================================================
%  FIGURE: Stark spectrum
%===============================================================================
%===============================================================================

\section{Laboratory spectroscopy}

\subsection{Analysis of the excited vibrational states}

An overview of the results of the spectroscopic analysis is provided in Table\,\ref{tab_specsum} and the 
determined spectroscopic constants necessary for generating linelists are given
in Table \ref{tab_const_dyads}, and \ref{tab_const_triad}-\ref{tab_const_minor}.

%===============================================================================
%  TABLE: dataset summary
%===============================================================================

\begin{table*}[ht]
\caption{Spectroscopic data sets for excited vibrational states of CH$_2$CHCN acquired in this work.}
\label{tab_specsum}
\renewcommand{\thefootnote}{\alph{footnote}}
%
%
 %{\small{\sf{\scriptsize
\begin{center}
\begin{tabular}{lk{5}cccccccc}

\hline\vspace{-0.2cm}\\
  \multicolumn{1}{c}{ excited state                }    &
  \multicolumn{1}{l}{ $E_{\rm vib}$$^a$            }    & 
  \multicolumn{1}{l}{ $\Delta E$$^b$               }    &  
  \multicolumn{1}{c}{ $N_{\rm fitted}$$^c$         }    &
  \multicolumn{1}{c}{  $N_{\rm unfitted}$$^d$      }    &
  \multicolumn{1}{c}{ $\sigma_{\rm fit}$$^e$       }    &
  \multicolumn{1}{c}{ $\sigma_{\rm rms}$$^f$       }    &
  \multicolumn{1}{c}{ $J$ range                    }    &
  \multicolumn{1}{c}{ $K_a$ range                  }    &
  \multicolumn{1}{c}{ frequency range$^g$          }    \\
  \multicolumn{1}{c}{                              }    &
  \multicolumn{1}{l}{  (cm$^{-1}$)                 }    & 
  \multicolumn{1}{l}{  (cm$^{-1}$)                 }    &  
  \multicolumn{1}{c}{                              }    &
  \multicolumn{1}{c}{                              }    &
  \multicolumn{1}{c}{  (MHz)                       }    &
  \multicolumn{1}{c}{                              }    &
  \multicolumn{1}{c}{                              }    &
  \multicolumn{1}{c}{                              }    &
  \multicolumn{1}{c}{  (GHz)                       }    \\
\vspace{-0.3cm}\\
\hline\\                                                                                                                                                                                     
$\varv_{10}        $       &    560.5          &    0           &    2135$^h$          &      55                &     0.324               &     1.446               & ~2 - ~99   &   0 -22      &  37.0 - 1893.4       \\
$\varv_{11}\varv_{15}$       &    562.9          &    2.391494(5) &    1837$^h$          &     136                &     0.382               &     1.872               & ~3 - 100   &   0 -20      &  39.0 - 1783.5       \\
\\                                                                                                                                                                                         
$2\varv_{15}$              &    663.5          &    0           &    1329$^i$          &      52                &     0.265               &     1.980               & ~1 - ~70   &   0 -17      &  18.6 - 1191.3       \\
$\varv_{14}$               &    681.8          &    18.31812(2) &    1287$^i$          &      53                &     0.228               &     1.467               & ~5 - ~70   &   0 -18      &  58.3 - 1891.1       \\
$3\varv_{11}$              &    686.6          &    23.16415(3) &    1250$^i$          &      81                &     0.309               &     2.329               & ~2 - ~69   &   0 -17      &  28.0 - 1196.5       \\
\\                                                                                                                                                                                                                 
$\varv_{10}\varv_{11}$       &    787.5          &    0           &     842$^j$          &       3                &     0.137               &     1.289               & ~3 - ~68   &   0 -12      &  37.1 - ~639.3       \\
$2\varv_{11}\varv_{15}$      &    793.9          &    6.44502(3)  &     860$^j$          &       7                &     0.164               &     1.551               & ~3 - ~69   &   0 -12      &  37.3 - ~640.0       \\
\\                                                                                                                                                                                                                   
$\varv_9$                  &    869.0          &                &      373             &       7                &     0.167               &     1.665               & ~1 - ~63   &   0 -7       &  18.5 - ~570.3       \\
4$\varv_{11}$              &    916.7          &                &      225             &      17                &     0.250               &     2.496               & ~3 - ~43   &   0 -5       &  37.4 - ~410.9       \\
\hline 
%\flushleft

\end{tabular}
\end{center}
%}}
%}

{\bfseries Notes.} $^{(a)}$Estimated vibrational energy (see text in Sect. \ref{vib_energies}).
$^{(b)}$Energy difference relative to the lowest level in the relevant polyad obtained from the perturbation analysis.
$^{(c)}$The number of distinct frequency fitted lines.
$^{(d)}$The number of confidently assigned lines rejected from the fit at the 10$\sigma$ cutoff criterion.
$^{(e)}$Deviation of fit for the vibrational subset.
$^{(f)}$Unitless deviation of fit for the vibrational subset.
$^{(g)}$Frequency coverage of transitions in the data set.
$^{(h),(i),(j)}$Transitions fitted jointly in a single fit accounting for interstate perturbations. 

\end{table*}
%===============================================================================

%-------------------------------------------------------------------------------
%  TABLE: diagonal spectroscopic constants for the dyads
%-------------------------------------------------------------------------------

\begin{table*}
%\section{\small{Online Tables and Figures}}
% \subfloat[][]{}
\caption{Spectroscopic constants in the diagonal blocks of the Hamiltonian for
the $\varv_{10}\Leftrightarrow \varv_{11}\varv_{15}$ and
the $\varv_{11}\varv_{10}\Leftrightarrow 2\varv_{11}\varv_{15}$ dyads of vibrational states in vinyl cyanide compared
with those for the ground state.}
\label{tab_const_dyads}
\renewcommand{\thefootnote}{\alph{footnote}}
%
%
%\begin{small}
\begin{center}
\begin{tabular}{lk{10}k{10}k{10}k{10}k{10}}
\hline\hline\vspace{-0.2cm}\\
     &  \multicolumn{1}{c}{ground state}          &
        \multicolumn{1}{c}{$\varv_{10}$}            &
        \multicolumn{1}{c}{$\varv_{11}\varv_{15}$}    &
        \multicolumn{1}{c}{$\varv_{11}\varv_{10}$}    &
        \multicolumn{1}{c}{$2\varv_{11}\varv_{15}$}      \\
\vspace{-0.2cm}\\
\hline
%                                  gs                   v10=1                       v11=1 v15=1               v11=1 v10=1              v11=2 v15=1    
%                                                                                                                                                     
                           &                     &                         &                        &                        &                           \\
$A$/MHz                    & 49850.69655(43)^a   &   49550.03(63)          &    49890.72(61)        &    48861.72(62)        &    49124.87(56)           \\
$B$/MHz                    &  4971.212565(37)    &    4965.6692(98)        &     4992.6723(70)      &     4984.979(32)       &     5011.494(25)          \\
$C$/MHz                    &  4513.828516(39)    &    4509.6228(13)        &     4531.6029(13)      &     4517.9357(31)      &     4540.0924(32)         \\
                           &                     &                         &                        &                        &                           \\
$\Delta_J$/kHz             &     2.244058(13)    &       2.20646(19)       &        2.26839(18)     &        2.24034(23)     &        2.28278(27)        \\
$\Delta_{JK}$/kHz   ~~~~~  &   -85.6209(35)      &     -89.854(83)         &      -80.615(83)       &      -88.79(17)        &      -63.97(17)           \\
$\Delta_K$/kHz             &  2715.4213(94)      &    2591.5(31)           &     2522.4(31)         &     2225.(16)          &     1842.(15)             \\
$\delta_J$/kHz             &     0.4566499(32)   &       0.44642(11)       &        0.465487(70)    &        0.46094(18)     &        0.47422(18)        \\
$\delta_K$/kHz             &    24.4935(22)      &      22.099(24)         &       25.225(14)       &       25.212(82)       &       24.683(96)          \\
                           &                     &                         &                        &                        &                           \\
$\Phi_{J}$/Hz              &     0.0064338(17)   &       0.006345(26)      &        0.006244(26)    &        0.006038(38)    &        0.005952(39)       \\
$\Phi_{JK}$/Hz             &    -0.00425(40)     &       0.0541(96)        &        0.0324(86)      &       -0.126(17)       &       -0.244(23)          \\
$\Phi_{KJ}$/Hz             &    -7.7804(39)      &      -5.74(11)          &       -5.18(11)        &        0.59(23)        &        1.52(22)           \\
$\Phi_{K}$/Hz              &   384.762(63)       &     399.73(71)          &      -86.8(11)         &      428.(396)         &    -1858.(389)            \\
$\phi_{J}$/Hz              &     0.00236953(79)  &       0.002405(22)      &        0.0021005(36)   &        0.002185(23)    &        0.002136(22)       \\
$\phi_{JK}$/Hz             &     0.14283(40)     &       0.1151(27)        &        0.1698(18)      &        0.145(13)       &        0.135(14)          \\
$\phi_{K}$/Hz              &    37.011(58)       &      51.4(12)           &       38.0(11)         &       17.1(27)         &       -5.6(38)            \\
                           &                     &                         &                        &                        &                           \\
$L_J$/mHz                  &    -0.000026315(71) &      -0.0000263(15)     &       -0.0000202(14)   &      [ 0.]             &      [ 0.]                \\

$L_{JJK}$/mHz              &    -0.001077(29)    &      -0.01178(86)       &       -0.00659(91)     &      [ 0.]             &      [ 0.]                \\
$L_{JK}$/mHz               &     0.4279(30)      &      -0.0703(85)        &      [ 0.]             &      [ 0.]             &      [ 0.]                \\
$L_{KKJ}$/mHz              &     0.012(12)       &       4.00(18)          &       -9.63(17)        &      [ 0.]             &      [ 0.]                \\
$L_{K}$/mHz                &   -61.41(17)        &     -55.6(29)           &      462.9(45)         &      [ 0.]             &      [ 0.]                \\
$l_J$/mHz                  &    -0.000011602(36) &      -0.0000165(13)     &      [ 0.]             &      [ 0.]             &      [ 0.]                \\
$l_{JK}$/mHz               &    -0.000956(20)    &     [ 0.]               &      [ 0.]             &      [ 0.]             &      [ 0.]                \\
$l_{KJ}$/mHz               &    -0.1436(46)      &      -1.79(11)          &       -0.86(12)        &      [ 0.]             &      [ 0.]                \\
$l_{K}$/mHz                &     8.91(18)        &     [ 0.]               &        9.21(43)        &      [ 0..]             &      [ 0.]                \\
                           &                     &                         &                        &                        &                           \\
$P_{KJ}$/mHz               &    -0.0000156(31)   &      -0.000147(14)      &      [ 0.]             &      [ 0.]             &      [ 0.]                \\
$P_{KKJ}$/mHz              &    -0.0001977(57)   &     [ 0.]               &      [ 0.]             &      [ 0.]             &      [ 0.]                \\
$P_{K}$/mHz                &     0.00867(15)     &       0.0286(23)        &       -0.3457(49)      &      [ 0.]             &      [ 0.]                \\
                           &                     &                         &                        &                        &                           \\
$\Delta E$$^b$/MHz         &                     &        0.0              &    71695.20(16)        &         0.0            &    193216.69(90)          \\
$\Delta E$/cm$^{-1}$       &                     &        0.0              &        2.391494(5)     &         0.0            &         6.44502(3)        \\
                           &                     &                         &                        &                        &                           \\
$N_{\rm lines}$$^c$        &                                                                                                                          
          \multicolumn{1}{l}{~~4490,0}           &                             
                               \multicolumn{1}{l}{~~~~2135,55~}            &     
                                                          \multicolumn{1}{l}{~~1837,136}             &  
                                                                                   \multicolumn{1}{l}{     842,3  }          &
                                                                                                              \multicolumn{1}{l}{     860,7  }          \\ 
$\sigma_{\rm fit}$$^d$/MHz &     0.144           &        0.324^e          &     0.382^e            &        0.137^f         &        0.164^f           \\
$\sigma_{\rm rms}$$^d$     &     0.713           &        1.446            &     1.872              &        1.289           &        1.551             \\
\vspace{-0.3cm}\\ 
\hline 
\\                      
\end{tabular}                                
\end{center}

{\bfseries Notes.} $^{(a)}$Round parentheses enclose standard errors in units
 of the last quoted digit of the value of the constant, square parentheses enclose assumed values.\\
$^{(b)}$The fitted vibrational energy difference relative to the lowest vibrational state in the respective dyad.\\
$^{(c)}$The number of distinct frequency fitted lines and the number of lines rejected at the 10$\sigma$ fitting criterion
of the SPFIT program.\\
$^{(d)}$Deviations of fit for the different vibrational subsets.\\
$^{(e)}$The coupled fit for the $\varv_{10}\Leftrightarrow \varv_{11}\varv_{15}$ dyad encompasses 3978 lines, 
at an overall $\sigma_{\rm fit}$ of 0.352 MHz and requires also the use of constants reported in 
Table\,\ref{tab_constoff_dyads}.\\
$^{(f)}$The coupled fit for the $\varv_{11}\varv_{10}\Leftrightarrow 2\varv_{11}\varv_{15}$ dyad encompasses 1702 lines, 
at an overall $\sigma_{\rm fit}$ of 0.151 MHz and requires also the use of constants reported in 
Table\,\ref{tab_constoff_dyads}.\\

%\end{small}                               
%
%
\end{table*}

Initial assignment was based on a combination of several techniques: inspection 
of Stark spectra such as that in Fig.\,\ref{fig_starkspec}, the use of the 
concept of harmonic behaviour of rotational constant changes on vibrational 
excitation (linear additivity of changes), and ab initio calculations of 
vibration-rotation constants. The final assignment of vibrational states is 
confirmed by the comparison of values of experimental vibration-rotation changes 
in rotational constants relative to the ground state with computed ab initio 
values, as listed in Table\,\ref{tab_vibrot}, that are available online.

Preliminary studies revealed a multitude of perturbations in rotational 
frequencies that necessitated the use of fits that accounted for interactions between 
vibrational states.   The grouping of energy levels visible in 
Fig.\,\ref{fig_vincne} suggested that above the last state studied in detail, 
namely 2$\varv_{11}$, it was possible to break the treatment down into three 
isolated polyads.  The symmetry classification of vibrational states ($A'$ and 
$A''$, $C_{s}$ point group) is marked in Fig.\,\ref{fig_vincne} and states of 
different symmetry need to be connected by $a$- and $b$-type Coriolis
interactions, while states of the same symmetry are coupled via $c$-type 
Coriolis and Fermi interactions. The Hamiltonian and the techniques of analysis 
used to deal with this type of problem have been described in detail in 
\citeauthor{kis12} (\citeyear{kis09}, \citeyear{kis12}). This type of analysis 
is far from trivial, but its eventual success for the polyads near 560, 680, and 
790 cm$^{-1}$ is confirmed in Table\,\ref{tab_specsum} by the magnitudes of the deviations of fit 
in relation to the numbers of fitted lines and their broad frequency coverage.
In the most extensive of the present analyses, that for the 
$\varv_{10}$=1$\Leftrightarrow$($\varv_{11}$=1,$\varv_{15}$=1) dyad the fit encompasses almost 4000 
lines and, in addition to $^aR$-type transitions includes $^bQ$- and $^bR$-types. 
We use the 10$\sigma$ cutoff criterion of SPFIT to prevent lines perturbed by 
factors outside the model from unduly affecting the fit and a moderate number of 
such lines (191) are rejected for this dyad.  These are confidently assigned 
lines, generally in high-$J$ tails of some transition sequences for higher 
values of $K_a$, but their incompatibility suggests that there is hope for a final 
global fit with coupling between the polyads. At the present stage 
the success of the perturbation fits is further reflected by additive 
vibrational changes in values of quartic centrifugal distortion 
constants, and by the relative changes in perturbation constants between 
the two dyads listed in Table\,\ref{tab_constoff_dyads}, which are 
similar to those found for the well studied case of ClONO$_2$ 
(\citeauthor{kis09a}, \citeyear{kis09a}).

\begin{figure}[!hb]
\centering
\resizebox{0.9\hsize}{!}{\includegraphics[angle=0,width=0.4\textwidth]{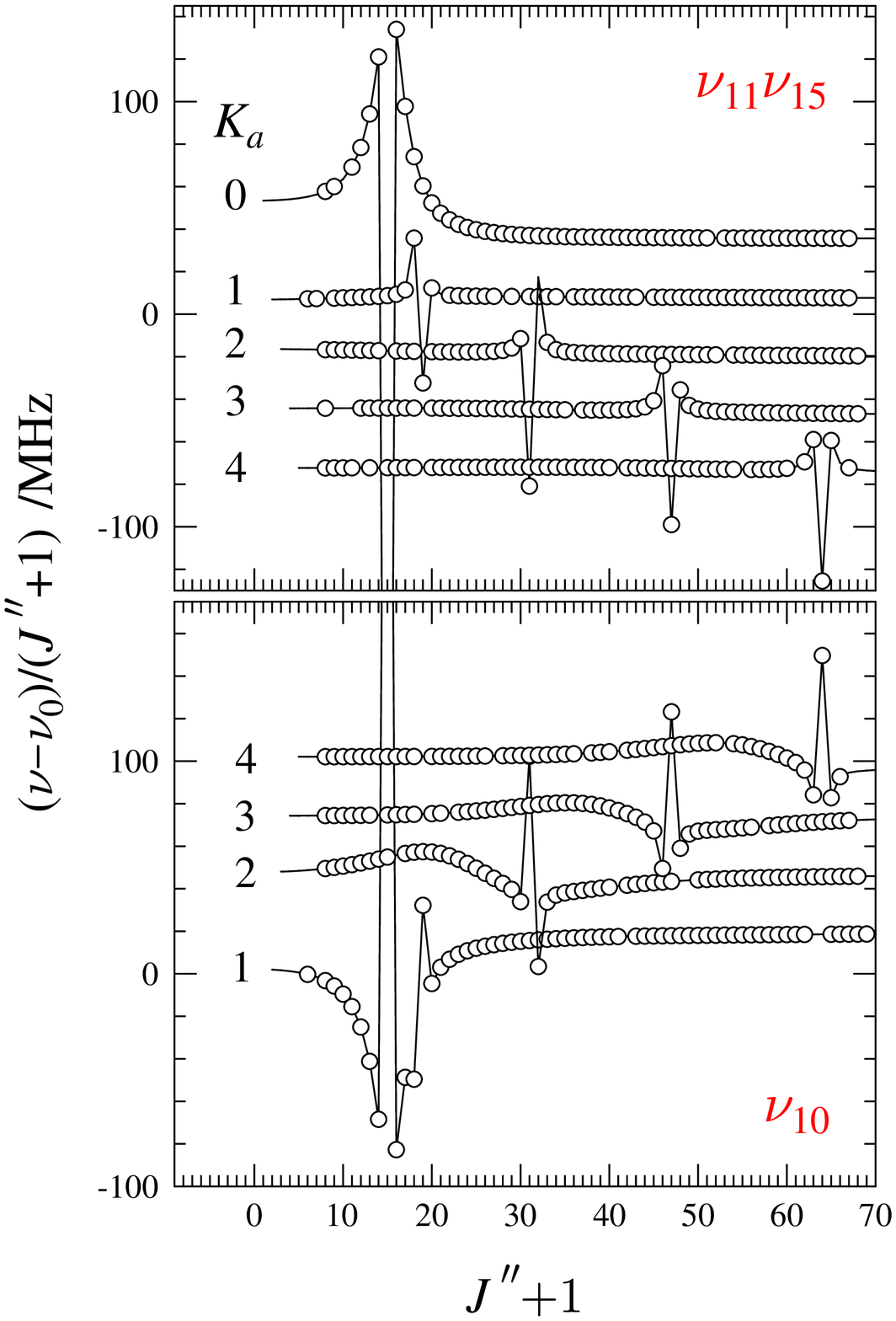}}
\caption{The effect of vibration-rotation perturbations on frequencies of the strongest 
rotational transitions in the $\varv_{10}\Leftrightarrow \varv_{11}\varv_{15}$ dyad of vibrational states.
The plotted quantities are scaled frequency differences relative to the same transitions
in the ground state.  Continuous lines are predictions from the final fit, circles mark assigned lines
and traces in each panel have added vertical shifts to improve clarity.}
\label{fig_vincnp}
\end{figure}

Unlike the situation in the ground state of vinyl cyanide (\citeauthor{kis09}, 
\citeyear{kis09}) the perturbations visible in the presently studied polyads are 
not a spectroscopic curiosity but affect the strongest, low-$K_a$, $^aR$-type 
transitions.  Such transitions occur in the mm- and submm-wave regions which 
are normally the choice for astrophysical 

studies. This effect is illustrated by the scaled plots in Fig.\,\ref{fig_vincnp} which, in 
the absence of perturbations, would have the form of near horizontal, very 
smoothly changing lines.  Perturbations lead to the marked spike shaped features 
in these plots.  Since evaluation of the Hamiltonian is made in separate blocks for 
each value of $J$ the perturbations affecting the two coupling states should have 
mirror image form, as seen in Fig.\,\ref{fig_vincnp}. 
The scaled nature of these plots hides the fact that perturbations to the frequencies of
many lines are considerable.  For example, the peak of the rightmost spike in 
Fig.\,\ref{fig_vincnp} corresponds to a perturbation shift of close to 50$\times$64 MHz,
namely 3.2 GHz.  The frequencies of $^aR$-transitions corresponding to the maximum 
perturbation peaks visible in Fig.\,\ref{fig_vincnp} are 154.1, 183.4, 301.4, 456.5 and 620.8 GHz for $\varv_{10}$,
and 131.9, 174.9, 290.3, 443.3, 604.9 GHz for $\varv_{11}\varv_{15}$. A significant number of
transitions around such peaks are also clearly perturbed. The perturbations are not
limited to frequency but also extend to intensities, which for pure rotation
transitions near the perturbation maxima are often significantly decreased.  The considerable
energy level mixing in such cases leads instead to appearance of transitions between
the perturbing vibrational states.  Such transitions could only be predicted accurately in the final stages
of the perturbation analysis, but were easily found in the compiled broadband laboratory spectrum, and are
explicitly identified in the data files. Fortunately, the linelists generated from perturbation fits 
with the use of the SPCAT program reflect both frequency and intensity perturbations.
Accounting for such effects at laboratory experimental accuracy 
is therefore the key to successful astrophysical studies.

Above the $\varv_{10}\varv_{11}\Leftrightarrow2\varv_{11}\varv_{15}$ dyad the density of 
vibrational states rapidly increases. The complexity of a thorough analysis 
appears to be  too forbidding at this stage but it is possible to check how 
successfully some of these states can be encompassed by single state, effective 
fits. The $\varv_9$ vibrational state seems to be the most isolated and its 
analysis could be taken up to $K_a$=7 and transition frequencies of 570 GHz. In contrast, 
the easy to locate 4$\varv_{11}$ state exhibited very incomplete sequences 
of transitions even at low values of $K_a$, so that its analysis could only be 
taken up to $K_a$=5.  The very fragmentary nature of line sequences for this 
state illustrates the limitations of single state approaches, but it 
nevertheless provides a useful starting point for any future work.
The complete results of fit and the primary data files for the SPFIT program for all coupled and
single-state effective fits 
are available online,\footnote{http://info.ifpan.edu.pl/{\~{ }}kisiel/data.htm} while the
predicted linelists will be incorporated in the JPL database.

\subsection {Vibrational energies}
\label{vib_energies}

In Table\,\ref{tab_specsum} we report a consistent set of vibrational energies 
for the studied excited states of vinyl cyanide, evaluated by taking advantage 
of results from the various perturbation analyses.  The values for $3\varv_{11}$ 
and 4$\varv_{11}$ are from $\varv_{11}$ and the anharmonicity coefficient 
$x_{11,11}$ from \citeauthor{kis12} (\citeyear{kis12}).  
The value for $\varv_{11}\varv_{15}$ comes from 
$\varv_{11}$ and $\varv_{15}$ augmented by $x_{11,15}$ calculated at the 
CCSD(T)/cc-PVDZ level that was benchmarked in \citeauthor{kis12} (\citeyear{kis12}) as the optimum 
level for evaluating this type of constant for vinyl cyanide. 
The remaining vibrational energies in the lower dyad and the 
triad are evaluated using the precise $\Delta E$ values from the perturbation 
analyses.  Finally $\varv_{10}\varv_{11}$ comes from $\varv_{10}$ and $\varv_{11}$ 
augmented by ab initio $x_{10,11}$.  A double check of this procedure is provided by an 
alternative evaluation for 2$\varv_{15}$ based on ab initio $x_{15,15}$, which gives a result 
within 0.5 cm$^{-1}$ of the more reliable tabulated value.
Only the vibrational energy for $\varv_9$ comes from the low resolution gas phase infrared spectrum
(\citeauthor{hal48}, \citeyear{hal48}).

%===============================================================================
%  FIGURE: lower dyad perturbations
%===============================================================================
%===============================================================================

\section{Astronomical detection of vinyl cyanide species}

Thanks to these new laboratory data we identified and detected the
$\varv_{10}$=1$\Leftrightarrow$($\varv_{11}$=1,$\varv_{15}$=1) vibrational modes of CH$_2$CHCN for the first
time in space. A consistent analysis of all detected species of
vinyl cyanide have been made in order to outline the knowledge of our
astrophysical environment. We also report the detection
of methyl isocyanide for the first time in Orion KL and a tentative detection of
vinyl isocyanide, and calculate abundance ratios between the cyanide species and
their corresponding isocyanide isomers.

\subsection{Observations and overall results}

\subsubsection{1D Orion-KL line survey}
\label{sect_obs}

The line survey was performed over three millimeter windows (3, 2, and 1.3 mm)
with the IRAM 30-m telescope (Granada, Spain). The observations were carried out
between September 2004 and January 2007 pointing toward the IRc2 source at
$\alpha$$_{2000.0}$ = 5$^h$ 35$^m$ 14.5$^s$, $\delta$$_{2000.0}$ = $-
5$$^{\circ}$ 22' 30.0''. All the observations were performed using the wobbler
switching mode with a beam throw in azimuth of $\pm$120''. 
System temperatures were in the range of 100-800 K from the
lowest to the highest frequencies. The intensity scale was calibrated using the
atmospheric transmission model (ATM, \citeauthor{cer85} \citeyear{cer85};
\citeauthor{par01a} \citeyear{par01a}). Focus and pointing were checked every
1-2 hours. Backends provided a spectrum of 1-1.25 MHz of spectral resolution. All
spectra were single side band reduced. For further information about
observations and data reduction see \citeauthor{Tercero10}
(\citeyear{Tercero10})\footnote{The data of the IRAM 30-m line survey of Orion-KL 
are available is ascii format on request to B. Tercero and J. Cernicharo and
will be available at the IRAM web page.}.

All figures are shown in main beam temperature ($T_{MB}$) that is related to
the antenna temperature ($T^{*}_{A}$) by the equation:
$T_{MB}$=$T^{*}_{A}$/$\eta_{MB}$, where $\eta_{MB}$ is the main beam efficiency
which depends on the frequency.

According to previous works, we characterize at least four different cloud
components overlapping in the beam in the analysis of low angular resolution line
surveys of Orion-KL (see e. g. \citeauthor{bla87} \citeyear{bla87};
\citeauthor{bla96} \citeyear{bla96}; \citeauthor{Tercero10}
\citeyear{Tercero10}; \citeauthor{Tercero11} \citeyear{Tercero11}): (i) a narrow
($\sim$4\,km\,s$^{-1}$ line-width)  component at
v$_{\rm LSR}$\,$\simeq$\,9\,km\,s$^{-1}$ delineating a north-to-south
\textit{extended ridge} or ambient cloud, a extended region with: $T_k$\,$\simeq$\,60 K,
$n$(H$_2$)\,$\simeq$\,10$^5$\,cm$^{-3}$; (ii) a compact ($d$$_{sou}$$\simeq$15'') and quiescent region,
the \textit{compact ridge}, (v$_{\rm LSR}$\,$\simeq$\,7-8\,km\,s$^{-1}$,
$\Delta$v\,$\simeq$\,3\,km\,s$^{-1}$, $T_k$\,$\simeq$\,150 K,
$n$(H$_2$)\,$\simeq$\,10$^6$\,cm$^{-3}$); (iii) the \textit{plateau} a mixture
of outflows, shocks, and interactions with the ambient cloud
(v$_{\rm LSR}$\,$\simeq$\,6-10\,km\,s$^{-1}$, $\Delta$v\,$\gtrsim$25\,km\,s$^{-1}$,
$T_k$\,$\simeq$\,150\,K, $n$(H$_2$)\,$\simeq$\,10$^6$\,cm$^{-3}$, and $d$$_{sou}$$\simeq$30'');
(iv) a \textit{hot core} component (v$_{\rm LSR}$\,$\simeq$\,5\,km\,s$^{-1}$,
$\Delta$v\,$\simeq$5-15\,km\,s$^{-1}$, $T_k$\,$\simeq$\,250 K,
$n$(H$_2$)\,$\simeq$\,5$\times$10$^7$\,cm$^{-3}$, and $d$$_{sou}$$\simeq$10''). Nevertheless, we found a more
complex structure of that cloud (density and temperature gradients of these
components and spectral features at a v$_{\rm LSR}$ of 15.5 and 21.5\,km\,
s$^{-1}$ related with the outflows) in our analysis of different families of
molecules (see e. g. \citeauthor{Tercero11} \citeyear{Tercero11};
\citeauthor{adm13} \citeyear{adm13}; \citeauthor{esp13a} \citeyear{esp13a}).

\subsubsection{2D survey observations}

We also carried out a two dimensional line survey with the same telescope in the
ranges 85$-$95.3, 105$-$117.4, and 200.4$-$298 GHz (N. Marcelino et al. private communication) 
during 2008 and 2010. This 2D survey consists of maps of
140$\times$140 arcsec$^2$ area with a sampling of 4 arcsec using On-The-Fly
mapping mode with reference position 10 arcminutes West of Orion-KL. The EMIR
heterodyne receivers were used for all the observations except for 220 GHz
frequency setting, for which the HERA receiver array was used. As backend we
used the WILMA backend spectrometer for all spectra (bandwidth of 4 GHz and 2
MHz of spectral resolution) and the FFTS (Fast Fourier Transform Spectrometer,
200\,kHz of spectral resolution) for
frequencies between 245$-$259, 264.4$-$278.6, and 289-298 GHz. Pointing and
focus were checked every 2 hours giving errors less than 3 arcsec. The data were
reduced using the GILDAS package\footnote{http://www.iram.fr/IRAMFR/GILDAS}
removing bad pixels, checking for image sideband contamination and emission from
the reference position, and fitting and removing first order baselines. Six
transitions of CH$_2$CHCN have been selected to study the spatial extent of
their emission with this 2D line survey.

\subsection{Results}
\label{sect_res}

\subsubsection{Detection of CH$_2$CHCN, its vibrationally excited states and its isotopologues in Orion-KL}
\label{sect_det}

Vinyl cyanide shows emission from a large number of rotational lines through the
frequency band 80-280 GHz. The dense and hot conditions of Orion-KL 
populate the low-lying energy excited states. Here, we present the first
interstellar detection of the $\varv_{10}$=1$\Leftrightarrow$($\varv_{11}$=1,$\varv_{15}$=1) vibrational
excited state.

\begin{figure*}[!ht] \centering
\includegraphics[angle=270,width=1.0\textwidth]{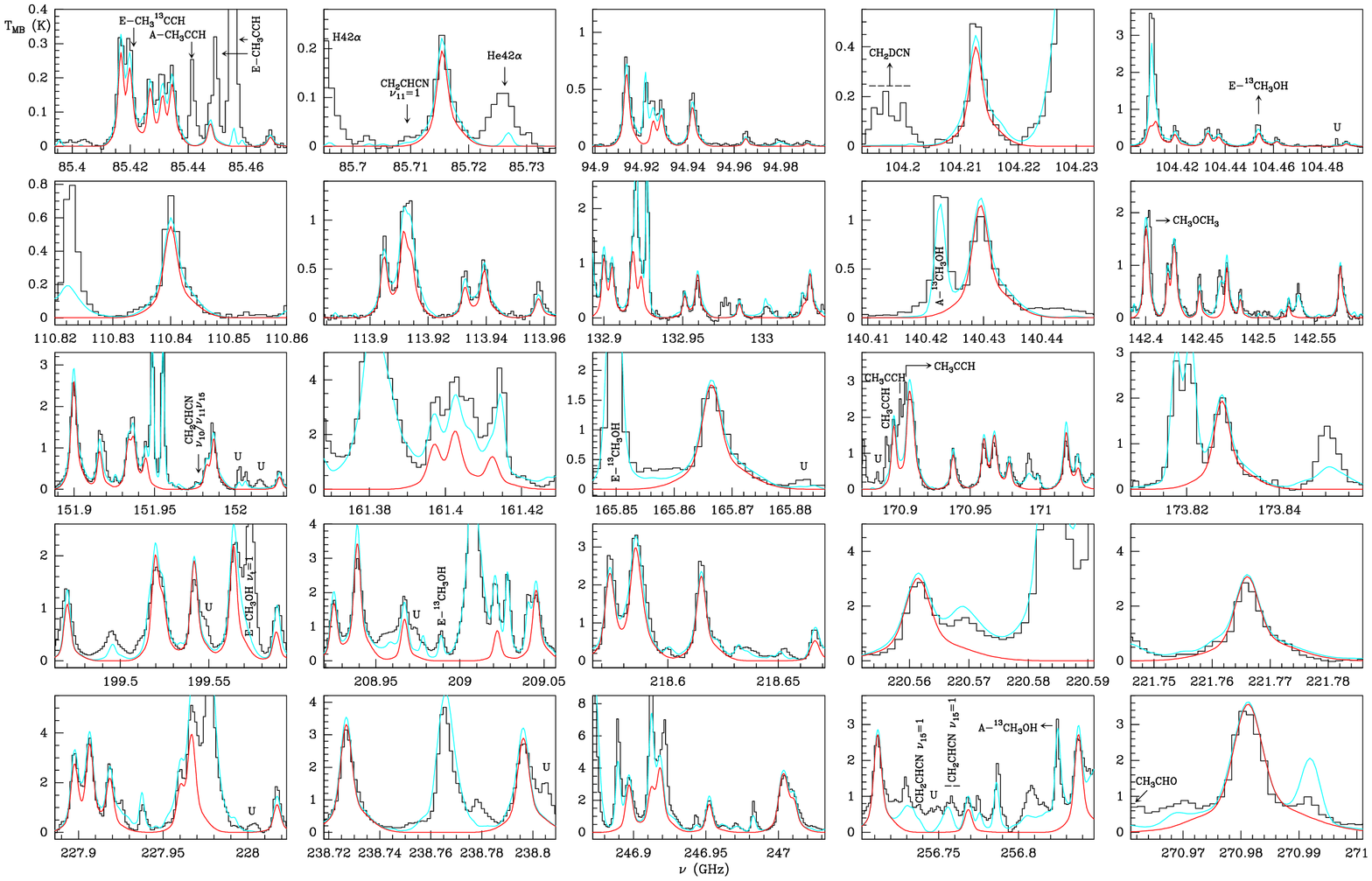}
\caption{Observed lines from Orion-KL (histogram spectra) and model (thin red
curves) of vinyl cyanide in the ground state. The cyan line corresponds to the model of
the molecules we have already studied in this survey (see text Sect. \ref{sec_cd})
including the CH$_2$CHCN species. A v$_{\rm LSR}$ of 5\,km s$^{-1}$ is assumed}. \label{fig_ground}
\end{figure*}

\begin{figure*}[ht]
\centering
\includegraphics[angle=270,width=1.0\textwidth]{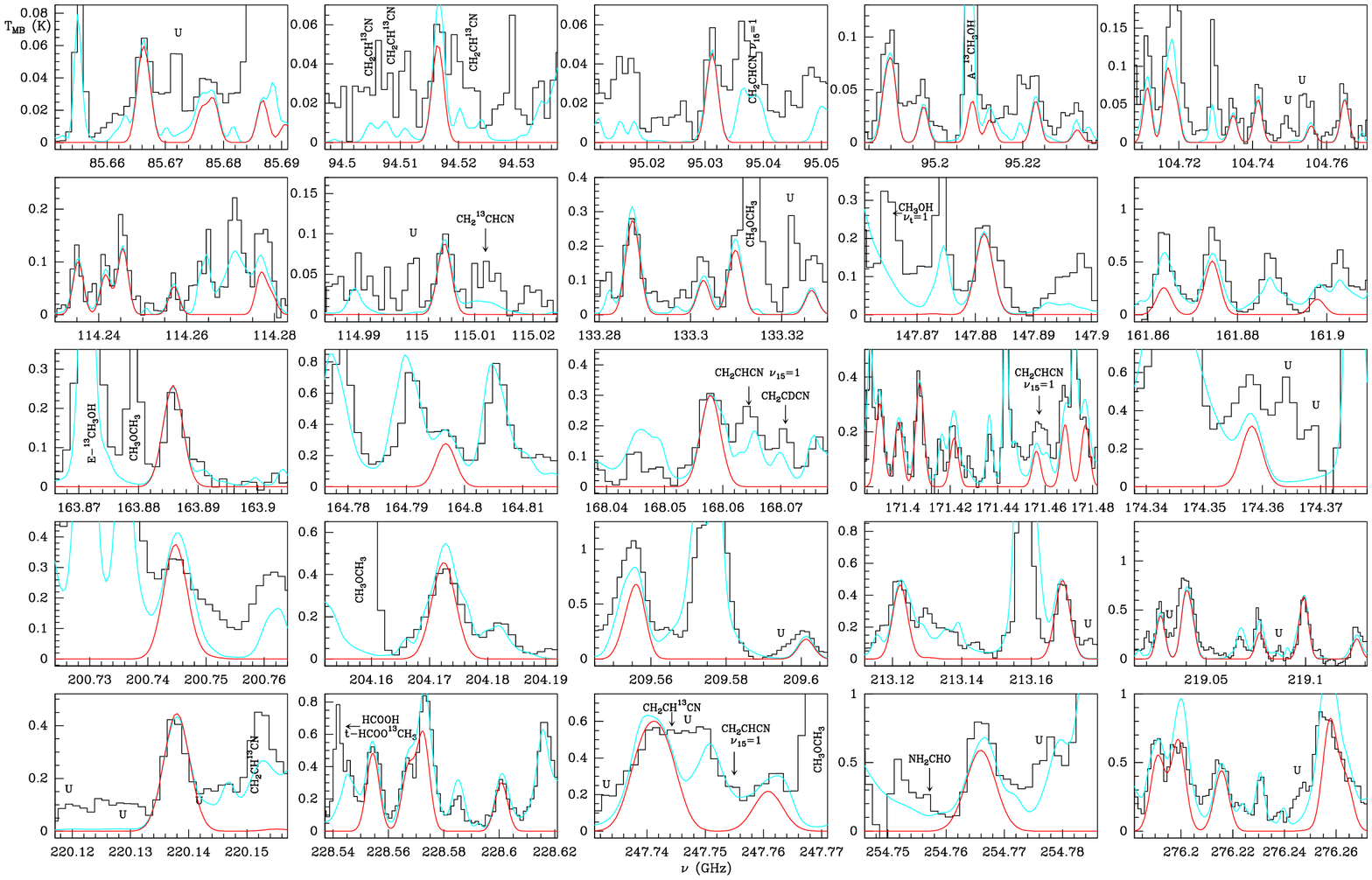}
\caption{Observed lines from Orion-KL (histogram spectra) and model (thin red
curves) of CH$_2$CHCN of $\varv_{11}$=1. The cyan line corresponds to the model of
the molecules we have already studied in this survey (see text Sect. \ref{sec_cd}) 
including the CH$_2$CHCN species.
A v$_{\rm LSR}$ of 5\,km s$^{-1}$ is assumed.}
\label{fig_v11}
\end{figure*}

\begin{figure*}[ht]
\centering
\includegraphics[angle=270,width=0.9\textwidth]{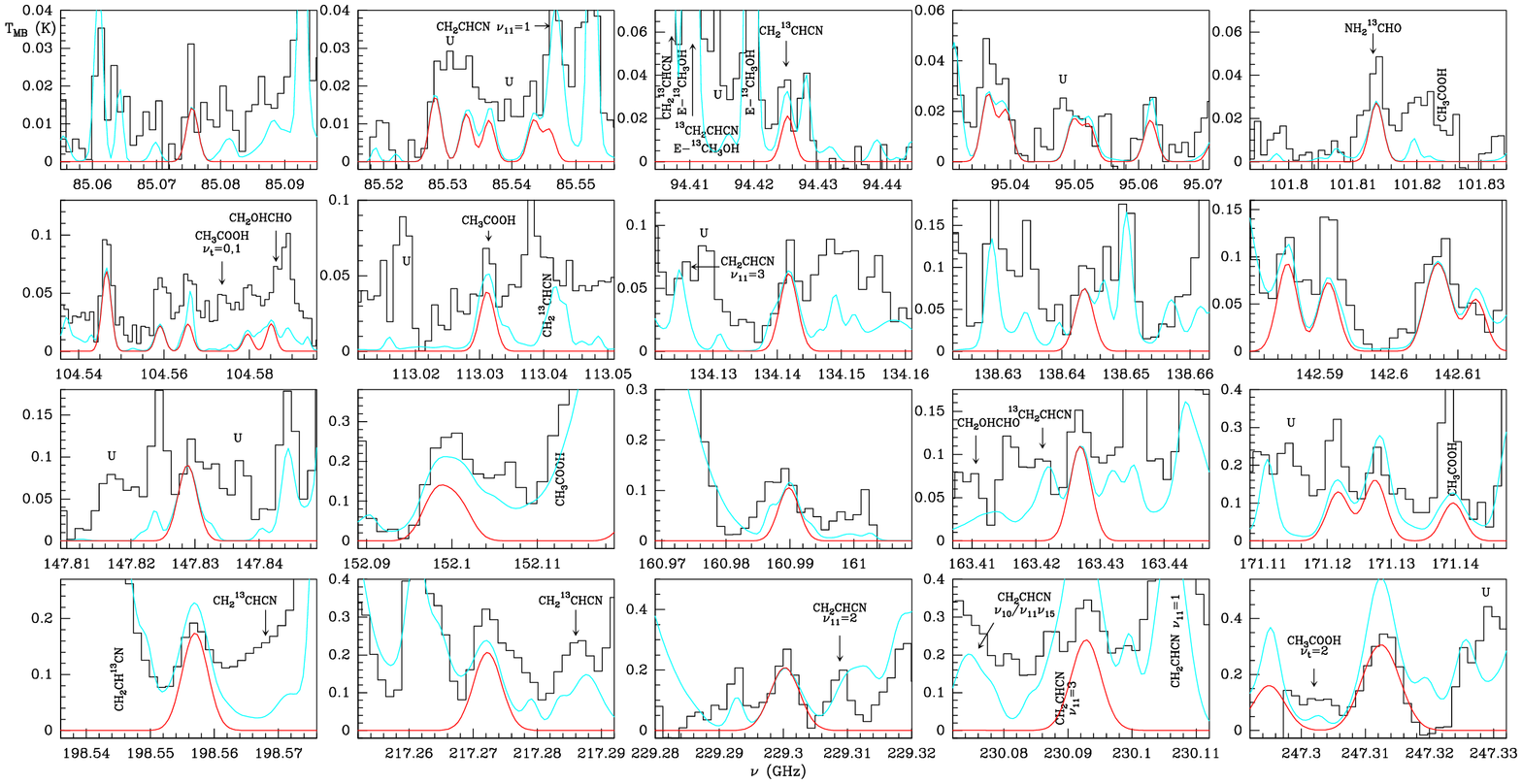}
\caption{Observed lines from Orion-KL (histogram spectra) and model (thin red
curves) for the $\varv_{15}$=1 vibrational state of CH$_2$CHCN. The cyan line corresponds to the model of
the molecules we have already studied in this survey (see text Sect. \ref{sec_cd})
including the CH$_2$CHCN species.
A v$_{\rm LSR}$ of 5\,km s$^{-1}$ is assumed.}
\label{fig_v15}
\end{figure*}

\begin{figure*}[ht]
\centering
\includegraphics[angle=270,width=0.8\textwidth]{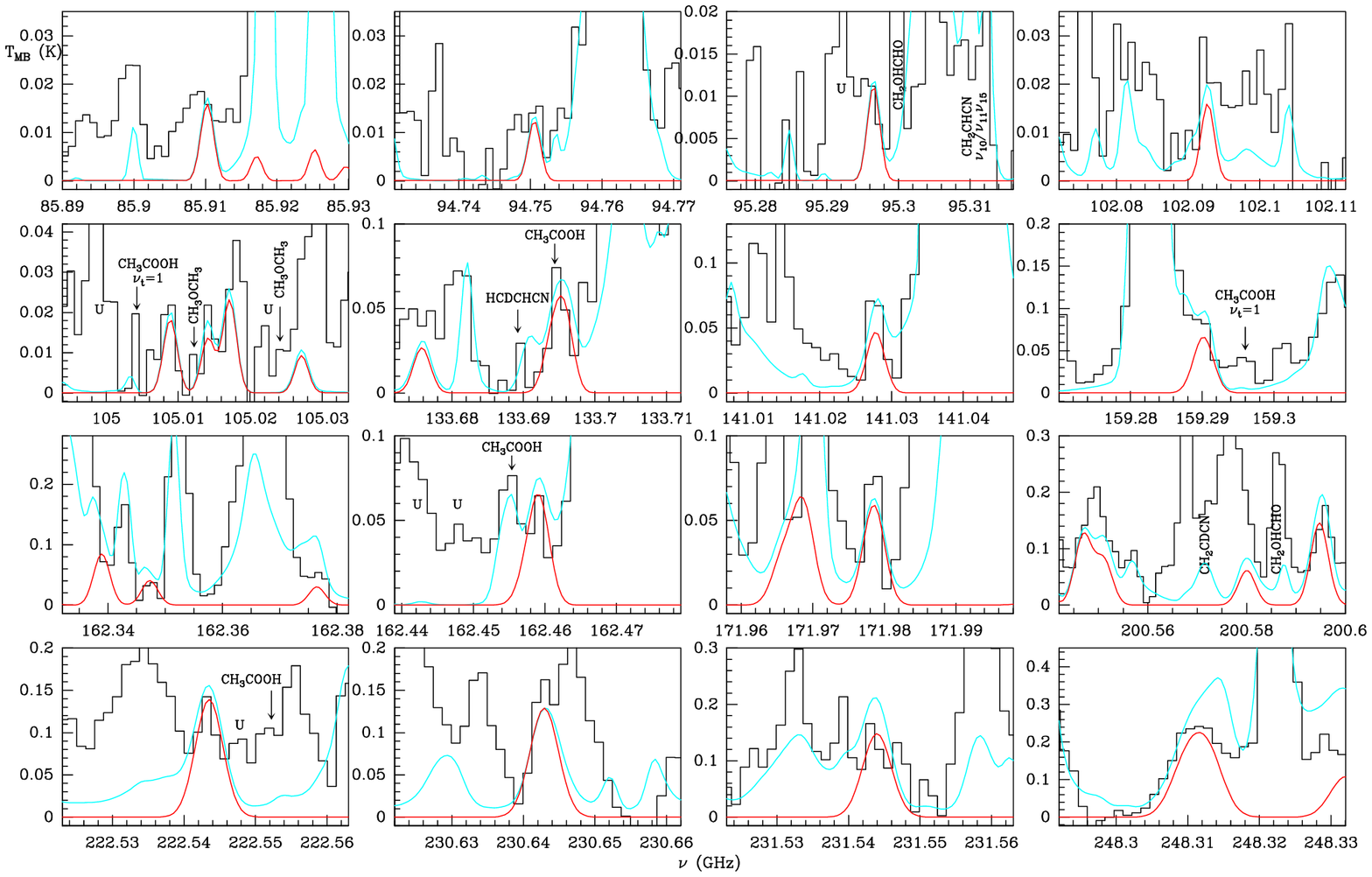}
\caption{Observed lines from Orion-KL (histogram spectra) and model (thin red
curves) for the $\varv_{11}$=2 vibrational state of CH$_2$CHCN. The cyan line corresponds to the model of
the molecules we have already studied in this survey (see text Sect. \ref{sec_cd})
including the CH$_2$CHCN species.
A v$_{\rm LSR}$ of 5\,km s$^{-1}$ is assumed.}
\label{fig_2v11}
\end{figure*}

\begin{figure*}[ht]
\centering
\includegraphics[angle=270,width=0.9\textwidth]{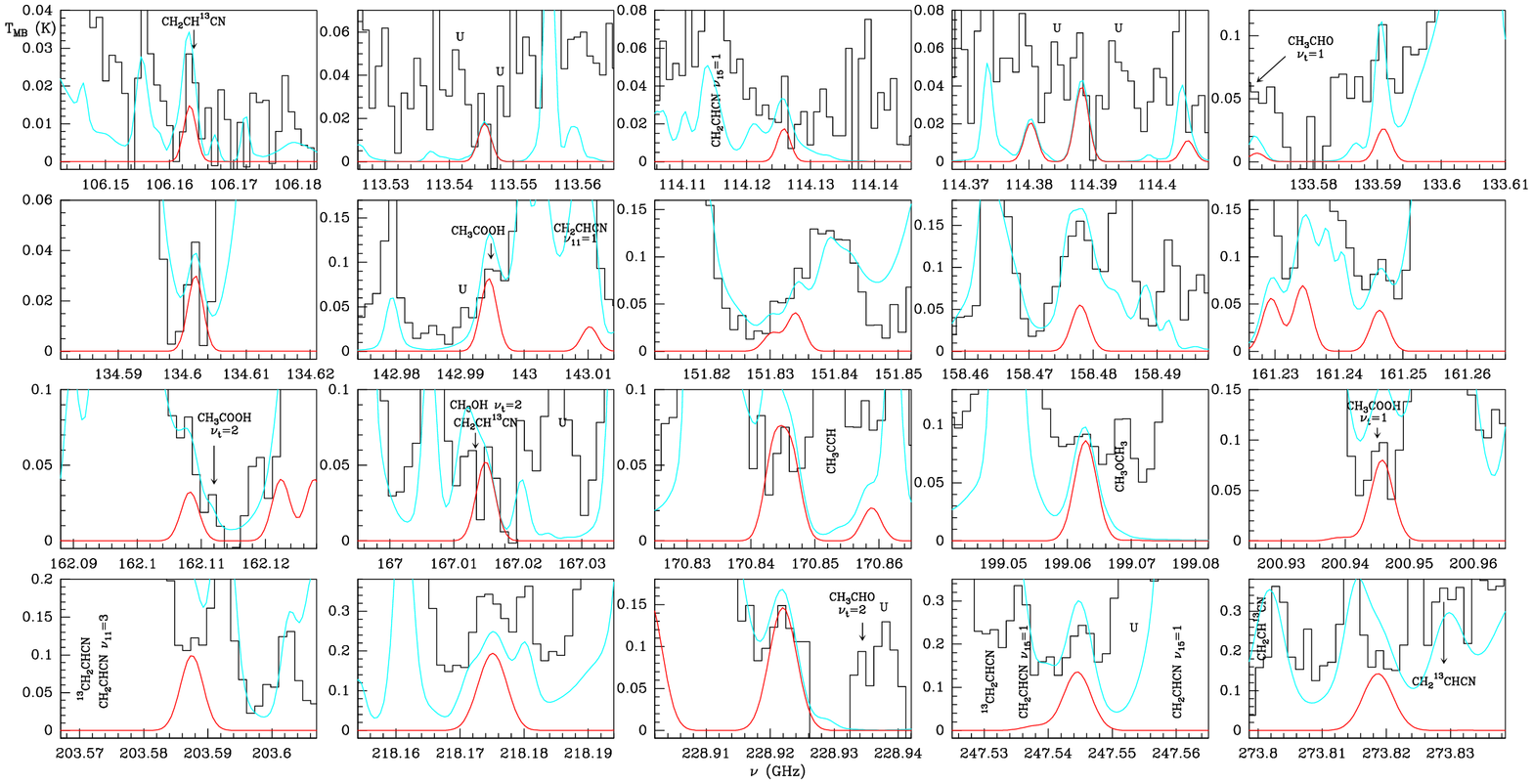}
\caption{Observed lines from Orion-KL (histogram spectra) and model (thin red
curves) for combined vibrationally excited states of CH$_2$CHCN in the
$\varv_{10}$=1$\Leftrightarrow$($\varv_{11}$=1,$\varv_{15}$=1) dyad. The cyan line corresponds to the model of
the molecules we have already studied in this survey (see text Sect. \ref{sec_cd}) including the CH$_2$CHCN species. 
A v$_{\rm LSR}$ of 5\,km s$^{-1}$ is assumed.}
\label{fig_comb}
\end{figure*}

Figures \ref{fig_ground}$-$\ref{fig_comb} and \ref{fig_3v11} (available online) show selected detected lines of the
g.s. of vinyl cyanide as well as five vibrationally excited states of the main
isotopologue CH$_2$CHCN: in plane C-C$\equiv$N bending mode ($\varv_{11}$=1,
228.1\,cm$^{-1}$ or 328.5\,K),   out of plane C-C$\equiv$N bending mode
($\varv_{15}$=1, 332.7\,cm$^{-1}$ or 478.6\,K), in plane C-C$\equiv$N bending mode
($\varv_{11}$=2, 457.2\,cm$^{-1}$ or 657.8\,K), combination state
$\varv_{10}$=1$\Leftrightarrow$($\varv_{11}$=1,$\varv_{15}$=1), 560.5/562.9\,cm$^{-1}$ or 806.4/809.9\,K), and in plane
C-C$\equiv$N bending mode ($\varv_{11}$=3, 686.6\,cm$^{-1}$ or 987.9\,K). The latter is in the detection limit, 
so we will not address the perturbations of this vibrational mode. 

\begin{figure*}[ht]
\centering
\includegraphics[angle=270,width=1.0\textwidth]{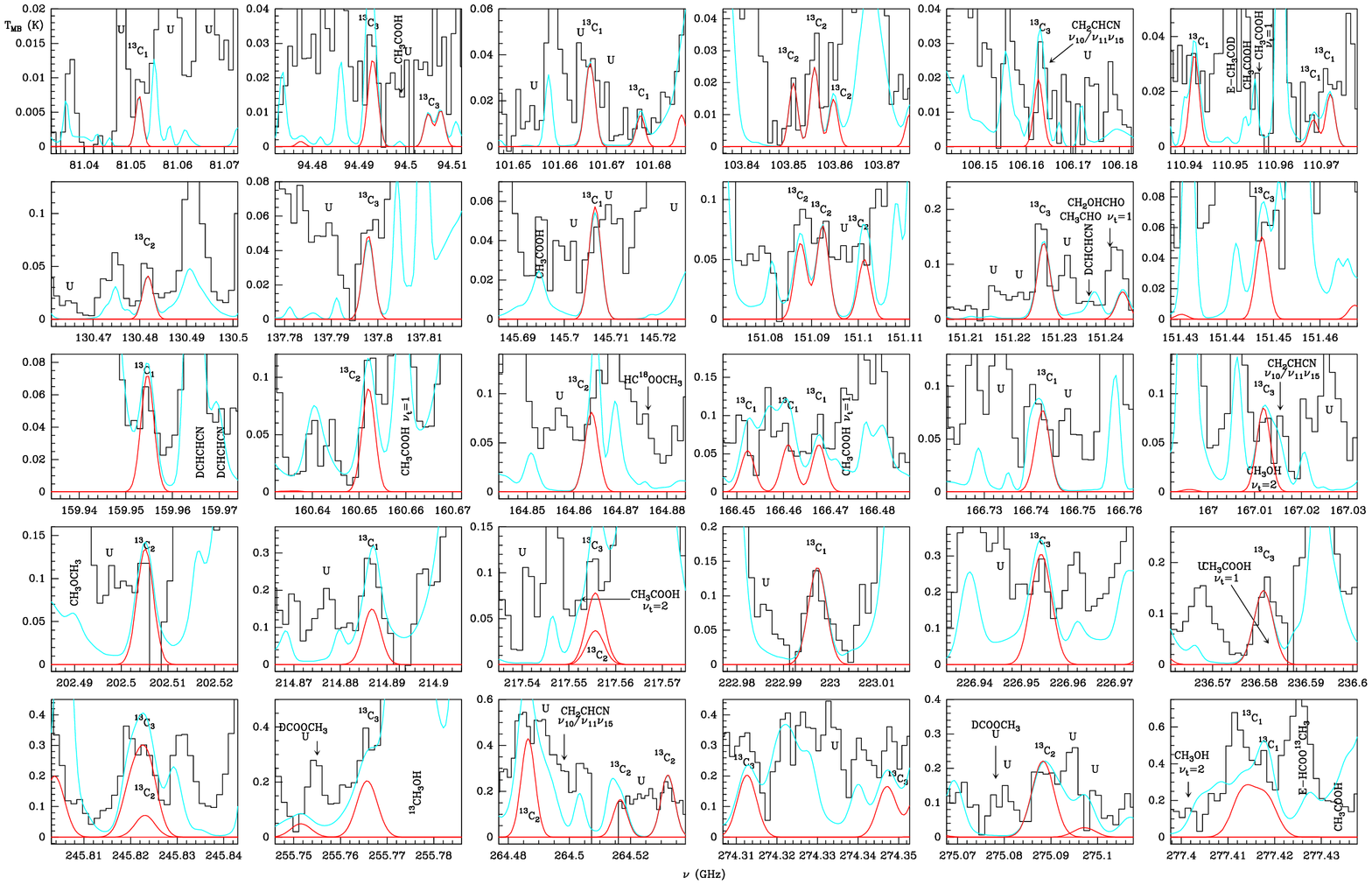}
\caption{Observed lines from Orion-KL (histogram spectra) and model (thin red
curves) of $^{13}$C isotopes for CH$_2$CHCN in the ground state. The subindex in
$^{13}$C$_i$ (i=1, 2, 3) corresponds to the position of the isotope in
the molecule ($^{i_{1}}$CH$_2$$^{i_{2}}$CH$^{i_{3}}$CN). The cyan line corresponds to the model of
the molecules we have already studied in this survey (see text Sect. \ref{sec_cd})
including the CH$_2$CHCN species. A v$_{\rm LSR}$ of 5\,km s$^{-1}$ is assumed.}
\label{fig_13C}
\end{figure*}

In addition, we detected the following isotopologues of vinyl cyanide in its ground
state: $^1$$^3$CH$_2$CHCN, CH$_2$$^1$$^3$CHCN, and CH$_2$CH$^1$$^3$CN (see Fig.
\ref{fig_13C}). For CH$_2$CHC$^1$$^5$N and the deuterated species of vinyl
cyanide, DCHCHCN, HCDCHCN, and CH$_2$CDCN (see Fig. \ref{fig_deu}), we only
provided a tentative detection in Orion-KL because of the small number of lines
with an uncertainty in frequency less than 2\,MHz (up to $K_a$=7,5,15 for
DCHCHCN, HCDCHCN, and CH$_2$CDCN, respectively), the weakness of the features,
and/or their overlap with other molecular species.

Tables \ref{tab_gs}$-$\ref{tab_13C} (available online) show observed and laboratory line
parameters for the ground state, the vibrationally excited states, and the
$^{13}$C-vinyl cyanide isotopologues. Spectroscopic constants were
derived from a fit with the MADEX code (\citeauthor{Cernicharo2012} \citeyear{Cernicharo2012}) to the
lines reported by \citeauthor{kis09} (\citeyear{kis09}, \citeyear{kis12}),
\citeauthor{caz88} (\citeyear{caz88}), and \citeauthor{col97} (\citeyear{col97});
For the $\varv_{10}$=1$\Leftrightarrow$($\varv_{11}$=1,$\varv_{15}$=1) state
spectroscopic constants are those derived in this work;
dipole moments were from \citeauthor{kra11} (\citeyear{kra11}). All these
parameters have been implemented in MADEX to obtain the predicted
frequencies and the spectroscopic line parameters.
We have displayed rotational
lines that are not strongly overlapped with lines from other species.
Observational parameters have been derived by Gaussian fits (using the GILDAS
software) to the observed line profiles that are not blended with other
features. For moderately blended and weak lines we show observed radial
velocities and intensities given directly from the peak channel of the line in
the spectra, so contribution from other species or errors in baselines could
appear for these values. Therefore, the main beam temperature for the weaker
lines ($T_{MB}$$<$0.1 K) must be considered as an upper limit.

From the derived Gaussian fits, we observe that vinyl cyanide lines reflect the spectral line parameters
corresponding to hot core/plateau components (v$_{\rm LSR}$ between 2-3\,km\,s$^{-1}$ for
the component of 20\,km\,s$^{-1}$ of line width, and 5-6\,km\,s$^{-1}$ for the
component of 6\,km\,s$^{-1}$ of line width). As shown by
\citeauthor{adm13} (\citeyear{adm13}) there is a broad component associated to the hot core that limits
the accuracy of the derived velocities for the hot core and this broad component.
Our velocity components for CH$_2$CHCN agree with those of CH$_3$CH$_2$CN obtained by
\citeauthor{adm13} (\citeyear{adm13}). Besides, for the vibrationally
excited states we found contribution of a narrow component with a v$_{\rm LSR}$ of
3-6\,km\,s$^{-1}$ and a line width of $\simeq$7\,km\,s$^{-1}$.

We rely on catalogs\footnote{Cernicharo private catalogs, CDMS
(\citeauthor{mul01} \citeyear{mul01}, \citeauthor{mul05} \citeyear{mul05}), and JPL
(\citeauthor{pic98} \citeyear{pic98})} to identify possible contributions
from other species overlapping the detected lines (\citeauthor{the12}
\citeyear{the12}), but it should be necessary to perform radiative transfer
modeling with all the known molecules in order to assess precisely how much the contamination from
other species influences the vinyl cyanide lines.

\begin{table*}[ht]
 %{\small{\sf{\scriptsize
\begin{center}
\caption{Number of identified lines of CH$_2$CHCN species.}
\begin{tabular}{|l|clll|}
\hline
 \label{tab_lines}
%\caption{continued.}\\
%\hline

Species             & Detectable  & Unblended  & Partially   & Totally\\
                    &             &            & blended     & blended\\
\hline
%\cline{1-5}
 & & & & \\
CH$_2$CHCN g. s. (a-type) & 350 & 204 (59\%) & 85 (24\%) & 61 (17\%) \\
\hline
 & & & & \\
CH$_2$CHCN $\varv_{11}$=1 & 307 & 111 (36\%) & 75 (25\%) & 121 (39\%) \\
\hline
 & & & & \\
CH$_2$CHCN $\varv_{11}$=2 & 253 & 59 (23\%) & 35 (14\%) & 159 (63\%) \\
\hline
 & & & & \\
CH$_2$CHCN $\varv_{11}$=3 & 245 & 30 (12\%) & 33 (14\%) & 182 (74\%) \\
\hline
 & & & & \\
CH$_2$CHCN $\varv_{15}$=1 & 287 & 68 (24\%) & 62 (22\%) & 157 (55\%) \\
\hline
CH$_2$CHCN  & & & & \\
$\varv_{10}$=1$\Leftrightarrow$($\varv_{11}$=1,$\varv_{15}$=1) & 474 & 65 (14\%) & 64 (14\%) & 345 (73\%) \\
\hline
 & & & & \\
($^{13}$C)-CH$_2$CHCN & 348 &  102 (29\%) & 115 (33\%) & 131 (38\%) \\
\hline
%\flushleft

\end{tabular}
\end{center}
%}}
\end{table*}

Table \ref{tab_lines} shows the number of lines of vinyl cyanide
identified in this work. Our identifications are based on a whole inspection of the data
and the modeled synthetic spectrum of the molecule we are studying (where we obtain the total
number of detectable lines) and all species already
identified in our previous papers. We consider blended lines those that are close enough to
other stronger features. Unblended features are those
which present the expected radial velocity (matching our model with the
peak channel of the line) (see e. g. lines at 115.00 and 174.36 GHz in Fig. \ref{fig_v11} or
line a 247.55 GHz in Fig. \ref{fig_comb}), and there are not another species at the same observed
frequency ($\pm$3 MHz) with significant intensity.
Partially blended lines are those which present either a mismatch in the
peak channel of the line (generally, these lines also present
a mismatch in intensity, see e. g. line at 152.0 GHz in Fig. \ref{fig_v15}) or significant
contribution from another species at the peak channel of the feature (see e. g.
line at 108.16 GHz in Fig. \ref{fig_comb}). If for the unblended frequencies we do not found the line we are
looking for, then we do not claim detection, so in the detected species
we do not accept missing lines. For species with quite strong lines (g.s., $\varv_{11}$=1, and
$\varv_{15}$=1), most of the totally and partially
blended lines are weaker due to the high energy of their transitions 
(see Tables \ref{tab_gs}, \ref{tab_v11}, and \ref{tab_v15}).
We observed a total number of $\simeq$640 unblended
lines of vinyl cyanide species. Considering also the moderately blended lines 
this number rises to $\simeq$1100. 
We detected lines of vinyl
cyanide in the g.s. with a maximum upper level energy value of about 
1400-1450\,K corresponding to a $J$$_{\rm max}$=30 and a ($K_a$)$_{\rm max}$=24. For the
vibrational states we observed transitions with maximum quantum rotational
numbers of ($K_a$)$_{\rm max}$=20,15,17,16,15 from the lowest energy vibrational state
to the highest (i.e. from $\varv_{11}$=1 to $\varv_{11}$=3) and the same
$J$$_{\rm max}$=30 value up to the maximum $E$$_{\rm upp}$ between 1300-1690\,K.

\subsubsection{CH$_2$CHCN maps}

\begin{figure*}[!ht]
\centering
\includegraphics[angle=0,width=0.8\textwidth]{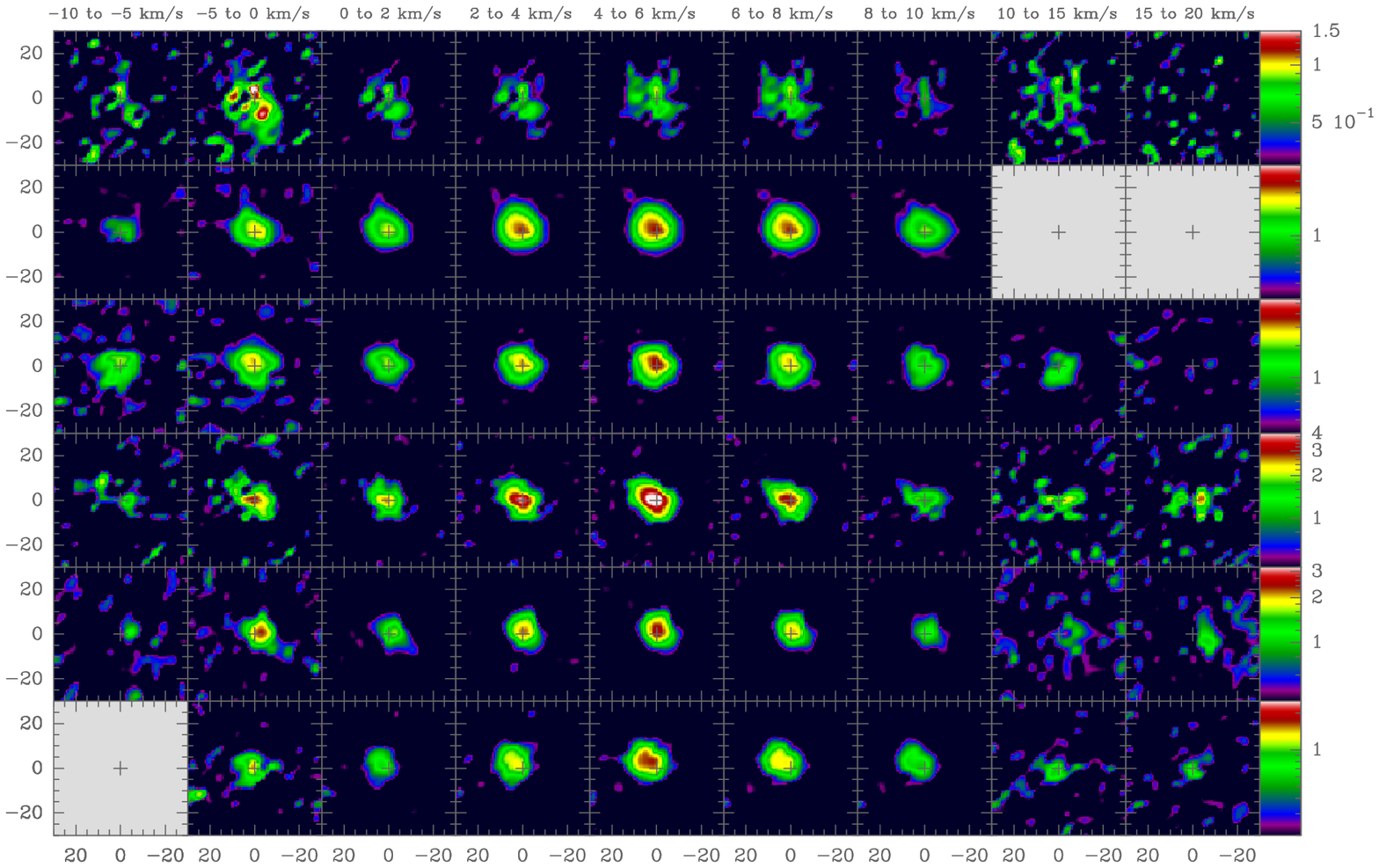}
\caption{Integrated intensity maps for 6 ground state transitions of vinyl cyanide. From line
1 (top row) to 6 (bottom row): 12$_{1,12}$-11$_{1,11}$ (110839.98\,MHz, 36.8\,K), 24$_{0,24}$-
23$_{0,23}$ (221766.03\,MHz, 134.5\,K), 24$_{2,23}$-23$_{2,22}$ (226256.88\,MHz,
144.8\,K), 26$_{1,26}$-25$_{1,25}$ (238726.808\,MHz, 157.4\,K), 26$_{2,25}$-
25$_{2,24}$ (244857.47\,MHz, 167.9\,K), and 24$_{10,15}$-23$_{10,14}$ and
24$_{10,14}$-23$_{10,13}$ (228017.34\,MHz, 352.0\,K) at different velocity
ranges (indicated in the top of each column). Three boxes have been blanked
because the emission at these velocities was blended with that from other well known species.
For each box axis are in units of
arcseconds ($\Delta$$\alpha$, $\Delta$$\delta$). Color logarithm scale is the
integrated intensity ($\int$T$^*$$_A$dv) in units of K\,km\,s$^{-1}$.}
\label{fig_map}
\end{figure*}

Figure \ref{fig_map} shows maps of the integrated emission of six transitions in
the g.s. of CH$_2$CHCN at different velocity ranges. From line 1 to 6 (top to bottom):
12$_{1,12}$-11$_{1,11}$ (110839.98\,MHz, $E_{\rm upp}$=36.8\,K),
24$_{0,24}$-23$_{0,23}$ (221766.03\,MHz, $E_{\rm upp}$=134.5\,K), 24$_{2,23}$-23$_{2,22}$
(226256.88\,MHz, $E_{\rm upp}$=144.8\,K), 26$_{1,26}$-25$_{1,25}$ (238726.81\,MHz,
$E_{\rm upp}$=157.4\,K), 26$_{2,25}$-25$_{2,24}$ (244857.47\,MHz,
$E_{\rm upp}$=167.9\,K), and 24$_{10,15}$-23$_{10,14}$ and 24$_{10,14}$-23$_{10,13}$
(228017.34\,MHz, $E_{\rm upp}$=352.0\,K). These maps reveal the emission from two
cloud components: a component at the position of the hot core at velocities from
2 to 8 km s$^{-1}$ and a component with a slight displacement of the intensity
peak at the extreme velocities. The intensity peak of the central velocities
coincides with that of the -CN bearing molecules found by \citeauthor{gue08},
\citeyear{gue08} (maps of one transition of ethyl and vinyl cyanide) and
\citeauthor{adm13} (\citeyear{adm13}) (maps of four transitions of ethyl
cyanide). We note a more compact structure in the maps of the transitions at
352.0\,K. Our maps do not show a more extended component found in the ethyl
cyanide maps by \citeauthor{adm13} (\citeyear{adm13}). We have obtained an
angular source size between 7''-10'' (in agreement with the hot core
diameter provided by different authors, see e. g. \citeauthor{cro14} \citeyear{cro14}, 
\citeauthor{nei13} \citeyear{nei13}, \citeauthor{beu08} \citeyear{beu08})
for central and extreme velocities assuming
emission within the half flux level and corrected for the size of the telescope
beam at the observed frequency. These integrated intensity maps allow us to
provide the offset position with respect to IRc2 and the source diameter parameters
needed for modeling the vinyl cyanide species (see Sect. \ref{model}).

\subsection{Rotational diagrams of CH$_2$CHCN (g.s., $\varv_{11}$=1,2, and $\varv_{15}$=1)}
\label{RD}

\begin{figure*}[ht]
\centering
\includegraphics[angle=0,width=0.8\textwidth]{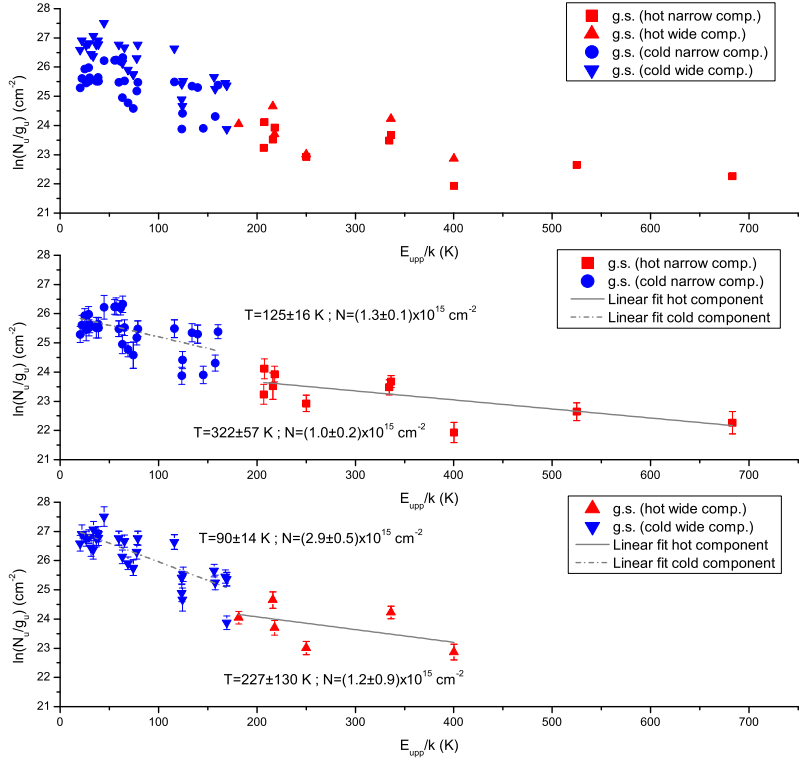}
\caption{Rotational diagram of CH$_2$CHCN in its ground state. The upper panel
displays the two components derived from the line profiles. The middle panel
shows two linear fits to the narrow component points; these linear regressions
yield temperatures and column densities of $T_{\rm rot}$=125$\pm$16\,K and
$N$=(1.3$\pm$0.1)$\times$10$^{15}$\,cm$^{-2}$ (Q$_{\rm rot}$=7.06$\times$10$^{3}$),
and $T_{\rm rot}$=322$\pm$57\,K and $N$=(1.0$\pm$0.2)$\times$10$^{15}$\,cm$^{-2}$
(Q$_{\rm rot}$=2.92$\times$10$^{4}$). Likewise, the bottom panel shows another two
linear fits to the points corresponding to the wide component. The results of
these fits are rotational temperatures of $T_{\rm rot}$=90$\pm$14\,K and
$T_{\rm rot}$=227$\pm$130\,K, and column densities of
(2.9$\pm$0.5)$\times$10$^{15}$\,cm$^{-2}$ (Q$_{\rm rot}$=4.31$\times$10$^{3}$) and
(1.2$\pm$0.9)$\times$10$^{15}$\,cm$^{-2}$ (Q$_{\rm rot}$=1.73$\times$10$^{4}$),
respectively.}
\label{fig_rot_gs}
\end{figure*}

\begin{figure*}[ht]
\centering
\includegraphics[angle=0,width=0.8\textwidth]{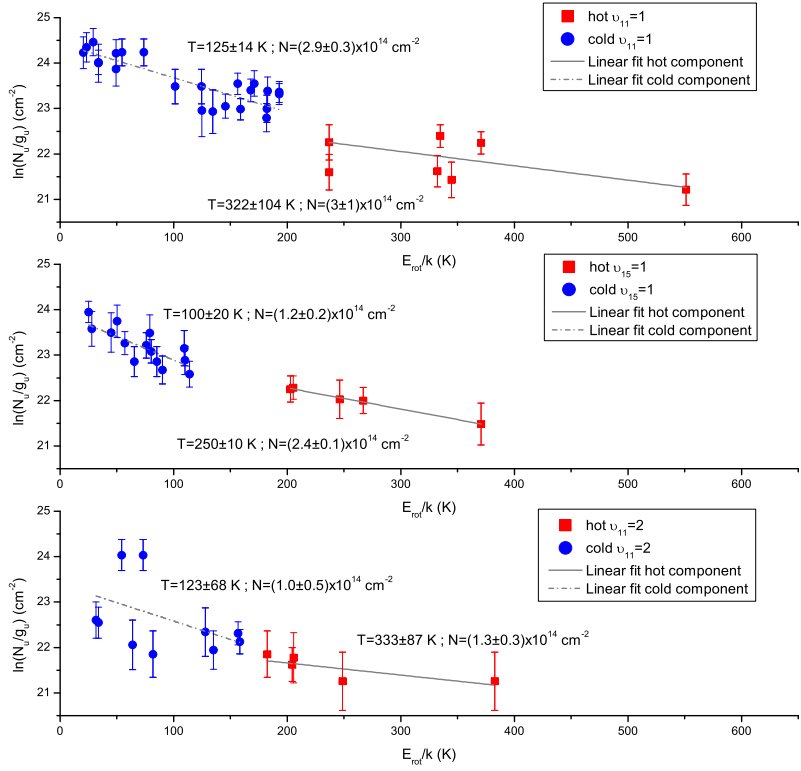}
\caption{Rotational diagrams for the 
vibrationally excited states of vinyl cyanide $\varv_{11}$=1, $\varv_{15}$=1, and
$\varv_{11}$=2 as function of rotational energy (upper level energy
corrected from the vibrational energy of each state)
sorted by increasing vibrational energy from top to bottom.}
\label{fig_rot_EE}
\end{figure*}

In order to obtain an estimate of the rotational temperature ($T_{\rm rot}$) for
different velocity components we made rotational diagrams which related the
molecular parameters with the observational ones (Eq. \ref{eq_RD}, see
eg. \citeauthor{gol99} \citeyear{gol99}), for CH$_2$CHCN in its ground state
(Fig. \ref{fig_rot_gs}) and for the lowest vibrationally excited states
$\varv_{11}$=1, 2, and $\varv_{15}$=1 (Fig. \ref{fig_rot_EE}).
Assumptions such as LTE approximation and
optically thin lines (see Sect. \ref{sect_opa}) are required in this analysis.
We have taken into account the effect of dilution of the telescope which was corrected by
calculation of the beam dilution factor (\citeauthor{dem08} \citeyear{dem08},
Eq. \ref{eq_BDF}):

\begin{equation}
\ln(\frac{N_u}{g_u})=\ln(\frac{8\pi k\nu^2W_{\rm obs}}{hc^3A_{ul}g_u})=\ln(\frac{N}{Q_{\rm rot}}) - \frac{E_{\rm upp}}{kT_{\rm rot}} + \ln b,
\label{eq_RD}
\end{equation}

\begin{equation}
b=\frac{\Omega_S}{\Omega_A}=\frac{\theta_S^2}{\theta_S^2 + \theta_B^2},
\label{eq_BDF}
\end{equation}
where $N_u$ is the column density of the considered vinyl cyanide species in
the upper state (cm$^{-2}$), $g_u$ is the statistical weight in the upper level,
$W_{\rm obs}$ (K\,cm\,s$^{-1}$) is the integrated line intensity
($W_{obs}$=$\int$$T_{{MB}_{obs}}$(v)dv), $A_{ul}$ is the  Einstein A-coefficient for
spontaneous emission, $N$ (cm$^{-2}$) is the total column density of the
considered vinyl cyanide species,  Q$_{\rm rot}$ is the rotational partition
function which depends on the rotational temperature derived from the diagrams,
$E_{\rm upp}$ (K) is the upper level energy, and $T_{\rm rot}$ (K) is the rotational
temperature. In Eq. \ref{eq_BDF},  b is the beam dilution factor, $\Omega_S$ and
$\Omega_A$ are the solid angle subtended by the source and under the main beam
of the telescope, respectively, and $\theta_S$ and $\theta_B$ are the angular
diameter of the source and the beam of the telescope, respectively. We note
that the factor b increases the fraction N$_u$/g$_u$ in Eq. \ref {eq_RD} and
yields a higher column density than if it were not considered.

For the
g.s. we used 117 transitions free of blending with upper level energies from
20.4 to 683.1\,K with two different velocity components, one with
v$_{\rm LSR}$=4-6\,km\,s$^{-1}$ and $\Delta$v=4-7\,km\,s$^{-1}$, and the second one with
v$_{\rm LSR}$=2-4\,km\,s$^{-1}$ and $\Delta$v=14-20\,km\,s$^{-1}$. For the
vibrationally excited states we considered 43 (40-550\,K), 24 (30-380\,K), and
33 (25-370\,K) transitions with line profiles that can be fitted to a single
velocity component (v$_{\rm LSR}$=4-6\,km\,s$^{-1}$ and $\Delta$v=5-7\,km\,
s$^{-1}$) for $\varv_{11}$=1, 2 and $\varv_{15}$=1, respectively.

\begin{figure*}[ht]
\centering
\includegraphics[angle=0,width=0.65\textwidth]{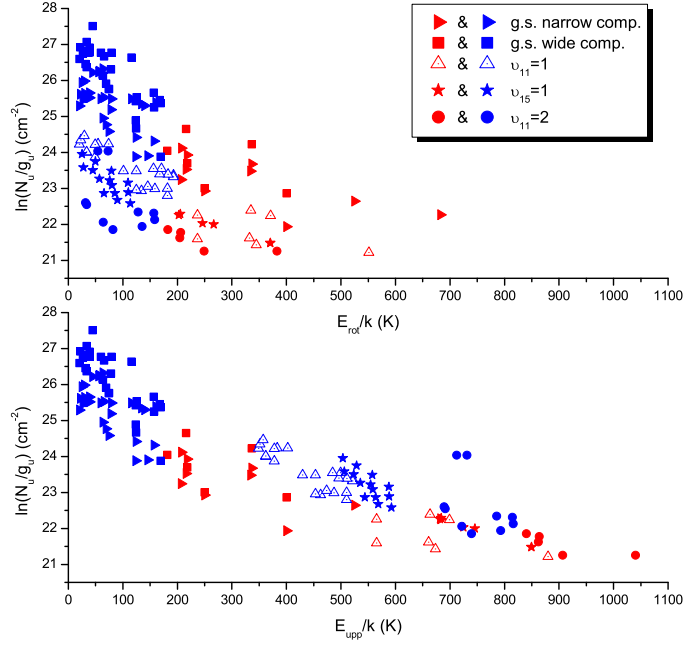}
\caption{Rotational diagram of CH$_2$CHCN in its ground and excited states shown
as function of rotational energy corrected from the vibrational energy in the
upper panel, while the bottom panel displays the ground state followed by
CH$_2$CHCN $\varv_{11}$=1, $\varv_{15}$=1, and $\varv_{11}$=2 excited states as a
function of the upper level energy.}
\label{fig_rot_upp_gs_EE}
\end{figure*}

The scatter in the rotational diagrams of CH$_2$CHCN g.s. is mainly due to the
uncertainty of fitting two Gaussian profiles to the lines with the CLASS software.
Rotational diagrams of the vibrationally excited states ($\nu$$_{11}$=1 and $\nu$$_{15}$=1)
are less scattered because there is only one fitted Gaussian to the line profile.
For the rotational diagram of the $\nu$$_{11}$=2 state, the scatter is mostly due to the
weakness of the observed lines for this species. We have done an effort in order 
to perform the diagrams with unblended lines; however,
some degree of uncertainty could come from non-obvious blends. 
The individual errors of the data points (error bars) are those derived by error propagation in the
calculated uncertainty of $ln$($N_u$/$g_u$), taking only into account the uncertainty of the
integrated intensity of each line ($W$) provided by CLASS and an error of 20\% for the source diameter.
The uncertainty of the final values of $T_{\rm rot}$ and $N$ 
has been calculated with the statistical errors given by the linear least squares fit 
for the slope and the intercept.

We assumed the same source diameter of 10'' for the emitting region of the two
components for the g.s. and the single component of the vibrationally excited
states. In Fig. \ref{fig_rot_gs}, the upper panel shows points in the diagram
related with the wide and narrow components for the CH$_2$CHCN g.s.
We observed two tendencies in the position of the data points up to/starting from
an upper state energy of $\simeq$200\,K. From the
narrow component, we derived two different rotational temperatures and column
densities, $T_{\rm rot}$=125$\pm$16\,K and
$N$=(1.3$\pm$0.1)$\times$10$^{15}$\,cm$^{-2}$, and $T_{\rm rot}$=322$\pm$57\,K and
$N$=(1.0$\pm$0.2)$\times$10$^{15}$\,cm$^{-2}$. Likewise, from the wide
component, we have determined cold and hot temperatures of about
$T_{\rm rot}$=90$\pm$14\,K and $T_{\rm rot}$=227$\pm$130\,K, and column densities of
$N$=(2.9$\pm$0.5)$\times$10$^{15}$\,cm$^{-2}$ and
$N$=(1.2$\pm$0.9)$\times$10$^{15}$\,cm$^{-2}$, respectively.

In order to quantify the uncertainty derived from the assumed
source size, we have also performed the rotational diagram of CH$_2$CHCN g. s.
adopting a source diameter of both 5'' and 15''.
The main effect of changing the source size on the rotational diagram is a
change in the slope and in the scatter.
Table \ref {tab_dr} shows the derived
values of $N$ and $T_{\rm rot}$ assuming different source sizes.
Therefore, as expected, derived rotational temperatures depend clearly on the assumed size
with a tendency to increase $T_{\rm rot}$ when increasing the source diameter.
The effect on the column density is less significant also due
to the correction on the partition function introduced by the change
in the rotational temperatures; in general, 
these values increased or decreased when
we decreased or increased the source size, respectively (see Table \ref {tab_dr}).

\begin{table*}[ht]
 %{\small{\sf{\scriptsize
\begin{center}
\caption{$N$ and $T_{\rm rot}$ from rotational diagrams of CH$_2$CHCN g. s. assuming different source sizes.}
\begin{tabular}{|l|ll|ll|}
\hline
 \label{tab_dr}
%\caption{continued.}\\
%\hline

                                    &\bfseries Hot narrow comp.  &\bfseries Cold narrow comp.  &\bfseries Hot wide comp.   &\bfseries Cold wide comp.\\
    &\multicolumn{2}{c}{v$_{LSR}$=4-6\,km\,s$^{-1}$ $\Delta$v=4-7\,km\,s$^{-1}$} \vline &\multicolumn{2}{c}{v$_{LSR}$=2-4\,km\,s$^{-1}$ $\Delta$v=14-20\,km\,s$^{-1}$}  \vline\\
\hline
%\cline{1-5}
$d$$_{\rm sou}$=5''                 &   $N$=(2.3$\pm$0.7)$\times$10$^{15}$\,cm$^{-2}$    &       $N$=(3.8$\pm$0.8)$\times$10$^{15}$\,cm$^{-2}$  &  $N$=(1.1$\pm$0.9)$\times$10$^{15}$\,cm$^{-2}$  &      $N$=(4.8$\pm$0.5)$\times$10$^{15}$\,cm$^{-2}$    \\
& $T_{\rm rot}$=(334$\pm$89)\,K & $T_{\rm rot}$=(100$\pm$20)\,K & $T_{\rm rot}$=(210$\pm$132)\,K & $T_{\rm rot}$=(71$\pm$5)\,K \\
\hline
$d$$_{\rm sou}$=10''                 &$N$=1.0$\pm$0.2)$\times$10$^{15}$\,cm$^{-2}$    &       $N$=(1.3$\pm$0.1)$\times$10$^{15}$\,cm$^{-2}$  &  $N$=(1.2$\pm$0.9)$\times$10$^{15}$\,cm$^{-2}$  &      $N$=(2.9$\pm$0.5)$\times$10$^{15}$\,cm$^{-2}$    \\
& $T_{\rm rot}$=(322$\pm$57)\,K & $T_{\rm rot}$=(125$\pm$16)\,K & $T_{\rm rot}$=(227$\pm$130)\,K & $T_{\rm rot}$=(90$\pm$14)\,K \\
\hline
$d$$_{\rm sou}$=15''                 &   $N$=(6.9$\pm$1.9)$\times$10$^{14}$\,cm$^{-2}$    &       $N$=(1.0$\pm$0.2)$\times$10$^{15}$\,cm$^{-2}$  &  $N$=(9$\pm$6)$\times$10$^{14}$\,cm$^{-2}$  &      $N$=(1.0$\pm$0.1)$\times$10$^{15}$\,cm$^{-2}$    \\
& $T_{\rm rot}$=(326$\pm$85)\,K & $T_{\rm rot}$=(166$\pm$55)\,K & $T_{\rm rot}$=(250$\pm$125)\,K & $T_{\rm rot}$=(100$\pm$10)\,K\\

\hline
%\flushleft

\end{tabular}
\end{center}
%}}
%}

\end{table*}

In Fig. \ref{fig_rot_EE}, the panels display the rotational diagrams of the
three vinyl cyanide excited states $\varv_{11}$=1, $\varv_{15}$=1, and $\varv_{11}$=2
sorted by vibrational energy from top to bottom. In the $x$-axis we show the
rotational energy which has been corrected from the vibrational energy in order
to estimate the appropriate column density. We also observed the same tendency
of the data points quoted above. The rotational temperature and the
column density conditions for the $\varv_{11}$=1 were $T_{\rm rot}$=125$\pm$14\,K and
(2.9$\pm$0.3)$\times$10$^{14}$\,cm$^{-2}$, and $T_{\rm rot}$=322$\pm$104\,K and
$N$=(3$\pm$1)$\times$10$^{14}$\,cm$^{-2}$. For the $\varv_{15}$=1 state we determine
$T_{\rm rot}$=100$\pm$20\,K and $N$=(1.2$\pm$0.2)$\times$10$^{14}$\,cm$^{-2}$, and
$T_{\rm rot}$=250$\pm$10\,K and $N$=(2.4$\pm$0.1)$\times$10$^{14}$\,cm$^{-2}$. For the
$\varv_{11}$=2 state were $T_{\rm rot}$=123$\pm$68\,K and
$N$=(1.0$\pm$0.5)$\times$10$^{14}$\,cm$^{-2}$, and $T_{\rm rot}$=333$\pm$87\,K and
$N$=(1.3$\pm$0.3)$\times$10$^{14}$\,cm$^{-2}$. Owing to the weakness of the
emission lines of the $\varv_{11}$=3 and $\varv_{10}$=1$\Leftrightarrow$($\varv_{11}$=1,$\varv_{15}$=1)
vibrational modes, we have not performed rotational diagrams for these species.

Figure \ref{fig_rot_upp_gs_EE} displays the combined rotational diagram for ground state 
of CH$_2$CHCN and 
$\varv_{11}$=1, $\varv_{15}$=1, and $\varv_{11}$=2 excited states. The
upper panel is referred to the rotational level energies of the vinyl cyanide
states, whereas the bottom panel shows the positions of the different rotational
diagrams in upper level energy when taking into account the vibrational energy for
the excited states.

Owing to the large range of energies and the amount of transitions in these
rotational diagrams we consider the obtained results ($T_{\rm rot}$) as a starting
point in our models (see Section \ref{model}).

\subsection{Astronomical modeling of CH$_2$CHCN in Orion-KL}

\subsubsection{Analysis: The Model}
\label{model}

From the observational line parameters derived in Sect. \ref{sect_det} (radial
velocities and line widths), the displayed
maps, and the rotational diagram results (two components $-$cold and hot$-$ for
each derived Gaussian fit to the line profiles), we consider that the emission of
CH$_2$CHCN species comes mainly from the four regions shown in Table \ref{tab_prop},
related with the hot core (those with $\Delta$$v$=6-7\,km\,s$^{-1}$) and plateau/hot core (those
with $\Delta$$v$=20\,km\,s$^{-1}$) components. \citeauthor{adm13} (\citeyear{adm13}) found that three components related
with the hot core were enough to properly fit their ethyl cyanide lines. The named "Hot Core 1" and "Hot Core 3" in \citeauthor{adm13} (\citeyear{adm13}) are
similar to our "Hot narrow comp." and "Cold wide comp." of Table \ref{tab_prop}, respectively. Interferometric
maps performed by \citeauthor{gue08} (\citeyear{gue08}) of ethyl and vinyl cyanide and those
of \citeauthor{Widicus2012} (\citeyear{Widicus2012}) of CH$_3$CH$_2$CN (the latter authors
affirm that in their observations CH$_3$CH$_2$CN, CH$_2$CHCN, and CH$_3$CN are cospatial)
show that the emission from these species comes from different cores at the position of the 
hot core and IRc7. The radial velocities found in the line profiles of vinyl cyanide (between 3-5\,km\,s$^{-1}$)
in this work together with the cited interferometric maps, could indicate that the four components of Table \ref{tab_prop}
are dominated by the emission of the hot core. For the
vibrationally excited states and for the isotopologues we found that two
components (both narrow components) are sufficient to reproduce the line profiles
(see Table \ref{tab_prop}). We note that for $\varv_{11}$=1 and $\varv_{15}$=1 we
need a somewhat higher value in the line width. This difference is probably due
to a small contribution of the wide component in these lines.

Spectroscopic (Sect. \ref{sect_exp}) and observational parameters $-$radial
velocity (v$_{\rm LSR}$); line width ($\Delta$v); temperature from rotational
diagrams (T$_{\rm rot}$); source diameter ($d_{\rm sou}$) and offsets from the maps$-$
were introduced in an excitation and radiative transfer code (MADEX) in order to
obtain the synthetic spectrum. We have taken into account the telescope dilution
and the position of the components with respect to the pointing position (IRc2).
LTE conditions have been assumed owing to the lack of collisional rates for
vinyl cyanide, which prevents a more detailed analysis of the emission of this
molecule. Nevertheless, we expect a good approximation to the physical and
chemical conditions due to the hot and dense nature of the considered
components. Rotational temperatures (which coincide with the excited and kinetic
temperatures in LTE conditions) have been slightly adapted from those of the
rotational diagrams in order to obtain the best fit to the line profiles. These
models allow us to obtain column density results for each species and components
independently. The sources of uncertainty that were described in
\citeauthor{Tercero10} (\citeyear{Tercero10}) have been considered. For the
CH$_2$CHCN g.s., $\varv_{11}$=1, and $\varv_{15}$=1 states we have adopted an
uncertainty of 30\%, while for the $^{13}$C isotopologues and the 
$\varv_{11}$=2, $\varv_{11}$=3,
and $\varv_{10}$=1$\Leftrightarrow$($\varv_{11}$=1,$\varv_{15}$=1) states we have adopted a 50\% 
uncertainty. Due to the weakness and/or high overlap with other molecular
species we only provided upper limits to the column densities of monodeuterated
vinyl cyanide and the $^{15}$N isotopologue.\\

\begin{table*}[ht]
 %{\small{\sf{\scriptsize
\begin{center}
\caption{Physico-chemical conditions of Orion-KL from ground and excited states of CH$_2$CHCN.}
\begin{tabular}{lcccc}
 \label{tab_prop} \\
%\caption{continued.}\\

\hline\hline
                                     &\bfseries Hot narrow comp.  &\bfseries Cold narrow comp.  &\bfseries Hot wide comp.   &\bfseries Cold wide comp.\\
\hline
$d$$_{\rm sou}$   ('')                              &        5                  &       10            &       5       &       10           \\
offset ('')                                         &        2                  &        2            &       0       &       0            \\
$\Delta$$v$$_{\rm FWHM}$  (km\,s$^{-1}$)                     &        6(7*)              &        6(7*)        &       20      &       20           \\
$v$$_{\rm LSR}$  (km\,s$^{-1}$)                     &        5                  &        5            &       3       &       3            \\
\hline                                             
\bfseries $T$$_{\rm rot}$ (K)                       &\bfseries 320              & \bfseries 100       & \bfseries 200 & \bfseries 90       \\
\hline\hline                                       
$N_{\rm CH_2CHCN (g.s.)}$ (cm$^{-2}$)               & (3.0$\pm$0.9)$\times$10$^{15}$       & (1.0$\pm$0.3)$\times$10$^{15}$ & (9$\pm$3)$\times$10$^{14}$ & (1.3$\pm$0.4)$\times$10$^{15}$ \\
$N_{\rm CH_2CHCN (\varv_{11}=1)}$(cm$^{-2}$)            & (9$\pm$3)$\times$10$^{14}$           & (2.5$\pm$0.8)$\times$10$^{14}$   & ... & ... \\
$N_{\rm CH_2CHCN (\varv_{11}=2)}$(cm$^{-2}$)            & (2$\pm$1)$\times$10$^{14}$           & (5$\pm$2)$\times$10$^{13}$ & ... & ... \\
$N_{\rm CH_2CHCN (\varv_{11}=3)}$(cm$^{-2}$)            & $\leq$(2$\pm$1)$\times$10$^{14}$     & $\leq$(5$\pm$2)$\times$10$^{13}$ & ... & ... \\
$N_{\rm CH_2CHCN (\varv_{15}=1)}$(cm$^{-2}$)            & (4$\pm$1)$\times$10$^{14}$       & (1.0$\pm$0.3)$\times$10$^{14}$ & ... & ... \\
$N_{\rm CH_2CHCN (\varv_{10}=1\Leftrightarrow(\varv_{11}=1,\varv_{15}=1))}$(cm$^{-2}$) & (4$\pm$2)$\times$10$^{14}$       & (8$\pm$4)$\times$10$^{13}$ & ... & ... \\
$N_{\rm ^{13}CH_2CHCN}$(cm$^{-2}$)                  & (4$\pm$2)$\times$10$^{14}$       & (5$\pm$2)$\times$10$^{13}$ & ... & ... \\
$N_{\rm CH_2^{13}CHCN}$(cm$^{-2}$)                  & (4$\pm$2)$\times$10$^{14}$       & (5$\pm$2)$\times$10$^{13}$ & ... & ... \\
$N_{\rm CH_2CH^{13}CN}$(cm$^{-2}$)                  & (4$\pm$2)$\times$10$^{14}$       & (5$\pm$2)$\times$10$^{13}$ & ... & ... \\
$N_{\rm CH_2CHC^{15}N}$(cm$^{-2}$)                  & $\leq$(1.0$\pm$0.5)$\times$10$^{14}$     & $\leq$(2$\pm$1)$\times$10$^{13}$ & ... & ... \\
$N_{\rm HCDCHCN}$(cm$^{-2}$)                        & $\leq$(4$\pm$2)$\times$10$^{14}$ & $\leq$(4$\pm$2)$\times$10$^{13}$ & ... & ... \\

$N_{\rm DCHCHCN}$(cm$^{-2}$)                        & $\leq$(4$\pm$2)$\times$10$^{14}$ & $\leq$(4$\pm$2)$\times$10$^{13}$ & ... & ... \\
$N_{\rm CH_2CDCN}$(cm$^{-2}$)                       & $\leq$(3$\pm$1)$\times$10$^{14}$     & $\leq$(3$\pm$1)$\times$10$^{13}$ & ... & ... \\
\hline
%\flushleft

\end{tabular}
\end{center}
%}}
%}

{\bfseries Note.} Physico-chemical conditions of Orion-KL from vinyl cyanide (see text
\ref{model}). * 7 km\,s$^{-1}$ is only considered for $\varv_{11}$=1 and
$\varv_{15}$=1 states.

\end{table*}

\subsubsection{Column densities}
\label{sec_cd}

The column densities that best reproduce the observations are shown in
Table \ref{tab_prop} and used for the model in Figs. \ref{fig_ground}$-$\ref{fig_13C} and \ref{fig_deu} (available online). Although the
differences between the intensity of the model and that of the observations are
mostly caused by blending with other molecular species, 
isolated vinyl cyanide lines confirm good agreement
between model and observations. We found small differences between the column
density values from the model and those from the rotational diagram, likely
because of the source diameters taken into account in the determination of the
beam dilution for the two components.

In Figs. \ref{fig_ground}$-$\ref{fig_13C}, \ref{fig_CH2CHNC}, \ref{fig_CH3NC}, and \ref{fig_3v11}-\ref{fig_NH2NC}, a model
with all already studied species in this survey is included (cyan line). 
The considered molecules and published works containing the detailed analysis
for each species are the following:
OCS, CS, H$_2$CS, HCS$^+$, CCS, CCCS species in \citeauthor{Tercero10} (\citeyear{Tercero10});
SiO and SiS species in \citeauthor{Tercero11} (\citeyear{Tercero11}); SO and SO$_2$ species in \citeauthor{esp13a} (\citeyear{esp13a});
HC$_3$N and HC$_5$N species in \citeauthor{esp13b} (\citeyear{esp13b}); CH$_3$CN in \citeauthor{tom14} (\citeyear{tom14});
CH$_3$COOCH$_3$ and t/g-CH$_3$CH$_2$OCOH in \citeauthor{Ter13} (\citeyear{Ter13});
CH$_3$CH$_2$SH, CH$_3$SH, CH$_3$OH, CH$_3$CH$_2$OH in \citeauthor{kol14} (\citeyear{kol14});
$^{13}$C-CH$_3$CH$_2$CN in \citeauthor{dem07} (\citeyear{dem07}); CH$_3$CH$_2$$^{15}$N, CH$_3$CHDCN, 
and CH$_2$DCH$_2$CN in \citeauthor{mar09} (\citeyear{mar09}); CH$_3$CH$_2$CN species in \citeauthor{adm13} (\citeyear{adm13}); 
$^{13}$C-HCOOCH$_3$ in \citeauthor{car09} (\citeyear{car09}); 
DCOOCH$_3$ and HCOOCH$_3$ in \citeauthor{mar10} (\citeyear{mar10}); 
$^{18}$O-HCOOCH$_3$ in \citeauthor{Tercero12} (\citeyear{Tercero12}); HCOOCH$_2$D in \citeauthor{cou13} (\citeyear{cou13});
$^{13}$C-HCOOCH$_3$ $\nu_t$=1, and HCOOCH$_3$ $\nu_t$=1 in \citeauthor{hay14} (\citeyear{hay14});
NH$_2$CHO $\nu_{12}$=1 and NH$_2$CHO in \citeauthor{mot12} (\citeyear{mot12});
CH$_2$CHCN species in this work; also HCOOCH$_3$ $\nu_t$=2 and CH$_3$COOH from
L\'opez et al. in preparation.

We obtained a total column density of vinyl cyanide in the ground state of
(6$\pm$2)$\times$10$^{15}$\,cm$^{-2}$. This value is a factor 7 higher than
the value in the Orion-KL hot core of \citeauthor{sch97} (\citeyear{sch97}), who
detected the vinyl cyanide g.s. in the frequency range from 325 to 360 GHz with
a column density (averaged over a beam of 10''-12'') of
8.2$\times$10$^{14}$\,cm$^{-2}$ and a $T_{\rm rot}$ of 96\,K. The difference
between both results is mostly due to our
more detailed model of vinyl cyanide which includes four components, two of them
with a source size of 5'' (a factor two lower than the beam size in \citeauthor{sch97} (\citeyear{sch97}).
\citeauthor{sut95} (\citeyear{sut95}) also derived a column density of
1$\times$10$^{15}$\,cm$^{-2}$ (beam size of 13.7'') toward the hot core position.
These authors found vinyl cyanide emission
toward the compact ridge position but at typical hot core  velocities.
Previous authors derived beam averaged column densities between
4$\times$10$^{13}$ and 2$\times$10$^{14}$\,cm$^{-2}$ (\citeauthor{joh84}
\citeyear{joh84}, \citeauthor{bla87} \citeyear{bla87}, \citeauthor{tur91}
\citeyear{tur91}, \citeauthor{ziu93} \citeyear{ziu93}).

The column density of CH$_2$CHCN $\varv_{11}$=1,
(1.0$\pm$0.3)$\times$10$^{15}$\,cm$^{-2}$, is four times smaller than that derived
for the ground state in the same components. Besides, we derived a column density of
(3$\pm$2)$\times$10$^{14}$, $\leq$(3$\pm$2)$\times$10$^{14}$,
(5$\pm$2)$\times$10$^{14}$, (5$\pm$2)$\times$10$^{14}$\,cm$^{-2}$, for the
$\varv_{11}$=2, $\varv_{11}$=3, $\varv_{15}$=1, and $\varv_{10}$=1$\Leftrightarrow$($\varv_{11}$=1,$\varv_{15}$=1)
states, respectively. \citeauthor{sch97} \citeyear{sch97} did not give column
density results for the tentative detection of $\varv_{11}$=1 and $\varv_{15}$=1
bending modes. We also obtained a column density of
(4$\pm$2)$\times$10$^{14}$\,cm$^{-2}$ for each $^{13}$C-isotopologue of
vinyl cyanide.\\

\subsubsection{Isotopic abundances}
\label{sec_isot}

It is now possible to estimate the isotopic abundance ratio
of the main isotopologue ($^{12}$C, $^{14}$N, $^1$H) with respect to $^{13}$C,
$^{15}$N, and D isotopologues, from the obtained column densities shown in
Table \ref{tab_prop}. For estimating these ratios,
we assume the same partition function for both the main and the rare isotopologues.

$^{12}$C/$^{13}$C: The column density ratio between the normal species and each
$^{13}$C isotopologue in Orion-KL, on taking into account the associated
uncertainties, varies between 4-20 for the hot narrow component and between 10-43
for the cold narrow component. The solar isotopic abundance
($^{12}$C/$^{13}$C=90, \citeauthor{and89} \citeyear{and89}) corresponds roughly
to a factor 2-22 higher than the value obtained in Orion. 
The $^{12}$C/$^{13}$C ratio indicates the degree of galactic chemical evolution, so
the solar system value could point out earlier epoch conditions of this region
(\citeauthor{wyc00} \citeyear{wyc00}; \citeauthor{sav02} \citeyear{sav02}). The
following previous estimates of the $^{12}$C/$^{13}$C ratio in Orion-KL from
observations of different molecules have been reported: 43$\pm$7 from CN (\citeauthor{sav02}
\citeyear{sav02}), 30-40 from HCN, HNC, OCS, H$_2$CO, CH$_3$OH
(\citeauthor{bla87} \citeyear{bla87}), 57$\pm$14 from CH$_3$OH
(\citeauthor{per07} \citeyear{per07}), 35 from methyl formate
(\citeauthor{car09} \citeyear{car09}, \citeauthor{hay14} \citeyear{hay14}),
45$\pm$20 from CS-bearing molecules (\citeauthor{Tercero10}
\citeyear{Tercero10}), 73$\pm$22 from ethyl cyanide (\citeauthor{adm13}
\citeyear{adm13}), and $\simeq$3-17 from cyanoacetylene in the hot
core (\citeauthor{esp13b} \citeyear{esp13b}). 
Taking into account the weakness of the $^{13}$C lines,
the derived ratios are compatible with a $^{12}$C/$^{13}$C ratio 
between 30-45 found by other authors.
Nevertheless, our results point out a possible chemical fractionation
enhancement of the $^{13}$C isotopologues of vinyl cyanide. The intensity
ratios derived in Sect. \ref{sect_opa} also indicate this possibility.
This ratio might be
underestimated if the lines from the g.s. were optically thick. However, our model
for the assumed sizes of the source yields values of $\tau$ (optical depth) that are much
lower than unity (see Sect. \ref{sect_opa}). In Sgr B2(N), \citeauthor{mul08} (\citeyear{mul08})
derived from their observations of CH$_2$CHCN a $^{12}$C/$^{13}$C ratio of
21$\pm$6.

$^{14}$N/$^{15}$N: We obtained an average lower limit value for
$N({\rm CH_2CHC^{14}N}$)/$N({\rm CH_2CHC^{15}N}$) of $\geq$33 for the two involved
components. In \citeauthor{adm13} (\citeyear{adm13}) (see Appendix B) we
provided a $^{14}$N/$^{15}$N ratio of 256$\pm$128 by means of ethyl cyanide, in
agreement with the terrestrial value (\citeauthor{and89} \citeyear{and89}) and
with the value obtained by \citeauthor{ada12} (\citeyear{ada12}) in the local
interstellar medium. The latter authors performed an evaluation of the
$^{14}$N/$^{15}$N ratio across the Galaxy (toward 11 molecular clouds) through
CN and HNC. They concluded that this ratio exhibits a positive gradient with
increasing distance from the Galactic center (in agreement with chemical
evolution models where $^{15}$N has a secondary origin in novae).

D/H: For a tentative detection of mono-deuterated forms of vinyl cyanide we
derived a lower limit D/H ratio of $\leq$0.12 (for HCDCHCN and DCHCHCN) and
$\leq$0.09 (for CH$_2$CDCN) for the hot narrow component, whereas we obtain
$\leq$0.04 (for HCDCHCN and DCHCHCN) and $\leq$0.03 (for CH$_2$CDCN) for the
cold component. Studies of the chemistry of deuterated species in hot cores
carried out by \citeauthor{rod96} (\citeyear{rod96}) conclude that the column
density ratio D-H remains practically unaltered during a large period of time
when D and H-bearing molecules are released to the gas phase from the ice
mantles of dust grains. These authors indicate that the observations of
deuterated molecules give insight into the processes occurring on the grain
mantles by inferring the fractionation of their parent molecules. Furthermore,
the fractionation also helps us to trace the physical and chemical conditions of
the region (\citeauthor{rou05} \citeyear{rou05}).   Values of this ratio were
given by \citeauthor{mar10} (\citeyear{mar10}) from observations of deuterated
methyl formate obtained $N$(DCOOH$_3$/HCOOCH$_3$)=0.04 for the hot core;
\citeauthor{Tercero10} (\citeyear{Tercero10}) estimated an abundance ratio of
$N$(HDCS)/$N$(H$_2$CS) being 0.05$\pm$0.02, also for the hot core component;
\citeauthor{nei13} (\citeyear{nei13}) provided a $N$(HDCO)/$N$(H$_2$CO) ratio in
the hot core of $\leq$0.005; \citeauthor{par01b} (\citeyear{par01b}) derived a
value between 0.004-0.01 in the plateau by means of $N$(HDO)/$N$(H$_2$O);
\citeauthor{per07} (\citeyear{per07}) also for $N$(HDO)/$N$(H$_2$O) derived
0.005, 0.001, and 0.03 for the large velocity plateau, the hot core, and the
compact ridge, respectively; and \citeauthor{sch92} (\citeyear{sch92}) provided
the DCN/HCN column density ratio of 0.001 for the hot core region.

\subsubsection{Line opacity}
\label{sect_opa}

The MADEX code gives us the line opacity for each transition
for the physical components assumed in Table \ref{tab_prop}.
Table \ref{tab_opa} shows the opacities for the four cloud
components shown in Table \ref{tab_prop} obtained by varying the source
diameter and the column density (the last in order to obtain
a good fit between the synthetic spectra and the observations).
When we decreased the source diameter, we have to increase the
column densities in order to properly fit the observations and,
therefore, the opacities of the lines increment. 
The extreme case where the hot and cold cloud components have diameters of 2'' and 5'', respectively, 
allow us to obtain the maximum total opacity of $\simeq$0.26 (sum of the opacity of all cloud components) for the 
30$_{0,30}$-29$_{0,29}$ transition at 275588 MHz. This value corresponds with a maximum correction of 
about 3-5\% for our column density results. In fact, column densities have to rise a factor 4
in order to obtain a total opacity of $\simeq$0.95 implying a large mismatch (a factor
$\simeq$3-4 in the line intensity) between model
and observations.

\begin{table*}[ht]
 %{\small{\sf{\scriptsize
\begin{center}
\caption{Line opacities}
\begin{tabular}{llc|llll|}
\label{tab_opa} \\
%\caption{continued.}\\

\cline{4-7}
\cline{4-7}
 &            &                        &Hot narrow comp.  &Cold narrow comp.  &Hot wide comp.   &Cold wide comp.\\
\cline{1-7}
 & & & $d$$_{\rm sou}$=10'' & $d$$_{\rm sou}$=15'' & $d$$_{\rm sou}$=10'' & $d$$_{\rm sou}$=15''\\
Transition & Freq. (MHz) & $E_{upp}$ (K) & $N$=1.6$\times$10$^{15}$\,cm$^{-2}$ & $N$=3.6$\times$10$^{14}$\,cm$^{-2}$ & $N$=8.2$\times$10$^{13}$\,cm$^{-2}$ & $N$=9.2$\times$10$^{14}$\,cm$^{-2}$ \\
\hline
%Transition & Freq. (MHz) & $E_{upp}$ (K) & & & & \\
%\cline {1-3}
11$_{0,11}$-10$_{0,10}$ & 103575.4 & 29.9& $\tau$=2.76$\times$10$^{-3}$ & $\tau$=8.99$\times$10$^{-3}$ & $\tau$=1.22$\times$10$^{-4}$ & $\tau$=8.70$\times$10$^{-3}$\\
14$_{3,11}$-13$_{3,10}$ & 133030.7 & 67.3& $\tau$=3.83$\times$10$^{-3}$ & $\tau$=9.72$\times$10$^{-3}$ & $\tau$=1.58$\times$10$^{-4}$ & $\tau$=9.03$\times$10$^{-3}$\\
18$_{0,18}$-17$_{0,17}$ & 167728.4 & 77.1& $\tau$=6.33$\times$10$^{-3}$ & $\tau$=1.51$\times$10$^{-2}$ & $\tau$=2.57$\times$10$^{-4}$ & $\tau$=1.39$\times$10$^{-2}$\\
23$_{0,23}$-22$_{0,22}$ & 212788.7 &123.8& $\tau$=8.89$\times$10$^{-3}$ & $\tau$=1.55$\times$10$^{-2}$ & $\tau$=3.31$\times$10$^{-4}$ & $\tau$=1.35$\times$10$^{-2}$\\
25$_{4,21}$-24$_{4,20}$ & 237712.0 &182.8& $\tau$=8.78$\times$10$^{-3}$ & $\tau$=1.02$\times$10$^{-2}$ & $\tau$=2.93$\times$10$^{-4}$ & $\tau$=8.39$\times$10$^{-3}$\\
28$_{0,28}$-27$_{0,27}$ & 257646.2 &181.4& $\tau$=1.10$\times$10$^{-2}$ & $\tau$=1.30$\times$10$^{-2}$ & $\tau$=3.68$\times$10$^{-4}$ & $\tau$=1.07$\times$10$^{-2}$\\
30$_{0,30}$-29$_{0,29}$ & 275588.1 &207.4& $\tau$=1.25$\times$10$^{-2}$ & $\tau$=1.15$\times$10$^{-2}$ & $\tau$=3.71$\times$10$^{-4}$ & $\tau$=9.20$\times$10$^{-3}$\\
\cline {1-7}
 & & & $d$$_{\rm sou}$=5'' & $d$$_{\rm sou}$=10'' & $d$$_{\rm sou}$=5'' & $d$$_{\rm sou}$=10''\\
Transition & Freq. (MHz) & $E_{upp}$ (K) & $N$=3.0$\times$10$^{15}$\,cm$^{-2}$ & $N$=1.0$\times$10$^{15}$\,cm$^{-2}$ & $N$=9.0$\times$10$^{14}$\,cm$^{-2}$ & $N$=1.3$\times$10$^{15}$\,cm$^{-2}$ \\
\hline
%Transition & Freq. (MHz) & $E_{upp}$ (K) & & & & \\
%\cline {1-3}
11$_{0,11}$-10$_{0,10}$ & 103575.4 & 29.9& $\tau$=5.07$\times$10$^{-3}$ & $\tau$=2.50$\times$10$^{-2}$ & $\tau$=1.37$\times$10$^{-3}$ & $\tau$=1.23$\times$10$^{-2}$\\

14$_{3,11}$-13$_{3,10}$ & 133030.7 & 67.3& $\tau$=7.06$\times$10$^{-3}$ & $\tau$=2.70$\times$10$^{-2}$ & $\tau$=1.78$\times$10$^{-3}$ & $\tau$=1.28$\times$10$^{-2}$\\
18$_{0,18}$-17$_{0,17}$ & 167728.4 & 77.1& $\tau$=1.16$\times$10$^{-2}$ & $\tau$=4.19$\times$10$^{-2}$ & $\tau$=2.89$\times$10$^{-3}$ & $\tau$=1.96$\times$10$^{-2}$\\
23$_{0,23}$-22$_{0,22}$ & 212788.7 &123.8& $\tau$=1.64$\times$10$^{-2}$ & $\tau$=4.30$\times$10$^{-2}$ & $\tau$=3.72$\times$10$^{-3}$ & $\tau$=1.91$\times$10$^{-2}$\\
25$_{4,21}$-24$_{4,20}$ & 237712.0 &182.8& $\tau$=1.62$\times$10$^{-2}$ & $\tau$=2.84$\times$10$^{-2}$ & $\tau$=3.30$\times$10$^{-3}$ & $\tau$=1.19$\times$10$^{-2}$\\
28$_{0,28}$-27$_{0,27}$ & 257646.2 &181.4& $\tau$=2.02$\times$10$^{-2}$ & $\tau$=3.61$\times$10$^{-2}$ & $\tau$=4.14$\times$10$^{-3}$ & $\tau$=1.51$\times$10$^{-2}$\\
30$_{0,30}$-29$_{0,29}$ & 275588.1 &207.4& $\tau$=2.14$\times$10$^{-2}$ & $\tau$=3.20$\times$10$^{-2}$ & $\tau$=4.18$\times$10$^{-3}$ & $\tau$=1.30$\times$10$^{-2}$\\
\cline {1-7}
 & & & $d$$_{\rm sou}$=2'' & $d$$_{\rm sou}$=5'' & $d$$_{\rm sou}$=2'' & $d$$_{\rm sou}$=5''\\
Transition & Freq. (MHz) & $E_{upp}$ (K) & $N$=8.0$\times$10$^{15}$\,cm$^{-2}$ & $N$=4.8$\times$10$^{15}$\,cm$^{-2}$ & $N$=2.5$\times$10$^{14}$\,cm$^{-2}$ & $N$=4.4$\times$10$^{15}$\,cm$^{-2}$ \\
\hline
%Transition & Freq. (MHz) & $E_{upp}$ (K) & & & & \\
%\cline {1-3}
11$_{0,11}$-10$_{0,10}$ & 103575.4 & 29.9& $\tau$=1.35$\times$10$^{-2}$ & $\tau$=1.20$\times$10$^{-1}$ & $\tau$=3.80$\times$10$^{-4}$ & $\tau$=4.16$\times$10$^{-2}$\\
14$_{3,11}$-13$_{3,10}$ & 133030.7 & 67.3& $\tau$=1.88$\times$10$^{-2}$ & $\tau$=1.30$\times$10$^{-1}$ & $\tau$=4.94$\times$10$^{-4}$ & $\tau$=4.32$\times$10$^{-2}$\\
18$_{0,18}$-17$_{0,17}$ & 167728.4 & 77.1& $\tau$=3.11$\times$10$^{-2}$ & $\tau$=2.01$\times$10$^{-1}$ & $\tau$=8.02$\times$10$^{-4}$ & $\tau$=6.64$\times$10$^{-2}$\\
23$_{0,23}$-22$_{0,22}$ & 212788.7 &123.8& $\tau$=4.36$\times$10$^{-2}$ & $\tau$=2.06$\times$10$^{-1}$ & $\tau$=1.03$\times$10$^{-3}$ & $\tau$=6.48$\times$10$^{-2}$\\
25$_{4,21}$-24$_{4,20}$ & 237712.0 &182.8& $\tau$=4.31$\times$10$^{-2}$ & $\tau$=1.37$\times$10$^{-1}$ & $\tau$=9.16$\times$10$^{-4}$ & $\tau$=4.01$\times$10$^{-2}$\\
28$_{0,28}$-27$_{0,27}$ & 257646.2 &181.4& $\tau$=5.39$\times$10$^{-2}$ & $\tau$=1.73$\times$10$^{-1}$ & $\tau$=1.15$\times$10$^{-3}$ & $\tau$=5.10$\times$10$^{-2}$\\
30$_{0,30}$-29$_{0,29}$ & 275588.1 &207.4& $\tau$=5.71$\times$10$^{-2}$ & $\tau$=1.54$\times$10$^{-1}$ & $\tau$=1.16$\times$10$^{-3}$ & $\tau$=4.40$\times$10$^{-2}$\\
\hline
%\flushleft

\end{tabular}
\end{center}
%}}
%}

{\bfseries Note.} Opacities for some lines of CH$_2$CHCN g.s. at different frequencies considering
different source diameters and column densities (see text, Sect. \ref{sect_opa}).

\end{table*}

\begin{figure*}[ht]
\centering
\includegraphics[angle=0,width=0.6\textwidth]{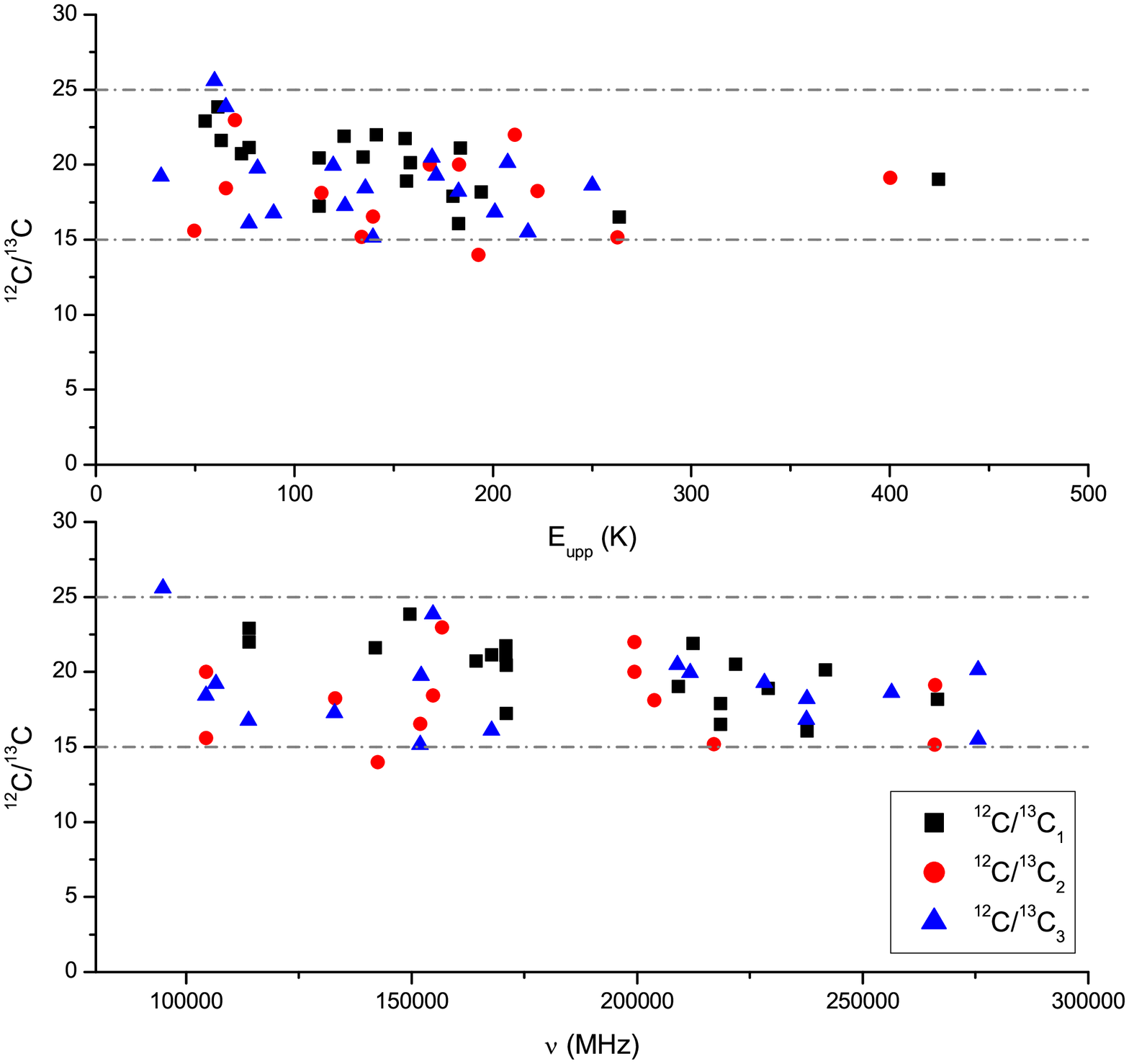}
\caption{$^{12}$C/$^{13}$C ratios of the observed line intensities for a given transition
in function of the upper level energy (top panel) and the frequency (bottom panel).}
\label{fig_12C-13C}
\end{figure*}

Figure \ref{fig_12C-13C} shows the $^{12}$C/$^{13}$C ratios of the observed 
line intensities for a given transition against its upper level energy and its frequency. 
As for the rotational diagrams, unblended lines have been used for deriving these ratios.  
We observe that most of these ratios are between 15 and 25 and we do not observe a clear
decline of this ratio neither with the increasing of upper state energy nor with the
increasing of the frequency. In case of optically thick lines, we should expect these 
large opacities for lines at the end of the 1.3 mm window (240-280 GHz) where the upper 
state energies are above 150 K even for transitions of $K$$_a$=0,1. Figure \ref{fig_12C-13C} 
suggests
that the CH$_2$CHCN g.s. lines have $\tau$$<$1. Nevertheless,
if the bulk of the emission comes from a very small region ($<$1''), opacities will be larger than 1.

From Fig. \ref{fig_12C-13C} we can estimate the average intensity
ratios for each $^{13}$C isotopologue being 20$\pm$6, 18$\pm$5, and 19$\pm$6
for $^{12}$C/$^{13}$C$_1$, $^{12}$C/$^{13}$C$_2$, and  $^{12}$C/$^{13}$C$_3$, respectively.
These results, together with the $^{12}$C/$^{13}$C column density ratio derived in Sect. \ref{sec_isot},
suggest possible chemical fractionation
enhancement of the $^{13}$C isotopologues of vinyl cyanide.

\subsubsection{Vibrational temperatures}
\label{vib_temp}

We can estimate the vibrational temperature between the different vibrational
modes of the vinyl cyanide according to:

\begin{equation}
\frac{N (\rm CH_2CHCN\;\;\varv_x)}{N (\rm CH_2CHCN)} = 
\frac{\exp \left( \begin{array}{c} - \frac{E_{\varv_x}}{T_{\rm vib}}\end{array} 
\right)}{f_{\nu}} ,
\end{equation}

where $\varv_x$ identifies the vibrational mode, E$_{\varv_x}$ is the energy of 
the corresponding vibrational state (328.5, 478.6, 657.8, 806.4/809.9, and 987.9\,K for $\varv_{11}$=1, $\varv_{15}$=1, $\varv_{11}$=2,
$\varv_{10}$=1$\Leftrightarrow$($\varv_{11}$=1,$\varv_{15}$=1), and $\varv_{11}$=3, respectively), T$_{\rm vib}$ is
the vibrational temperature, $f_{\nu}$ is the vibrational partition function,
$N$(CH$_2$CHCN $\varv_x$) is the column density of the vibrational state, and
$N$(CH$_2$CHCN) is the total column density of vinyl cyanide. Taking into
account the relation $N$(CH$_2$CHCN)=$N_{\rm g.s.}$$\times$$f_{\nu}$ 
and assuming the same partition function for these species in
the ground and in the vibrationally excited states,
we only need the energy of each vibrational state and the calculated column density to
derive the vibrational temperatures.  The vibrational temperature ($T_{\rm vib}$) is
given as a lower limit, since the vibrationally excited gas emitting region
may not coincide with that of the ground state.

From the column density results, the $T_{\rm vib}$ in the hot narrow component for
each vibrationally excited level were $\simeq$268$\pm$80\,K, $\simeq$246$\pm$74\,K,
$\simeq$265$\pm$132\,K, $\simeq$402$\pm$201\,K, and $\simeq$385$\pm$192\,K for
$\varv_{11}$=1, $\varv_{15}$=1, $\varv_{11}$=2, $\varv_{10}$=1$\Leftrightarrow$($\varv_{11}$=1,$\varv_{15}$=1), and
$\varv_{11}$=3, respectively. In the same way, the $T_{\rm vib}$ in the cold narrow
component for each vibrationally excited level were $\simeq$237$\pm$71\,K,
$\simeq$208$\pm$62\,K, $\simeq$220$\pm$110\,K, $\simeq$324$\pm$162\,K, and
$\simeq$330$\pm$165\,K for $\varv_{11}$=1, $\varv_{15}$=1, $\varv_{11}$=2,
$\varv_{10}$=1$\Leftrightarrow$($\varv_{11}$=1,$\varv_{15}$=1), and $\varv_{11}$=3, respectively. The average
vibrational temperature for $\varv_{11}$=1,2, and $\varv_{15}$=1, from both narrow
components was 252$\pm$76\,K, 242$\pm$121\,K, and 227$\pm$68\,K, respectively. In
the case of $\varv_{10}$=1$\Leftrightarrow$($\varv_{11}$=1,$\varv_{15}$=1) and $\varv_{11}$=3, the derived
$T_{\rm vib}$ is larger than the $T_{\rm rot}$ in the hot narrow component (320\,K),
which could suggest an inner and hotter region for the emission of these
vibrationally excited states of vinyl cyanide. Moreover, a tendency to increase
the vibrational temperature with the vibrational energy of the considered state
is observed. Vibrational transitions imply ro-vibrational states that may be excited by 
dust IR photons or collisions with the most abundant molecules in the cloud. Nevertheless,
collisional rates are needed to evaluate the excitation mechanisms. The observed differences
between $T_{\rm rot}$ and $T_{\rm vib}$ indicate either a far-IR pumping of the highly excited
vibrational levels or the presence of a strong temperature gradient towards the inner regions.
Some internal heating might be reflected in temperature and density
gradients due to processes such as, for example, star formation.

\subsection{Detection of isocyanide species}
\label{isoCN}

We searched for the isocyanide counterparts of vinyl, ethyl, and methyl cyanide,
cyanoacetylene, and cyanamide in our line survey. In this section we report the 
first detection towards Orion-KL of methyl isocyanide and a tentative
detection of vinyl isocyanide. The first to sixth
columns of Table \ref{tab_comparison} show the cyanide and isocyanide molecules
studied in Orion-KL, their column density values in the components where we
assumed emission from the isocyanides, the column density ratio between the
cyanide and its isocyanide counterpart, the same ratio obtained by previous
authors in Sgr B2 and TMC-1 sources, and the difference of the bond energies
between the -CN and -NC isomers.

 \begin{table*}[ht]
\begin{center}
\caption{Column densities of the isocyanide species and $N$(-CN)/$N$(-NC) ratios.}
% {\small{\sf{\scriptsize
\begin{tabular}[0.1\textwidth]{lcrrrc}
\label{tab_comparison} \\
%\caption{continued.}\\
\hline\hline
 Molecule   &\bfseries N$_{TOTAL}$ (cm$^{-2}$)&\multicolumn{3}{c}{[$N(-NC)/N(-CN)$]}& Isomerization energy \\\cline{3-5}
                   &                       & Orion-KL                            &Sgr B2                         & TMC-1                      &  (cm$^{-1}$)        \\

\hline
\bfseries {CH$_2$CHCN}    & (4$\pm$1)$\times10$$^{15}$     &                                 &                              &           &           \\
\bfseries {CH$_2$CHNC}    & $\leq$(4$\pm$2)$\times10$$^{14}$   & $\leq$(1.0$\pm$0.5)$\times$10$^{-1}$       &5$\times$10$^{-3}$($a$)       &           &   8658($a$)\\
\bfseries {CH$_3$CN}      & (3.2$\pm$0.9)$\times10$$^{16}$     &                                 &                              &           &           \\
\bfseries {CH$_3$NC}      & (6.0$\pm$3.0)$\times10$$^{13}$     &(2$\pm$1)$\times$10$^{-3}$       &2$\times$10$^{-2}$($a$)&$\geq$9$\times$10$^{-2}$($c$)& 9486($a$)\\
                          &                                    &                                 &(3-5)$\times$10$^{-2}$($b$)   &           &            \\
\bfseries {CH$_3$CH$_2$CN}& (7$\pm$2)$\times10$$^{16}$         &                                 &              &                           &            \\
\bfseries {CH$_3$CH$_2$NC}&$\leq$(2.0$\pm$0.6)$\times$10$^{14}$&$\leq$(3$\pm$2)$\times$10$^{-3}$ &$\leq$3$\times$10$^{-1}$($a$) &           &   8697($a$)\\
\bfseries {HCCCN}         & (4$\pm$1)$\times10$$^{15}$         &                                 &              &                           &            \\
\bfseries {HCCNC}         &$\leq$(3$\pm$1)$\times10$$^{13}$    &$\leq$(8$\pm$4)$\times$10$^{-3}$ &              &(2-5)$\times$10$^{-2}$($d$)&   6614($d$)\\
                          &                                    &                                 &              &8$\times$10$^{-3}$($e$)    &            \\
\bfseries {HNCCC}         &$\leq$(3$\pm$1)$\times10$$^{13}$    &$\leq$(8$\pm$4)$\times$10$^{-3}$ &              &(2-6)$\times$10$^{-3}$($f$)&  17745($d$)\\
                          &                                    &                                 &              &1$\times$10$^{-3}$($e$)    &            \\
\bfseries {NH$_2$CN}      &$\leq$(3$\pm$1)$\times$10$^{13}$    &                                 &              &                           &            \\
\bfseries {NH$_2$NC}      &$\leq$(5$\pm$2)$\times$10$^{13}$    & \multicolumn{1}{c}{...} & \multicolumn{1}{c}{...} & \multicolumn{1}{c}{...}&  18537($g$)\\

\hline
\hline

%\hline
%\flushleft

\end{tabular}
%}}
%}
\end{center}
{\bfseries Note.} Derived column densities for the cyanide and isocyanide
species (Col. 2). Columns 3-5 show the ratio between the cyanide and its
isocyanide isomer in this work and that derived from other authors in Sgr B2 and
TMC-1. Col. 6 gives the energy difference for the isomerization between the
isocyanide species and its corresponding cyanide. ($a$) \citeauthor{rem05}
(\citeyear{rem05}). ($b$) \citeauthor{cer88} (\citeyear{cer88}). ($c$)
\citeauthor{irv84} (\citeyear{irv84}). ($d$) \citeauthor{kaw92a}
(\citeyear{kaw92a}). ($e$) \citeauthor{ohi98} (\citeyear{ohi98}). ($f$)
\citeauthor{kaw92b} (\citeyear{kaw92b}). ($g$) \citeauthor{tur75}
(\citeyear{tur75}).
\end{table*}

\begin{figure*}[ht]
\centering
\includegraphics[angle=270,width=0.9\textwidth]{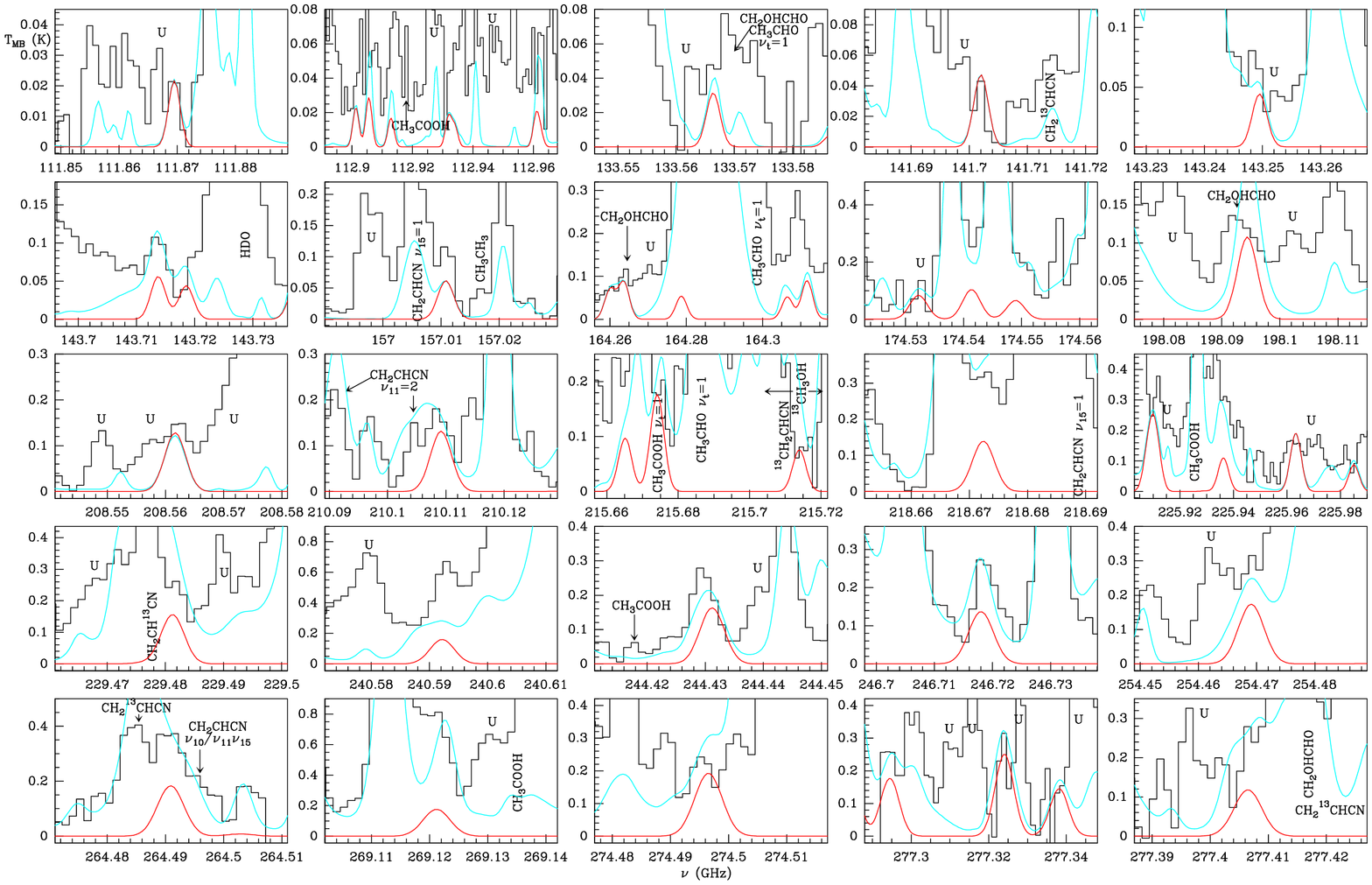}
\caption{Observed lines from Orion-KL (histogram spectra) and model
 (thin red curves) of vinyl isocyanide in its ground state. The cyan line corresponds to the model of
the molecules we have already studied in this survey (see text Sect. \ref{sec_cd})
including the CH$_2$CHCN species. We consider the detection as tentative.
A v$_{\rm LSR}$ of 5\,km s$^{-1}$ is assumed.}
\label{fig_CH2CHNC}
\end{figure*}

Vinyl isocyanide (CH$_2$CHNC) is an isomer of the unsaturated hydrocarbon vinyl
cyanide. The structure differences between the vinyl cyanide and isocyanide are
due to the CNC and CCN linear bonds and their energies, where CCN displays
shorter bond distances. The bonding energy difference between vinyl cyanide and
isocyanide is 8658 cm$^{-1}$ (24.8 kcal mol$^{-1}$) (\citeauthor{rem05}
\citeyear{rem05}), with the cyanide isomer being more stable than the
isocyanide. We have tentatively detected vinyl isocyanide in our line survey (Fig.
\ref{fig_CH2CHNC}), with 28 unblended lines and 26 partially blended lines 
from a total of 96 detectable lines.
This detection is just above the confusion
limit. In Table \ref{tab_VyISOCN} we show spectroscopic and observational
parameters of detected lines of vinyl isocyanide 
(rotational constants were derived fitting all experimental data
from \citeauthor{bol70} \citeyear{bol70}, \citeauthor{yam75} \citeyear{yam75}, and
\citeauthor{bes82} \citeyear{bes82}; the dipole moments
were from \citeauthor{bol70} \citeyear{bol70}).
For modeling this molecule we
assume the same physical conditions as those found for the vinyl cyanide species
(we consider both narrow components). We derived a column density of
$\leq$(3$\pm$2)$\times$10$^{14}$ cm$^{-2}$ (hot narrow component) and
$\leq$(5$\pm$3)$\times$10$^{13}$\,cm$^{-2}$ (cold narrow component). We estimate a
$N$(CH$_2$CHNC)/$N$(CH$_2$CHCN) ratio of $\leq$0.10$\pm$0.05, while \citeauthor{rem05}
(\citeyear{rem05}) derived a ratio of about $\leq$0.005 toward Sgr B2 with an
upper limit for the vinyl isocyanide column density of
$\leq$1.1$\times$10$^{13}$ cm$^{-2}$.

\begin{figure*}[ht]
\centering
\includegraphics[angle=270,width=0.60\textwidth]{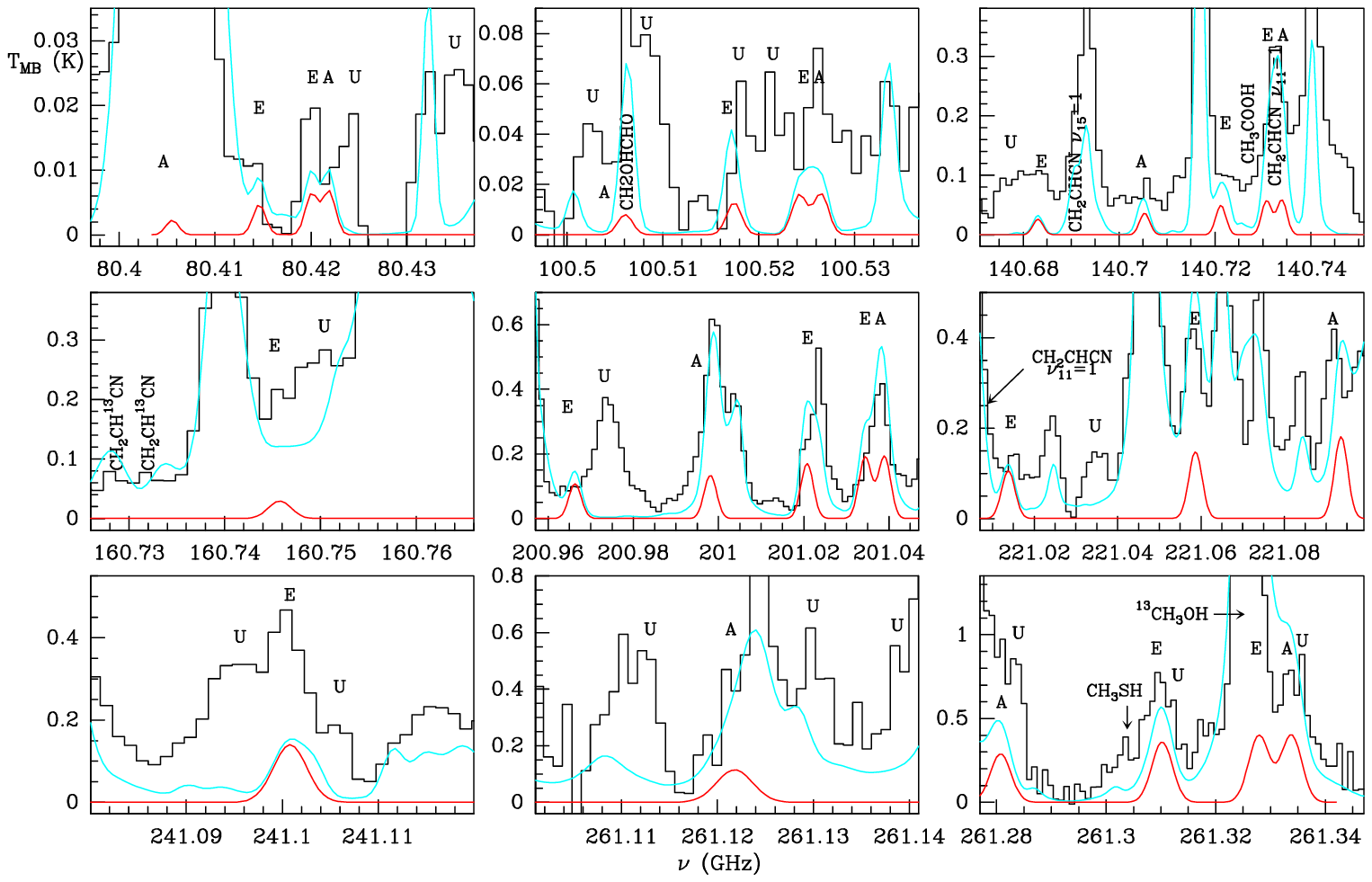}
\caption{Observed lines from Orion-KL (histogram spectra) and model
 (thin red curves) of methyl isocyanide in its ground state.
 The cyan line corresponds to the model of
the molecules we have already studied in this survey (see text Sect. \ref{sec_cd})
including the CH$_2$CHCN species. 
A v$_{\rm LSR}$ of 5\,km s$^{-1}$ is assumed.}
\label{fig_CH3NC}
\end{figure*}

Methyl cyanide (CH$_3$CN) is a symmetric rotor molecule whose internal rotor
leads to two components of symmetry A and E. The column densities of the ground
state obtained for both A and E sub-states using an LVG model were derived 
by \citeauthor{tom14} (\citeyear{tom14}) in Orion-KL. They fitted
separately different series of $K$-ladders transitions ($J$=6-5, $J$=12-11, $J$=13-12, $J$=14-13).
We averaged the model results for these four series at the IRc2 position deriving
a column density of 3.1$\times$10$^{16}$\,cm$^{-2}$ and a kinetic temperature of $\simeq$265\,K.
The CH$_3$CN molecule has a
metastable isomer named methyl isocyanide (CH$_3$NC) that has been found in
dense interstellar clouds (Sgr B2) by \citeauthor{cer88} (\citeyear{cer88}) and
\citeauthor{rem05} (\citeyear{rem05}). The bonding energy difference between
methyl cyanide and isocyanide is 9486 cm$^{-1}$ (27.1 kcal mol$^{-1}$)
(\citeauthor{rem05} \citeyear{rem05}). We observed methyl isocyanide in
Orion-KL for the first time (Fig. \ref{fig_CH3NC} available online). For modeling the weak lines of methyl
isocyanide we assume a hot core component ($T$=265\,K,
$d_{\rm sou}$=10", offset=3", v$_{\rm LSR}$=5\,km\,s$^{-1}$,
$\Delta$$v$=5\,km\,s$^{-1}$) consistent with those derived by \citeauthor{tom14} (\citeyear{tom14}).
Rotational constants were derived from a fit to
the data reported by \citeauthor{bau70} (\citeyear{bau70}),
\citeauthor{pra11} (\citeyear{pra11}). The constants $H$$_J$, $L$$_J$, and $L$$_{JKKK}$
have been fixed to the values derived by \citeauthor{pra11} (\citeyear{pra11}).
The constants $A$ and $D$$_K$ were from \citeauthor{pli95} (\citeyear{pli95}).
Dipole moment was that of \citeauthor{gri00} (\citeyear{gri00}).
We derived a
column density of (3.0$\pm$0.9)$\times10$$^{13}$ cm$^{-2}$ for each A and E
symmetry sub-states. We determined a $N$(CH$_3$NC)/$N$(CH$_3$CN) ratio of 0.002
(assuming the three hot core components of those showed above, 
which is a factor 15-25 lower than the value obtained by
\citeauthor{cer88} (\citeyear{cer88}) toward Sgr B2. \citeauthor{def85}
(\citeyear{def85}) by means of chemical models predicted this ratio in dark
clouds in the range of 0.1-0.4.

Ethyl cyanide (CH$_3$CH$_2$CN) is a heavy asymmetric rotor with a rich spectrum.
In our previous paper (\citeauthor{adm13} \citeyear{adm13}), three cloud
components were modeled in LTE conditions in order to determine the column
density\footnote{We found a typographical error of a factor of 2 difference in the column
density of the hot core component 1 for the ground and excited states in our
previous paper (\citeauthor{adm13} \citeyear{adm13}) hence the isotopic
abundance has to be modified. We have attached the tables of column densities
and that of isotopic abundance for CH$_3$CH$_2$CN in Appendix B.} of this
molecule. We obtained a total column density of (7$\pm$2)$\times10$$^{16}$
cm$^{-2}$ for this species.

The bonding energy difference between ethyl cyanide and isocyanide is 8697
cm$^{-1}$ (24.9 kcal mol$^{-1}$) (\citeauthor{rem05} \citeyear{rem05}). 
The spectroscopic parameters used for ethyl isocyanide (CH$_3$CH$_2$NC)
were obtained from recent measurements in Lille up to 1 THz by Margul\`es et al. (2014, in preparation).
For CH$_3$CH$_2$NC we provide an upper limit to its column density
of (2$\pm$1)$\times$10$^{14}$ cm$^{-2}$. Then, we derived a
$N$(CH$_3$CH$_2$NC)/$N$(CH$_3$CH$_2$CN) ratio of 0.003. This value is 100-fold
lower than the upper limit value obtained by \citeauthor{rem05}
(\citeyear{rem05}) toward Sgr B2 of $\leq$0.3.

We observe that the upper limit for the CH$_2$CHNC column density is 5-fold higher than the value of
methyl isocyanide, and holds a similar order of magnitude relationship with the upper limit
column density of the tentatively detected ethyl isocyanide.

Cyanoacetylene (HCCCN) is a linear molecule with a simple spectrum.
Its lines emerge from diverse parts of the
cloud (\citeauthor{esp13b} \citeyear{esp13b}), although mainly from the hot
core. The model of the HCCCN lines was set up using LVG conditions. The
authors determined a total column density of (3.5$\pm$0.8)$\times10$$^{15}$
cm$^{-2}$.

Isocyanoacetylene (HCCNC) is a stable isomer of cyanoacetylene and has an
energy barrier of 6614 cm$^{-1}$ (18.9 kcal mol$^{-1}$). Owing to high overlap
problems in our data we only found one line of HCCNC free of blending at
99354.2\,MHz. In order to obtain an upper limit for its column density we assumed
the same physical components as those of \citeauthor{esp13b}
(\citeyear{esp13b}). Spectroscopic parameters were derived fitting the lines reported by 
\citeauthor{gua92} (\citeyear{gua92}); dipole moment was taken from \citeauthor{gri00} (\citeyear{gri00}).
We obtained an upper limit to the HCCNC column density of
$\leq$(3$\pm$1)$\times10$$^{13}$\,cm$^{-2}$. We estimated an upper limit for the
$N$(HCCNC)/$N$(HCCCN) ratio of $\leq$0.008. HCCNC was observed for the first
time toward TMC-1 (three rotational lines in the frequency range 40-90\,GHz) by
\citeauthor{kaw92a} (\citeyear{kaw92a}). They obtained a $N$(HCCNC)/$N$(HCCCN)
ratio in the range 0.02-0.05 in that dark cloud, which is around 2-6 times
higher than our upper limit. \citeauthor{ohi98} (\citeyear{ohi98}) provided an
upper limit value of $\leq$0.001 also in TMC-1. This molecule has also been
detected in the envelope of the carbon star IRC+10216 by \citeauthor{gen97}
(\citeyear{gen97}).

The other carbene-type isomer of HCCCN is 3-imino-1,2-propadienylidene (HNCCC)
that was predicted to have a relative energy of about 17744.6 cm$^{-1}$ with
respect to HCCCN (\citeauthor{kaw92b} \citeyear{kaw92b}). We provided a
tentative detection of this isomer in our survey (Fig. \ref{fig_HNCCC} available online). Its
column density, (3$\pm$1)$\times10$$^{13}$\,cm$^{-2}$, has been obtained by assuming
the same cloud components as those of \citeauthor{esp13b} (\citeyear{esp13b}). 
Spectroscopic parameters were derived from a fit to lines reported by
\citeauthor{kaw92b} (\citeyear{kaw92b}) and \citeauthor{hir93} (\citeyear{hir93}), and
three lines observed in IRC+10216 with an accuracy of 0.3 MHz. 
Dipole moment was that of \citeauthor{bot92} (\citeyear{bot92}).
We derived a $N$(HNCCC)/$N$(HCCCN) upper limit ratio of 0.008.
\citeauthor{kaw92a} (\citeyear{kaw92a}) obtained a $N$(HNCCC)/$N$(HCCCN) ratio
in the range 0.002-0.006 in TMC-1.

After the detection of cyanamide (NH$_2$CN) by \citeauthor{tur75}
(\citeyear{tur75}), \citeauthor{cum86} (\citeyear{cum86}), and
\citeauthor{bel13} (\citeyear{bel13}) in Sgr B2, we report a tentative detection
of cyanamide (NH$_2$CN) in Orion-KL (see Fig. \ref{fig_NH2CN} available online). 
Frequencies, energies, and line intensities for O$^+$-NH$_2$CN and O$^-$-NH$_2$CN 
were those published in the JPL catalog (based on the works of \citeauthor{Read1986}
\citeyear{Read1986} and \citeauthor{Birk1988}
\citeyear{Birk1988}). We estimated a column density 
$\leq$(3$\pm$1)$\times10$$^{13}$ cm$^{-2}$ (O$^+$+O$^-$) 
by assuming that its lines are coming only from one component (hot
core) at 200\,K (v$_{\rm LSR}$=5\,km\,s$^{-1}$, $\Delta$$v$=5\,km\,s$^{-1}$,
$d_{\rm sou}$=10", offset=2"). NH$_2$CN has an isomer differing only in the CN group,
so that, the isomerization energy between the cyanamide and
isocyanamide (NH$_2$NC) is 18537 cm$^{-1}$ (\citeauthor{vin80}
\citeyear{vin80}).
%This high value leads to support the possible existence of isolated NH$_2$NC (\citeauthor{vin80} \citeyear{vin80}).
In this work, we also provided only an upper limit column density of isocyanamide (O$^+$+O$^-$) 
being $\leq$(5$\pm$1)$\times10$$^{13}$ cm$^{-2}$. Spectroscopic parameters 
were derived fitting the rotational lines reported by \citeauthor{scha81} (\citeyear{scha81}).
Dipole moment was that of \citeauthor{ich82} (\citeyear{ich82}) 
from ab-initio calculations.

In Table \ref{tab_comparison}, we give values of interconversion energies
between cyanide and isocyanide molecules. These interconversion barriers are
rather high, and it is unlikely that under astronomical environments such as the
hot cores the isocyanide isomers are produced by rearrangement of the
corresponding cyanide species. \citeauthor{rem05} (\citeyear{rem05}), proposed
that non-thermal processes (such as shocks or enhanced UV flux in the
surrounding medium) may be the primary route to the formation of interstellar
isocyanides by the conversion of the cyanide to its isocyanide counterpart.
Nevertheless, other formation routes have to be explored in order to explain
their presence in environments dominated by thermal processes. Dissociative
recombination reactions on the gas phase probably lead to the formation of the
cyanide or isocyanide molecules. Depending on the structure of the protonated
hydrocarbon and the branching ratios of the dissociative recombination pathway,
H$_2$C$_3$N$^+$ might produce cyanoacetylene and isocyanoacetylene, and similarly 
C$_2$H$_4$N$^+$ could yield methyl cyanide and methyl isocyanide
(\citeauthor{gre79} \citeyear{gre79}). \citeauthor{def85} (\citeyear{def85})
found that the calculated ratio of the formation of the protonated precursor
ions (CH$_3$CNH$^+$ and CH$_3$NCH$^+$) was in agreement with the detection of
CH$_3$NC in dark clouds (\citeauthor{irv84} \citeyear{irv84}).
In the same way, the recombination reaction of C$_2$H$_6$N$^+$ could give ethyl
isocyanide (\citeauthor{bou92} \citeyear{bou92}). Once the isocyanides are
formed they remain as metastable species due to the high barrier quoted above
supporting the possible existence of isolated isocyanides (\citeauthor{vin80}
\citeyear{vin80}).
On the other hand, a recent experimental study of the interaction of the
diatomic radical CN and the $\pi$-system C$_2$H$_4$ confirms that the possible
pathway to CH$_2$CHNC becomes negligible even at temperatures as high as
1500\,K (\citeauthor{bal00} \citeyear{bal00}, \citeauthor{leo12}
\citeyear{leo12}). Since the cyanide molecules are strongly related to the dust
chemistry (\citeauthor{bla87} \citeyear{bla87}, \citeauthor{cha92}
\citeyear{cha92}, \citeauthor{cas93} \citeyear{cas93}, \citeauthor{rod01}
\citeyear{rod01}, \citeauthor{gar08} \citeyear{gar08}, \citeauthor{bel13}
\citeyear{bel13}), we also can infer a probable origin for the isocyanides
from reactions on grain mantles.

\section{Discussion}
\label{disc}

\subsection{Abundances and column density ratios between the cyanide species}

Table \ref{tab_ratios} shows the ground state abundances in the hot core (or hot
core + plateau) component of the studied species in this work, and the column
density ratios between vinyl cyanide and other cyanide molecules. Results
provided by different authors in Orion-KL, in the well studied star forming
region Sgr B2, in the star forming complex G34.3+0.2 (hot core), and in the dark
molecular cloud TMC-1,
are also given in this table.

For Orion-KL our study covers a wide frequency range allowing detailed
modeling of the molecular emission. Moreover, the molecular abundances obtained
from other authors, and shown in Table \ref{tab_ratios}, are often obtained with
different telescopes and different assumptions on the size and physical
conditions of Orion-KL. For this reason, these abundances are given in Table
\ref{tab_ratios} for comparison purposes, but we will focus on the results
obtained in this work that have been derived from a common set of assumptions,
sizes, and physical conditions for Orion-KL.

In order to estimate molecular abundances for the cyanide and isocyanide species
we assume that the column density of H$_2$ ($N_{\rm H_2}$) is
4.2$\times10$$^{23}$\,cm$^{-2}$ for the hot core,
2.1$\times10$$^{23}$\,cm$^{-2}$ for the plateau, and 7.5$\times10$$^{22}$\,cm$^{-2}$
for the compact ridge
and for the extended ridge, as derived by \citeauthor{Tercero10}
(\citeyear{Tercero10}).

The total abundance for the CH$_2$CHCN ground state, derived from all the
components (hot core + mix hot core-plateau) (see Table \ref{tab_prop}), was
$X(N_{\rm CH_2CHCN}/N_{\rm H_2})$=(2.0$\pm$0.6)$\times10$$^{-8}$. By means of the
derived vibrational temperatures we can estimate the vibrational partition
function that follows the equation (\ref{eq_VPF}) for a Boltzmann distribution in both narrow components 
(1.7 and 1.5 for hot and cold narrow components,
respectively) and correct the ground state column density to the total one (see Sect. \ref{vib_temp}).
Taking into account these results for the vibrational partition function, we
obtained $X_{CH_2CHCN}$$\simeq$(3.1$\pm$0.9)$\times10$$^{-8}$.

\begin{equation}
\label{eq_VPF}
f_{\nu} = 1 + \sum_{x=1}^{n}d_x \exp(\begin{array}{c} -
  \frac{E_{\varv_{x}}}{T_{\rm vib}} \end{array})  \,
\end{equation}

where $d_x$ is the degeneracy of the vibrational mode $x$, and for low T$_{\rm vib}$ leads to f$_{\nu}$ $\simeq$ 1.

Assuming the column density values of CH$_3$CN of \citeauthor{tom14} (\citeyear{tom14}),
the abundance for CH$_3$CN ground state in the hot core
component was $\simeq$(1.0$\pm$0.3)$\times10$$^{-7}$. In order to estimate the
correction to the column density of CH$_3$CN from excited vibrational states, we
have derived the column density of this molecule in its $\varv_{8}$=1 lower energy
state (525.2\,K) and found a
$N({\rm CH_3CN,}\varv_{8}=1)$$\simeq$1.4$\times10$$^{15}$\,cm$^{-2}$ and
$T_{\rm vib}$$\simeq$159\,K (considering only the hot core and plateau components).
Hence, the vibrational partition function is $\sim$1.04 and
$X_{\rm Total}$$\simeq$$X_{\rm ground}$ for methyl cyanide.

For ethyl cyanide we use the column density results of \citeauthor{adm13}
(\citeyear{adm13}) (see Appendix B). We determine the
$X(N_{\rm CH_3CH_2CN}/N_{\rm H_2})$ ratio being (1.8$\pm$0.5)$\times10$$^{-7}$ for the
ground state. Assuming the vibrational temperatures obtained in
\citeauthor{adm13} (\citeyear{adm13}) $-$$\simeq$160$\pm$50\,K,
$\simeq$185$\pm$55\,K, and $\simeq$195$\pm$95\,K, for $\varv_{13}$/$\varv_{21}$
($E_{u}$=306.3/315.4\,K), $\varv_{20}$ ($E_{u}$=531.2\,K), and $\varv_{12}$
($E_{u}$=763.4\,K)$-$, the estimated vibrational partition function is 1.4, so
we derived an abundance ratio $X$$\simeq$(2.5$\pm$0.8)$\times10$$^{-7}$ for
ethyl cyanide.

\citeauthor{esp13b} (\citeyear{esp13b}) derived an abundance of
7.3$\times10$$^{-9}$ for HC$_3$N (hot core + plateau) in the ground state.

Assuming a mean vibrational temperature of 360\,K in the hot core calculated by
these authors, the vibrational partition function from $\varv_{5}$ ($E_{u}$=954.48), $\varv_{6}$ ($E_{u}$=718.13), 
and $\varv_{7}$=1 ($E_{u}$=320.45),2 ($E_{u}$=642.67) is $\sim$2.6 for the hot core components. Applying this
correction in the hot core, we obtained a total abundance of
$X_{\rm HC_3N}$$\simeq$1.3$\times$10$^{-8}$ (hot core + plateau).

For NH$_2$CN we determine a molecular abundance $X(N_{\rm NH_2CN}/N_{\rm H_2})$ of $\leq$(7$\pm$2)$\times10$$^{-11}$.

The column density ratio between the unsaturated hydrocarbon CH$_2$CHCN and
other -CN bearing molecules such as CH$_3$CH$_2$CN, CH$_3$CN, HC$_3$N, and
NH$_2$CN, $N$(CH$_2$CHCN)/$N$(X-CN), could be used to track a possible different
evolutionary state described by different chemical models and to provide inputs
for the chemical modeling of the cloud.

We obtain an abundance ratio $N$(CH$_2$CHCN)/$N$(X-CN)$<$1 for the saturated
cyanide molecules (CH$_3$CH$_2$CN and CH$_3$CN); these ratios related with
methyl and ethyl cyanide obtained by different authors are lower (in general) in
Orion-KL than in the galactic center (Sgr B2). We also note that the relative
abundance of CH$_3$CH$_2$CN with respect to vinyl cyanide in Orion is 2 times
larger than that found in the hot core G34.3+0.2. In contrast, in the dark cloud
TMC-1
vinyl cyanide is more abundant than the saturated -CN hydrocarbons.

The relative abundance between CH$_2$CHCN and HC$_3$N follows an opposite
tendency than that of methyl and ethyl cyanide: HC$_3$N is less abundant than
CH$_2$CHCN in Orion-KL and Sgr B2, whereas in TMC-1,  
HC$_3$N is, at least, two orders of magnitude more abundant than CH$_2$CHCN. Nevertheless, we
want to remark that in Table \ref{tab_ratios} we only address the hot core (or
hot core + plateau) abundances in our work. For vinyl and ethyl cyanide and
cyanamide these abundances correspond with the total abundance in the ground
state in Orion-KL. However, cyanoacetylene appears in all the Orion-KL
components, so its total abundance is higher than that of vinyl cyanide taking
into account the whole Orion-KL region covered by our observations.

Finally, we find an lower limit of 286 for $X$(CH$_2$CHCN)/$X$(NH$_2$CN).

The formation routes of the cyanide molecules in several environments have been
extensively discussed by different authors. As ethyl cyanide was mainly detected
in hot core regions (\citeauthor{joh84} \citeyear{joh84};
\citeauthor{sut85} \citeyear{sut85}; \citeauthor{bla87} \citeyear{bla87}), the
grain surface production (by hydrogenation of HC$_3$N) seemed to be the main
formation mechanisms for this molecule (\citeauthor{bla87} \citeyear{bla87};
\citeauthor{cha92} \citeyear{cha92}; \citeauthor{cas93} \citeyear{cas93}). On
the other hand, as well as in hot cores (where appeared correlated with ethyl
cyanide emission), vinyl cyanide has been detected in the dark could TMC-1
(\citeauthor{mat83} \citeyear{mat83}), indicating that gas phase production
was also important for forming these species. Chemical models of
\citeauthor{cha92} (\citeyear{cha92}) and \citeauthor{cas93} (\citeyear{cas93})
in hot cores predicted that CH$_2$CHCN forms in gas phase reactions involving
CH$_3$CH$_2$CN. \citeauthor{cas93} (\citeyear{cas93}) derived a relation between
the abundance ratio $X$(CH$_2$CHCN)/$X$(CH$_3$CH$_2$CN) and the evolutionary
stage of the core. This ratio has been used by several authors as a chemical
clock to estimate the cloud age (\citeauthor{fon08} \citeyear{fon08};
\citeauthor{mul08} \citeyear{mul08}). \citeauthor{bel09} (\citeyear{bel09})
performed a detailed analysis of the cyanide species detected in Sgr B2,
including chemical models based on that of \citeauthor{gar08}
(\citeyear{gar08}). Taking into account the observed relative abundances between
these species, they conclude that the main formation mechanism for alkyl
cyanides is probably the sequential addition of CH$_2$ or CH$_3$ radicals to CN,
CH$_2$CN and C$_2$H$_4$CN on the grain surfaces. Formation of methyl cyanide is
dominated by reactions on the grains by addition of CH$_3$ and CN radicals, but
it may also be formed by gas-phase processes after the evaporation of HCN. Vinyl
cyanide is formed predominantly in the gas-phase through the reaction of CN with
C$_2$H$_4$. Then CH$_2$CHCN accretes onto the grains being a potential
precursor, together with HC$_3$N, of ethyl cyanide and $n$-propyl cyanide. After the
evaporation of the ice mantles, vinyl cyanide is efficiently formed again in the
gas phase.

\begin{table*}
\begin{center}
\caption{Column density ratios and molecular abundances.}
%\begin{tabular}[0.1\textwidth]{lcccccccc}
 %{\small{\sf{\scriptsize
\begin{tabular}{lccccccccc}
\label{tab_ratios} \\
%\caption{continued.}\\
\hline\hline
 Molecule   &\multicolumn{2}{c}{\underline{Orion-KL}} &  \multicolumn{2}{c}{\underline{Sgr B2}}&  \multicolumn{2}{c}{\underline{G34.3+0.2}}  &  \multicolumn{2}{c}{\underline{TMC-1}}     \\%\cline{2-3} \cline{4-5}\cline{6-7}\cline{8-9}
        &  X  &R &  X  &R &  X  &R &  X  &R  \\
\hline
                          &                & & & & & & & \\
\bfseries {CH$_2$CHCN}    & (2.0$\pm$0.6)$\times10$$^{-8}$$\dagger$ &...&6.2$\times10$$^{-8}$($j$)&...&3.0$\times10$$^{-10}$($n$)&... &1.0$\times10$$^{-9}$($p$)&...   \\
                          & 1.5$\times10$$^{-9}$($a$)               &        & 6.0$\times10$$^{-8}$($k$) & &        &             &            \\
                          & 1.8$\times10$$^{-9}$($b$)               &           &                &     & &&\\
%\bfseries {CH$_2$CHNC}    & (3.0$\pm$1)$\times10$$^{-10}$            &...        &                &          &          &             &            &          &    &\\
                          &                &                                     &                &          &          &             &            &          \\
\bfseries {CH$_3$CN}      &(1.0$\pm$0.3)$\times10$$^{-7}$$\dagger$($d$)& 0.20$\dagger$($d$)     &3.0$\times10$$^{-8}$($k$)&0.40($e$)& ... & ... &7.5$\times10$$^{-10}$($p$)& 1.3($p$) \\
                          &4.0$\times10$$^{-9}$($a$)&0.39($a$)         &  &2.0($k$)               &          &             &            &          \\
                          & 7.8$\times10$$^{-9}$($b$)&0.23($b$)         &  &3.4($l$)               &          &             &            &          \\
                          &5.1$\times10$$^{-9}$($c$)& 1.7($e$)        &  &0.40($m$)
																			&          &             &            &          \\
                          &&0.18-1.8($f$)                               &       & 0.37($o$)               &          &             &            &          \\
                          &&0.48($g$)                                   &        &                 &          &             &            &          \\
                          &&                              &       &                 &          &             &            &          \\
%\bfseries {CH$_3$NC}      & (7$\pm$2)$\times10$$^{-11}$            &...          &...             &          &...       &             &...         &          &... &\\
                          &                &                                     &                &          &          &             &            &          \\
\bfseries {CH$_3$CH$_2$CN}&(1.8$\pm$0.5)$\times10$$^{-7}$$\dagger$($h$)& 0.11$\dagger$($h$) &6.0$\times10$$^{-10}$($k*$)&0.72($e$)&1.0$\times10$$^{-8}$($n$)&0.20($n$)& ...  &$>$2.00($q$)\\
                          &3.0$\times10$$^{-9}$($a$)&0.50($a$)    &             & 11($l$)              &          &             &            &          \\
                          &9.8$\times10$$^{-9}$($b$)&0.18($b$)    &             & 0.67($m$)               &          &             &            &          \\
                          &&0.15($e$)                              &             &  1.7($n$)             &          &             &            &          \\
                          &&0.14($f$)                              &             & 0.40($o$)               &          &             &            &          \\
                          &&0.06($g$)                              &     &          &          &             &            &          \\
                          &&                        &       &          &          &             &            &          \\
                          &&                                       &                 &          &          &             &            &          \\
%\bfseries {CH$_3$CH$_2$NC}& $\leq$(5$\pm$2)$\times10$$^{-10}$            &...    &...             &          &...       &             &...         &          &... &\\
                          &                &                                     &                &          &          &             &            &          \\
\bfseries {HCCCN}         &(7$\pm$2)$\times10$$^{-9}$$\dagger$($i$)& 2.9$\dagger$($i$)   &5.0$\times10$$^{-9}$($k$) &0.13($e$)& ... & ... &7.5$\times10$$^{-8}$($p$)&
0.01($p$)\\
                          &1.8$\times10$$^{-9}$($a$)&0.86($a$)      & & 12($k$) &          &             &            &          \\
                          &1.6$\times10$$^{-9}$($b$)& 1.1($b$)              &  & 61($o$)               &          &             &            &          \\
                          &1.8$\times10$$^{-9}$($c$)&2.11($e$)               & &               &          &             &            &          \\
                          &                         &0.6-1.5($f$)             &  &            &          &             &            &          \\
                          &&0.16($g$)                                           &                &          &          &             &            &          \\
                          &&                                     &                &          &          &             &            &          \\
%\bfseries {HCCNC}         &$\leq$(1.1$\pm$0.5)$\times10$$^{-10}$& ...            &...             &          &...       &             &...         &          &... &\\
                          &                &                                     &                &          &          &             &            &          \\
%\bfseries {HNCCC}         &$\leq$(1.1$\pm$0.5)$\times10$$^{-10}$& ...            &...             &          &...       &             &...         &          &... &\\
%                          &                &                                     &                &          &          &             &            &          &    &\\
\bfseries {NH$_2$CN}      &$\leq$(7$\pm$2)$\times10$$^{-11}$$\dagger$& $\geq$286$\dagger$          &9.0$\times10$$^{-11}$($k*$)&1.4($e$) &  ...     &  ...           & ...           & ...         \\
                          &&30($e$)                                          &&  14($o$)             &          &             &            &          \\
                          &                &                            & &         &          &             &            &          \\
%                          &                &                            & &                       &          &             &            &          &    &\\
%\bfseries {NH$_2$NC}      &$\leq$(6$\pm$3)$\times10$$^{-11}$& ...                &...             &          &...       &             &...         &          &... &\\
%                          &                & & & & & & & &  &\\
\hline
\hline
%\flushleft

\end{tabular}
%}}
%}
\end{center}
{\bfseries Note.} Abundances ($X$) and column density ratios between vinyl
cyanide and some studied cyanides ($R$) in Orion-KL and other sources.
$\dagger$ This work.
($a$) \citeauthor{sut95} (\citeyear{sut95}), hot core, telescope beam $\sim$13.7''. 
($b$) \citeauthor{bla87} (\citeyear{bla87}), hot core, telescope beam $\sim$30''. 
($c$) \citeauthor{per07} (\citeyear{per07}), hot core, source size 10''. 
($d$) \citeauthor{tom14} (\citeyear{tom14}), hot core (different components between 5-10'') + plateau (10''). 
($e$) \citeauthor{tur91} (\citeyear{tur91}). 
($f$) \citeauthor{joh84} (\citeyear{joh84}). 
($g$) \citeauthor{sch97} (\citeyear{sch97}). 
($h$) \citeauthor{adm13} (\citeyear{adm13}), hot core (4-10'') and mix hot core-plateau (25''). 
($i$) \citeauthor{esp13b} (\citeyear{esp13b}) hot core (7-10'') and plateau (20''). 
($j$) \citeauthor{mul08} (\citeyear{mul08}). 
($k$) \citeauthor{num00} (\citeyear{num00}) small source-size averaged. 
($k*$) \citeauthor{num00} (\citeyear{num00}) beam averaged. 
($l$) \citeauthor{rem05} (\citeyear{rem05}). 
($m$) \citeauthor{bel09} (\citeyear{bel09}). 
($n$) \citeauthor{meh96} (\citeyear{meh96}). 
($o$) \citeauthor{bel13} (\citeyear{bel13}). 
($p$) \citeauthor{ohi98} (\citeyear{ohi98}). 
($q$) \citeauthor{min91} (\citeyear{min91}).
\end{table*}

\subsection{Chemical model}

We have investigated the observed column densities of CH$_2$CHCN, CH$_3$CN,
CH$_3$CH$2$CN, HCCCN, and CN using a time and depth dependent gas-grain chemical
model, UCL\_CHEM. UCL\_CHEM is a two-phase model which follows the collapse of a
prestellar core (Phase I), followed by the warming and evaporation of grain
mantles (Phase II). In Phase II we increase the dust and gas temperature up to
300 K, to simulate the presence of a nearby infrared source in the core. For the
hot core component we model both a 10 M$_{\sun}$ and 15 M$_{\sun}$ star, with a
final density of 10$^7$ cm $^{-3}$. During the collapse, atoms and molecules
collide with, and freeze on to, grain surfaces. The depletion efficiency is
determined by the fraction of the gas-phase material that is frozen on to the
grains, which is dependent on the density, the sticking coefficient and other
properties of the species and grains (see \citeauthor{raw92} \citeyear{raw92}).
In our modelling we have explored the uncertainty in grain properties and
sticking coefficients by varying the depletion percentage. Initial atomic
abundances are taken from \citeauthor{sof01} (\citeyear{sof01}), as in
\citeauthor{vit04} (\citeyear{vit04}). Gas-phase reaction rate coefficients are
taken from the UMIST database of \citeauthor{woo07} (\citeyear{woo07}), however,
some have been updated with those from the KIDA database (\citeauthor{wak09}
\citeyear{wak09}). We also include some simple grain-surface reactions (mainly
hydrogenation) as in \citeauthor{vit04} (\citeyear{vit04}). While COMs
(complex organic molecules) may also
form via surface reactions involving heavier (than hydrogen) species 
(e.g. \citeauthor{gar08} \citeyear{gar08}), the mobility of most heavy species on grains has 
not been experimentally investigated; hence, for this qualitative analysis, 
we chose to adopt a simpler model where only the most efficient surface 
reactions occur: in this way we can give a lower limit to the formation 

of COMs which may be augmented by more complex reactions should they occur. In Phase I
non-thermal desorption is considered as in \citeauthor{rob07} (\citeyear{rob07}).

\begin{figure*}[!ht]
\centering
\includegraphics[angle=270,width=0.8\textwidth]{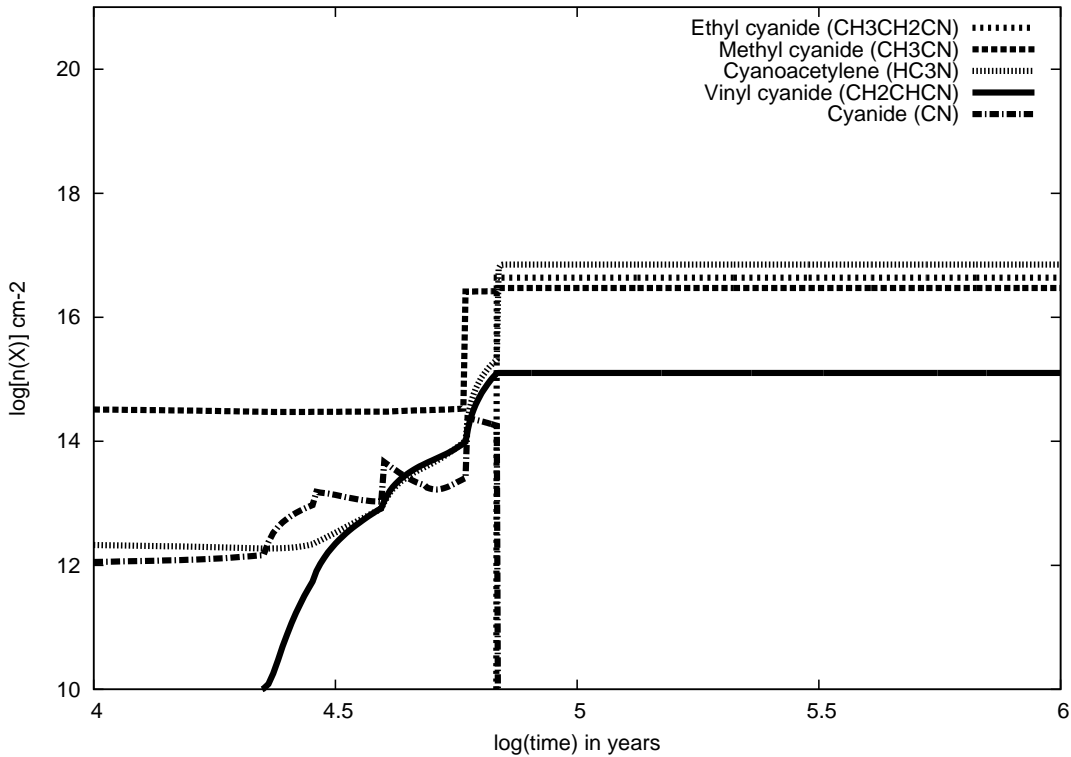}
\caption{Time evolution of the column densities of CH$_2$CHCN, CH$_3$CN, CH$_3$CH$_2$CN, 
HC$_3$N, and CN for a hot core chemical model.}
\label{fig_chem_model}
\end{figure*}

Within our grid of models we find that models where we simulate a 10 M$_{\sun}$ star and
100\% of CO frozen onto grain surfaces most accurately reproduce the observed column
densities of CH$_2$CHCN, CH$_3$CN,  and CH$_3$CH$_2$CN.
Figure \ref{fig_chem_model}  shows the column density as a function of time
during phase II for this model. The column density produced by the model for
HCCCN is an order of magnitude higher than the observed value. Whilst our models
simulate both gas phase and grain surface reactions for all of these species,
the grain surface reactions are essential in order to reproduce the observed
column densities. We therefore conclude that we are missing some grain surface
destruction routes for HCCCN and consequently overproduced this species in our
models. Moreover, the deep decreased
of the CN abundance when CH$_3$CH$_2$CN appears is observationally
confirmed by the lack of the hot core component in the CN lines even
at the HIFI frequencies (\citeauthor{cro14} \citeyear{cro14}).

For details of the  same surface chemistry approach see 
\citeauthor{vit04} (\citeyear{vit04}) and \citeauthor{tom14} (\citeyear{tom14}).

\subsection{Further issues for CH$_2$CHCN}

Further observations with telescopes with higher sensitivity and spatial
resolution, such as ALMA (Atacama Large Millimeter/submillimeter Array), could
provide additional detections of other vibrationally excited states above
600\,cm$^{-1}$, such as the outstanding states in the 3$\varv_{11}$/2$\varv_{15}$/$\varv_{14}$ 
triad of states near 680\,cm$^{-1}$ for which the spectroscopy is reported presently. In
this study, we found that the $\varv_{11}$=3 (987.9\,K or 686.6\,cm$^{-1}$)
vibrational mode was near the detection limit so we could not reliably address other vibrational 
components of the 3$\varv_{11}$/2$\varv_{15}$/$\varv_{14}$ triad. We also note that \citeauthor{bel13}
(\citeyear{bel13}) have recently detected the combination state
$\varv_{15}$=$\varv_{11}$=1 (809.9 K) but has not yet reported detection of $\varv_{10}$ (806.4 K) state
toward Sgr B2(N).
On the other hand, the $\varv_{15}$=2 (960.2\,K) excited state detected by these
authors might be detected only at the limit in our study.

In the present work we extended the laboratory coverage of the rotational spectrum of 
vinyl cyanide and the analysis of its vibrationally excited states to provide ample
basis for detection of transitions from further excited vibrational states at even higher 
vibrational energies.  Laboratory basis for detecting states up to as high as $\varv_9$ (1250\,K) is now available.
On the other hand, as implied by Fig.\,\ref{fig_vincne} and results for 4$\varv_{11}$, considerable 
further spectroscopic analysis is required for satisfactory understanding of states above $\varv_9$.

\section{Summary}
\label{summ}

Vinyl cyanide is one of the most abundant molecules in Orion-KL and a possible
precursor of alanine. This study of the vinyl cyanide species contributes to improve
the knowledge of the physical and chemical conditions of this high-mass star
forming region. We have performed an identification of the ground state of CH$_2$CHCN and of its
vibrationally excited states (up to 988\,K) in the Orion-KL Nebula thanks to
a new spectroscopic laboratory analysis using Stark modulation and
frequency-modulated spectrometers. Our results are based on rotational diagrams,
integrated-frequency maps, and Gaussian fits in order to optimize the physical
and chemical parameters that simulate the best synthetic spectrum of CH$_2$CHCN (using MADEX)
that fits the observation conditions of the Orion-KL region in an accurate way. 
We have found $N$(CH$_2$CHCN)$\simeq$(6$\pm$2)$\times$10$^{15}$\,cm$^{-2}$ 
from four cloud components of hot core/plateau nature\,(320-90 K). A
total abundance of (3.1$\pm$0.9)$\times$10$^{-8}$ for vinyl cyanide is
provided in this work. We have detected the CH$_2$CHCN $\varv_{11}$=2,3
vibrational modes for the first time in Orion-KL and the CH$_2$CHCN
$\varv_{10}$=1$\Leftrightarrow$($\varv_{11}$=1,$\varv_{15}$=1) excited state for the first time in the space.
We have seen that these species together with those of the three monosubstituted
$^{13}$C and the $^{15}$N isotopologues, and the tentative detection of the
three monodeuterated species of vinyl cyanide, contribute with more than 1100
observed lines in the 80-280 GHz domain covered by the Orion line survey. We
highlight the importance for spectroscopic catalogs to introduce vibrationally
excited species in the astronomical detections.

The column density ratios between the vinyl cyanide g.s. and the vibrationally
excited states 
have been used in order to obtain temperatures at which the vibrational modes
are excited, and to correct the ground column density from the vibrational
partition function. The high vibrational temperature ($T_{\rm vib}$$>$$T_{\rm rot}$) for
the states $\varv_{10}$=1$\Leftrightarrow$($\varv_{11}$=1,$\varv_{15}$=1) and $\varv_{11}$=3 suggests a
temperature gradient toward the inner regions of the hot core. To infer the
population mechanism of the vibrationally excited states (collisions and/or
infrared radiation) collisional rates are needed.

Owing to the importance of isomerism for understanding in a more
precise way the formation of interstellar molecules, we have included the study
of ethyl isocyanide, methyl isocyanide, isocyanoacetylene,
3-imino-1,2-propadienylidene, and isocyanamide in our work. We have provided the detection
of methyl isocyanide for the first time in Orion-KL, and tentative detections for the rest.

Finally, we have investigated the studied column densities of CH$_2$CHCN, CH$_3$CN,
CH$_3$CH$_2$CN, and HCCCN using a time dependent gas-grain chemical model (UCL\_CHEM)
reproducing reasonably well the observed column densities for these molecules, 
although with an overestimation for that of HCCCN; this is probably due  to the efficiency
for its formation on the grains being too high: a detailed investigation of the formation 
and destruction route for this species in chemical models is beyond the scope of this work; 
more quantitative models ought to be able to reproduce this molecule by investigating the 
efficiency of the formation of HCCCN on the grains.

\begin{acknowledgements}
We thank the anonymous referee who provided comments that improved this manuscript.
We thank INTA-CSIC, and the Spanish MINECO and Junta de Castilla y Le\'on for funding support from grants
the CONSOLIDER program "ASTROMOL" CSD2009-00038, AYA2009-07304, AYA2012-32032, CTQ2010-19008 and VA175U13. 
C.B. thanks also the Spanish MINECO for the FPI grant (BES-2011-047695).
The IFPAN authors acknowledge a grant from the Polish National Science Centre, decision
number DEC/2011/02/A/ST2/00298. Portions of this research were carried out at the Jet Propulsion Laboratory,
California Institute of technology, under contract with the National Aeronautics and Space Administration. 
\end{acknowledgements}

{}

\clearpage

\Online
%%%%%%%%%%%%%%%%%%%%%%%%%%%%%%%%%%%%%%%%%%%%%%%%%%%%%%%%%%%%%%%%%%%%%%%%%%%%%%%%
%%%%%%%%%%%%%%%%%%%%%%%%%%%%%%%%%%%%%%%%%%%%%%%%%%%%%%%%%%%%%%%%%%%%%%%%%%%%%%%%
%  ONLINE APPENDIX A: spectroscopy
%%%%%%%%%%%%%%%%%%%%%%%%%%%%%%%%%%%%%%%%%%%%%%%%%%%%%%%%%%%%%%%%%%%%%%%%%%%%%%%%

\begin{appendix}{}

%-------------------------------------------------------------------------------
%  TABLE: diagonal spectroscopic constants for the triad
%-------------------------------------------------------------------------------

\begin{table*}
\section{\small{Online Tables and Figures}}
\caption{Spectroscopic constants in the diagonal blocks of the Hamiltonian for
the $2\varv_{15}\Leftrightarrow \varv_{14} \Leftrightarrow 3\varv_{11}$ 
 triad of vibrational states in vinyl cyanide compared
with those for the ground state.}
\label{tab_const_triad}
\renewcommand{\thefootnote}{\alph{footnote}}
%
%
%\begin{small}
\begin{center}
\begin{tabular}{lk{10}k{10}k{10}k{10}}
\hline\hline\vspace{-0.2cm}\\
     &  \multicolumn{1}{c}{ground state}          &
        \multicolumn{1}{c}{$2\varv_{15}$}           &
        \multicolumn{1}{c}{$\varv_{14}$}            &
        \multicolumn{1}{c}{$3\varv_{11}$}           \\
\vspace{-0.2cm}\\
\hline
%                                  gs                   v15=2                       v14=1                     v11=3             
%                                                                                                                               
                           &                     &                        &                        &                             \\
$A$/MHz                    & 49850.69655(43)^a   &    51864.336(36)       &     50344.872(21)      &    47990.329(47)            \\
$B$/MHz                    &  4971.212565(37)    &     4977.66513(90)     &      4974.53973(78)    &     5028.9277(13)           \\
$C$/MHz                    &  4513.828516(39)    &     4532.01579(78)     &      4519.08450(62)    &     4537.6921(11)           \\
                           &                     &                        &                        &                             \\
$\Delta_J$/kHz             &     2.244058(13)    &        2.23613(25)     &         2.26543(26)    &        2.31540(32)          \\
$\Delta_{JK}$/kHz   ~~~~~  &   -85.6209(35)      &     -101.725(13)       &       -81.531(12)      &      -62.131(12)            \\
$\Delta_K$/kHz             &  2715.4213(94)      &     4021.39(65)        &      2752.53(50)       &     1416.57(67)             \\
$\delta_J$/kHz             &     0.4566499(32)   &        0.44884(16)     &         0.45991(16)    &        0.48731(24)          \\
$\delta_K$/kHz             &    24.4935(22)      &       26.477(90)       &        27.126(82)      &       32.642(96)            \\
                           &                     &                        &                        &                             \\
$\Phi_{J}$/Hz              &     0.0064338(17)   &        0.006223(44)    &         0.006097(45)   &        0.006248(56)         \\
$\Phi_{JK}$/Hz             &    -0.00425(40)     &        0.129(16)       &         0.063(15)      &        0.094(17)            \\
$\Phi_{KJ}$/Hz             &    -7.7804(39)      &      -25.21(11)        &        -7.03(10)       &        4.74(10)             \\
$\Phi_{K}$/Hz              &   384.762(63)       &     1391.8(35)         &       395.8(16)        &     -317.0(41)              \\
$\phi_{J}$/Hz              &     0.00236953(79)  &        0.002184(22)    &         0.002218(23)   &        0.002359(33)         \\
$\phi_{JK}$/Hz             &     0.14283(40)     &        0.126(16)       &         0.088(16)      &        0.252(17)            \\
$\phi_{K}$/Hz              &    37.011(58)       &       66.2(24)         &        42.9(21)        &       27.1(21)              \\
                           &                     &                        &                        &                             \\
$L_J$/mHz                  &    -0.000026315(71) &      [ 0.]             &       [ 0.]            &      [ 0.]                  \\
$L_{JJK}$/mHz              &    -0.001077(29)    &      [ 0.]             &       [ 0.]            &      [ 0.]                  \\
$L_{JK}$/mHz               &     0.4279(30)      &      [ 0.]             &       [ 0.]            &      [ 0.]                  \\
$L_{KKJ}$/mHz              &     0.012(12)       &        9.07(38)        &         3.62(38)       &        3.64(22)             \\
$L_{K}$/mHz                &   -61.41(17)        &     -658.9(84)         &       -77.6(88)        &      161.3(94)              \\
$l_J$/mHz                  &    -0.000011602(36) &      [ 0.]             &       [ 0.]            &      [ 0.]                  \\
$l_{JK}$/mHz               &    -0.000956(20)    &      [ 0.]             &       [ 0.]            &      [ 0.]                  \\
$l_{KJ}$/mHz               &    -0.1436(46)      &      [ 0.]             &       [ 0.]            &        0.988(40)            \\
$l_{K}$/mHz                &     8.91(18)        &       16.3(10)         &         8.03(88)       &      -25.43(69)             \\
                           &                     &                        &                        &                             \\
$P_{KJ}$/mHz               &    -0.0000156(31)   &      [ 0.]             &       [ 0.]            &      [ 0.]                  \\
$P_{KKJ}$/mHz              &    -0.0001977(57)   &      [ 0.]             &       [ 0.]            &      [ 0.]                  \\
$P_{K}$/mHz                &     0.00867(15)     &      [ 0.]             &       [ 0.]            &      [ 0.]                  \\
                           &                     &                        &                        &                             \\
$\Delta E$$^b$/MHz         &                     &        0.0             &   549163.34(55)        &    694443.66(90)            \\
$\Delta E$/cm$^{-1}$       &                     &        0.0             &       18.31812(2)      &        23.16415(3)          \\
                           &                     &                        &                        &                             \\
$N_{\rm lines}$$^c$        &                                                                                                                          
          \multicolumn{1}{l}{~~4490,0}           &                             
                               \multicolumn{1}{l}{~~~~~1329,52}           &     
                                                          \multicolumn{1}{l}{~~1287,53}             &  
                                                                                   \multicolumn{1}{l}{~~~~~1250,81}               \\
$\sigma_{\rm fit}$$^d$/MHz &     0.144          &        0.265^e          &     0.228^e            &        0.309^e              \\
$\sigma_{\rm rms}$$^d$     &     0.713          &        1.980            &     1.467              &        2.329                \\
\vspace{-0.3cm}\\ 
\hline 
\\                      
\end{tabular}                                
\end{center}

{\bfseries Notes.} $^{(a)}$Round parentheses enclose standard errors in units
 of the last quoted digit of the value of the constant, square parentheses enclose assumed values.\\
$^{(b)}$The fitted vibrational energy difference relative to the lowest vibrational state in the triad.\\
$^{(c)}$The number of distinct frequency fitted lines and the number of lines rejected at the 10$\sigma$ fitting criterion
of the SPFIT program.\\
$^{(d)}$Deviations of fit for the different vibrational subsets.\\
$^{(e)}$The coupled fit for the complete triad encompasses 3866 lines, 
at an overall $\sigma_{\rm fit}$ of 0.269 MHz and requires also the use of constants reported in 
Table\,\ref{tab_constoff_triad}.

%\end{small}                               
%
%
\end{table*}

\clearpage

%-------------------------------------------------------------------------------
%  TABLE: off-diagonal spectroscopic constants for the two dyads  
%-------------------------------------------------------------------------------

\begin{table*}
\caption{Spectroscopic constants in the off-diagonal blocks of the Hamiltonian for
the $\varv_{10}\Leftrightarrow \varv_{11}\varv_{15}$ 
and  $\varv_{11}\varv_{10}\Leftrightarrow 2\varv_{11}\varv_{15}$
dyads of vibrational states in vinyl cyanide.}
\label{tab_constoff_dyads}
\renewcommand{\thefootnote}{\alph{footnote}}
%
%
%\begin{small}
\begin{center}
\begin{tabular}{lk{10}k{10}lk{10}k{10}}

\hline\vspace{-0.2cm}\\
     &  \multicolumn{1}{c}{$\varv_{10}\Leftrightarrow \varv_{11}\varv_{15}$ }  &
        \multicolumn{1}{c}{$\varv_{11}\varv_{10}\Leftrightarrow 2\varv_{11}\varv_{15}$}          & &
        \multicolumn{1}{c}{$\varv_{10}\Leftrightarrow \varv_{11}\varv_{15}$ }    &
        \multicolumn{1}{c}{$\varv_{11}\varv_{10}\Leftrightarrow 2\varv_{11}\varv_{15}$}         \\
\vspace{-0.3cm}\\
\hline
%                             01 lower dyad        01 upper dyad                               01 lower dyad        01 upper dyad 
%                                                                       
                        &                     &                    &                      &                    &                           \\
 $G_a$/MHz              &   1623.(13)         &   1557.(34)        &    $G_b$/MHz         &   643.32(46)       &       929.1(30)           \\
 $G_a^{J}$/MHz          &      0.2192(48)     &      0.4834(45)    &    $G_b^J$/MHz       &     0.01663(17)    &         0.00766(38)       \\
 $G_a^{K}$/MHz          &      6.432(76)      &      3.67(10)      &                      &                    &                           \\
 $G_a^{JK}$/MHz         &      0.0001240(23)  &                    &                      &                    &                           \\
                        &                     &                    &                      &                    &                           \\
 $F_{bc}$/MHz           &      4.151(90)      &      9.447(88)     &    $F_{ac}$/MHz      &    -27.14(29)      &       -14.20(69)          \\
                        &                     &                    &    $F_{ac}^{J}$/MHz  &     -0.0001184(26) &                           \\
                        &                     &                    &                      &                    &                           \\
\vspace{-0.3cm}\\ 
\hline 
\\                      
\end{tabular}                                
\end{center}

{\bfseries Notes.} $^{(a)}$Round parentheses enclose standard errors in units
 of the last quoted digit of the value of the constant, and only the constants with non-zero values are listed.\\
$^{(b)}$These constants complement those in the diagonal blocks
listed in Table~\ref{tab_const_dyads}.\\

%\end{small}                               
\vspace{3.cm} 
\end{table*}

%-------------------------------------------------------------------------------
%  TABLE: off-diagonal spectroscopic constants for the triad
%-------------------------------------------------------------------------------

\begin{table*}
\caption{Spectroscopic constants in the off-diagonal blocks of the Hamiltonian for
the $2\varv_{15}\Leftrightarrow \varv_{14} \Leftrightarrow 3\varv_{11}$ 
 triad of vibrational states in vinyl cyanide.}
\label{tab_constoff_triad}
\renewcommand{\thefootnote}{\alph{footnote}}
%
%
%\begin{small}
\begin{center}
\begin{tabular}{lk{10}k{10}lk{10}}

\hline\vspace{-0.2cm}\\
     &  \multicolumn{1}{c}{ $2\varv_{15}\Leftrightarrow \varv_{14}$    }    &
        \multicolumn{1}{c}{ $\varv_{14} \Leftrightarrow 3\varv_{11}$   }    & &
        \multicolumn{1}{c}{ $2\varv_{15}\Leftrightarrow 3\varv_{11}$}        \\
\vspace{-0.3cm}\\
\hline
%                                 01                   12                                        02               
%                                                                       
                        &                            &                    &                       &                        \\
 $G_a$/MHz              &     5852.21(58)            &   223.63(15)       &     $G_c$/MHz         &    25.8(27)            \\
 $G_a^{J}$/MHz          &        0.01958(32)         &                    &                       &                        \\
 $G_a^{K}$/MHz          &       -2.790(18)           &    -0.1703(17)     &     $F_{ab}$/MHz      &     3.59(22)           \\
 $G_a^{JJ}$/MHz         &       -0.000000324(61)     &                    &                       &                        \\
 $G_a^{JK}$/MHz         &       -0.0000335(23)       &                    &     $W$               &   14179.2(17)          \\
 $G_a^{KK}$/MHz         &        0.001503(51)        &                    &                       &                        \\
                        &                            &                    &     $W_\pm$/MHz       &     0.8521(24)         \\  
  $F_{bc}$/MHz          &                            &    0.4557(30)      &     $W_\pm^J$/MHz     &     0.00000540(82)     \\
                        &                            &                    &     $W_\pm^K$/MHz     &     0.00847(15)        \\
 $G_b$/MHz              &      -9.50(21)             &   18.180(61)       &     $W_\pm^{JK}$/kHz  &    -0.000000700(44)    \\
 $G_b^K$/MHz            &      -0.4173(94)           &                    &     $W_\pm^{KK}$/kHz  &     0.00000523(16)     \\

                        &                            &                    &                       &                        \\
\vspace{-0.3cm}\\ 
\hline 
\\                      
\end{tabular}                                
\end{center}

{\bfseries Notes.} $^{(a)}$Round parentheses enclose standard errors in units
 of the last quoted digit of the value of the constant, and only the constants with non-zero values are listed.\\
$^{(b)}$These constants complement those in the diagonal blocks
listed in Table~\ref{tab_const_triad}.\\

%\end{small}                               
%
%
\end{table*}
\clearpage

%-------------------------------------------------------------------------------
%  TABLE: spectroscopic constants for v9 and 4v11
%-------------------------------------------------------------------------------

\begin{table*}
\caption{Spectroscopic constants in the effective rotational Hamiltonian for
the $\varv_{9}$ and $4\varv_{11}$ excited vibrational states of vinyl cyanide.}
\label{tab_const_minor}
\renewcommand{\thefootnote}{\alph{footnote}}
%
%
%\begin{small}
\begin{center}
\begin{tabular}{lk{10}k{10}}
\hline\hline\vspace{-0.2cm}\\
     &  \multicolumn{1}{c}{$\varv_9$     }      &
        \multicolumn{1}{c}{$4\varv_{11}$ }       \\
\vspace{-0.2cm}\\
\hline
%                                  v9                   4v11                   
%                                                                              
                           &                     &                              \\
$A$/MHz                    & 49828.53(89)^a      &     47495.5(22)              \\
$B$/MHz                    &  4953.0854(34)      &      5047.5125(98)           \\
$C$/MHz                    &  4501.1911(18)      &      4545.1778(85)           \\
                           &                     &                              \\
$\Delta_J$/kHz             &     2.2025(18)      &         2.3223(66)           \\
$\Delta_{JK}$/kHz   ~~~~~  &   -92.279(81)       &       -56.27(34)             \\
$\Delta_K$/kHz             &  2500.(135)         &       -87.(529)              \\
$\delta_J$/kHz             &     0.4392(11)      &         0.4766(46)           \\
$\delta_K$/kHz             &    11.17(53)        &        28.9(48)              \\
                           &                     &                              \\
$\Phi_{J}$/Hz              &     0.00469(72)     &         0.0368(68)           \\
$\Phi_{JK}$/Hz             &    -2.22(36)        &         5.3(31)              \\
$\Phi_{KJ}$/Hz             &   -28.6(17)         &       -32.(13)               \\
$\Phi_{K}$/Hz              &   [ 0.]             &    104900.(42879)            \\
$\phi_{J}$/Hz              &     0.00145(38)     &        -0.0185(36)           \\
$\phi_{JK}$/Hz             &    -0.93(18)        &         7.6(18)              \\
$\phi_{K}$/Hz              &  -159.(48)          &      1131.(532)              \\
                           &                     &                              \\
$N_{\rm lines}$$^b$        &    373,7            &      225,17                  \\                                                                    
$\sigma_{\rm fit}$/MHz     &     0.167           &        0.250                 \\
$\sigma_{\rm rms}$         &     1.665           &        2.496                 \\
\vspace{-0.3cm}\\ 
\hline 
\\                      
\end{tabular}                                
\end{center}

{\bfseries Notes.} $^{(a)}$Round parentheses enclose standard errors in units
 of the last quoted digit of the value of the constant, square parentheses enclose assumed values.  
$^{(b)}$The number of distinct frequency fitted lines and the number of confidently assigned 
lines rejected at the 10$\sigma$ fitting criterion of the SPFIT program.\\

%\end{small}                               
%
%
\end{table*}
\clearpage

%-------------------------------------------------------------------------------
%  TABLE: vibrational changes in rotational constants
%-------------------------------------------------------------------------------

\begin{table*}
\caption{Vibrational changes in rotational constants$^a$ for the studied 
 excited vibrational states in vinyl cyanide.}
\label{tab_vibrot}
\renewcommand{\thefootnote}{\alph{footnote}}
%
%
%\begin{small}
\begin{center}
\begin{tabular}{lk{10}k{7}k{7}lk{10}k{10}}

\hline\vspace{-0.2cm}\\
         &  \multicolumn{1}{c}{ exp.                }    &
            \multicolumn{1}{l}{ calc. I$^b$         }    & 
            \multicolumn{1}{l}{ calc. II$^c$        }    & &
            \multicolumn{1}{c}{ exp.                }    &
            \multicolumn{1}{c}{ estimated$^d$       }        \\
\vspace{-0.3cm}\\
\hline
%                                 01                   12                                        02               
%                                                                       
              &                      &                 &               &                       &                 &                  \\
$\varv_{11}$    &    -680.2411(15)^d   &    -782         &  -674.20      &  $\varv_{11}\varv_{15}$   &   40.02(61)     &   271            \\
              &      19.56011(10)^d  &      19.14      &    20.28      &                       &   21.4597(70)   &    22.69         \\
              &       8.13824(10)^d  &       8.00      &     8.51      &                       &   17.7744(13)   &    17.31         \\
              &                      &                 &               &                       &                 &                  \\
$\varv_{15}$    &     951.3883(18)     &    1002         &   900.77      &  $2\varv_{15}$          &  2013.674(35)   &  1903            \\
              &       3.13085(12)    &       5.94      &     3.94      &                       &     6.45356(89) &     6.26         \\
              &       9.16787(11)    &      10.28      &     8.95      &                       &    18.2016(10)  &    18.34         \\
              &                      &                 &               &                       &                 &                  \\
$\varv_{10}$    &    -300.67(63)       &    -369         &  -326.14      &  $3\varv_{11}$          & -1860.399(47)   & -2041.           \\
              &      -5.5434(98)     &      -5.49      &    -5.79      &                       &    57.7162(12)  &    58.69         \\
              &      -4.2057(13)     &      -3.98      &    -4.20      &                       &    23.8494(13)  &    24.41         \\
              &                      &                 &               &                       &                 &                  \\
$\varv_{14}$    &     494.2055(11)     &     606         &   566.66      &  $\varv_{10}\varv_{11}$   &  -988.98(62)    &  -981            \\
              &       3.32747(40)    &       2.37      &     3.10      &                       &    13.766(32)   &    14.02         \\
              &       5.25568(31)    &       4.43      &     4.88      &                       &     4.1072(31)  &     3.93         \\
              &                      &                 &               &                       &                 &                  \\
$\varv_{9}$     &     -22.17(89)       &     -37         &    39.53      &  $2\varv_{11}\varv_{15}$  &  -725.83(56)    &  -409            \\
              &     -18.1272(34)     &     -17.52      &   -18.47      &                       &    40.2814(98)  &    42.25         \\
              &     -12.6374(11)     &     -12.35      &   -13.27      &                       &    26.2639(85)  &    25.44         \\
              &                      &                 &               &                       &                 &                  \\
              &                      &                 &               &  $4\varv_{11}$          & -2355.2(22)     & -2721            \\
              &                      &                 &               &                       &    76.2999(98)  &    78.24         \\
              &                      &                 &               &                       &    31.3493(85)  &    32.55         \\
              &                      &                 &               &                       &                 &                  \\
\vspace{-0.3cm}\\ 
\hline 
\\                      
\end{tabular}                                
\end{center}

{\bfseries Notes.} $^{(a)}$The tabulated values for each state are differences relative to the ground state constants:
$(A_v-A_0)$, $(B_v-B_0)$, and $(C_v-C_0)$, all in MHz.\\
$^{(b)}$From anharmonic force field calculation at the MP2/6-311++G(d,p) level.\\
$^{(c)}$From anharmonic force field calculation at the CCSD(T)/6-31G(d,p) level.\\        
$^{(d)}$Estimated from experimental changes listed in the second column by assuming their additivity.\\

%\end{small}                               
%
%
\vspace{1.cm} 
\end{table*}

%-------------------------------------------------------------------------------
%-------------------------------------------------------------------------------

\begin{figure*}[ht]
\centering
\includegraphics[angle=270,width=0.8\textwidth]{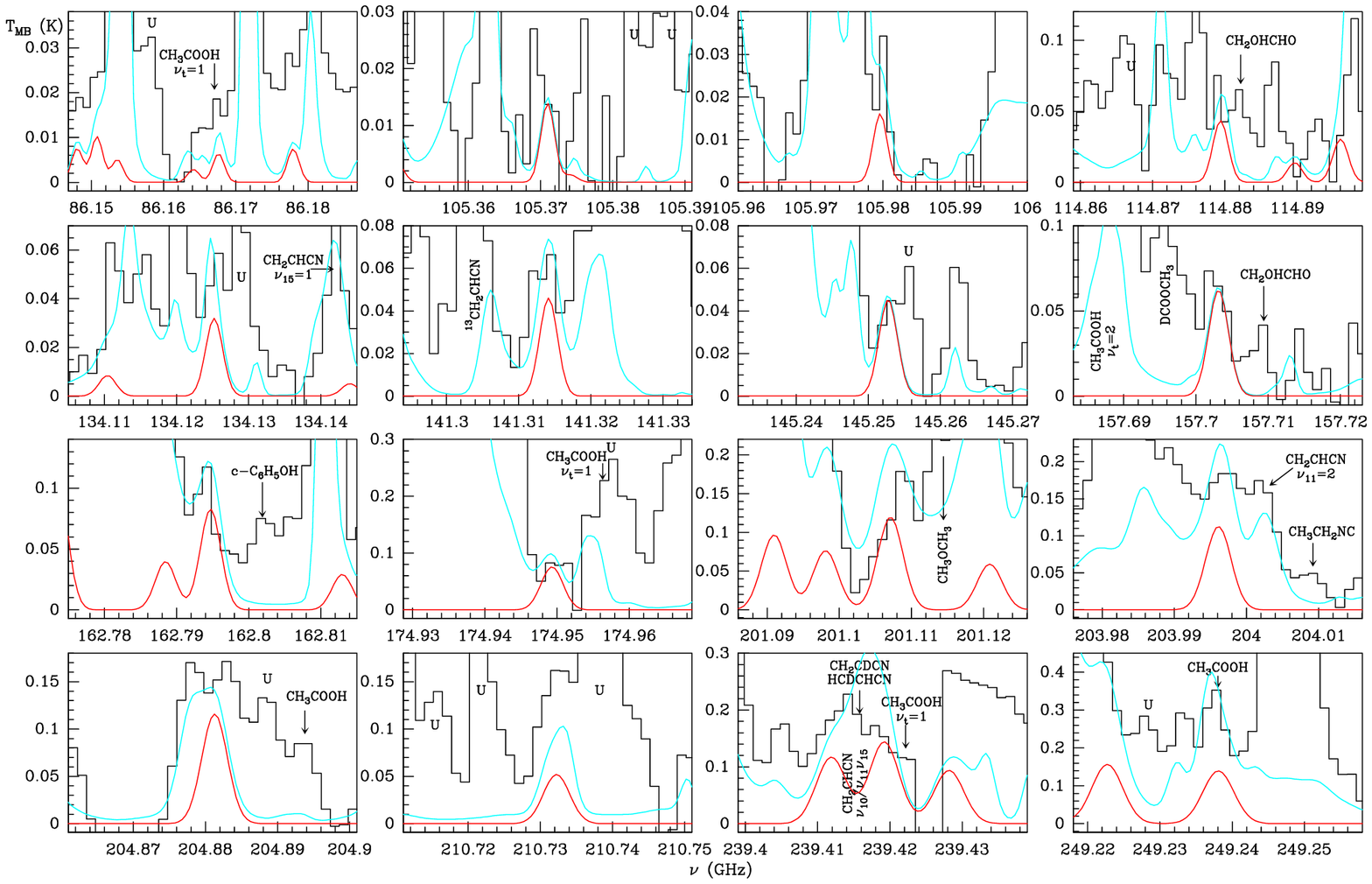}
\caption{Observed lines from Orion-KL (histogram spectra) and model (thin red
curves) of CH$_2$CHCN of $\varv_{11}$=3. The cyan line corresponds to the model of
the molecules we have already studied in this survey (see text Sect. \ref{sec_cd})
including the CH$_2$CHCN species.
A v$_{\rm LSR}$ of 5\,km s$^{-1}$ is assumed.}
\label{fig_3v11}
\end{figure*}

\begin{figure*}[ht]
\centering
\includegraphics[angle=270,width=0.9\textwidth]{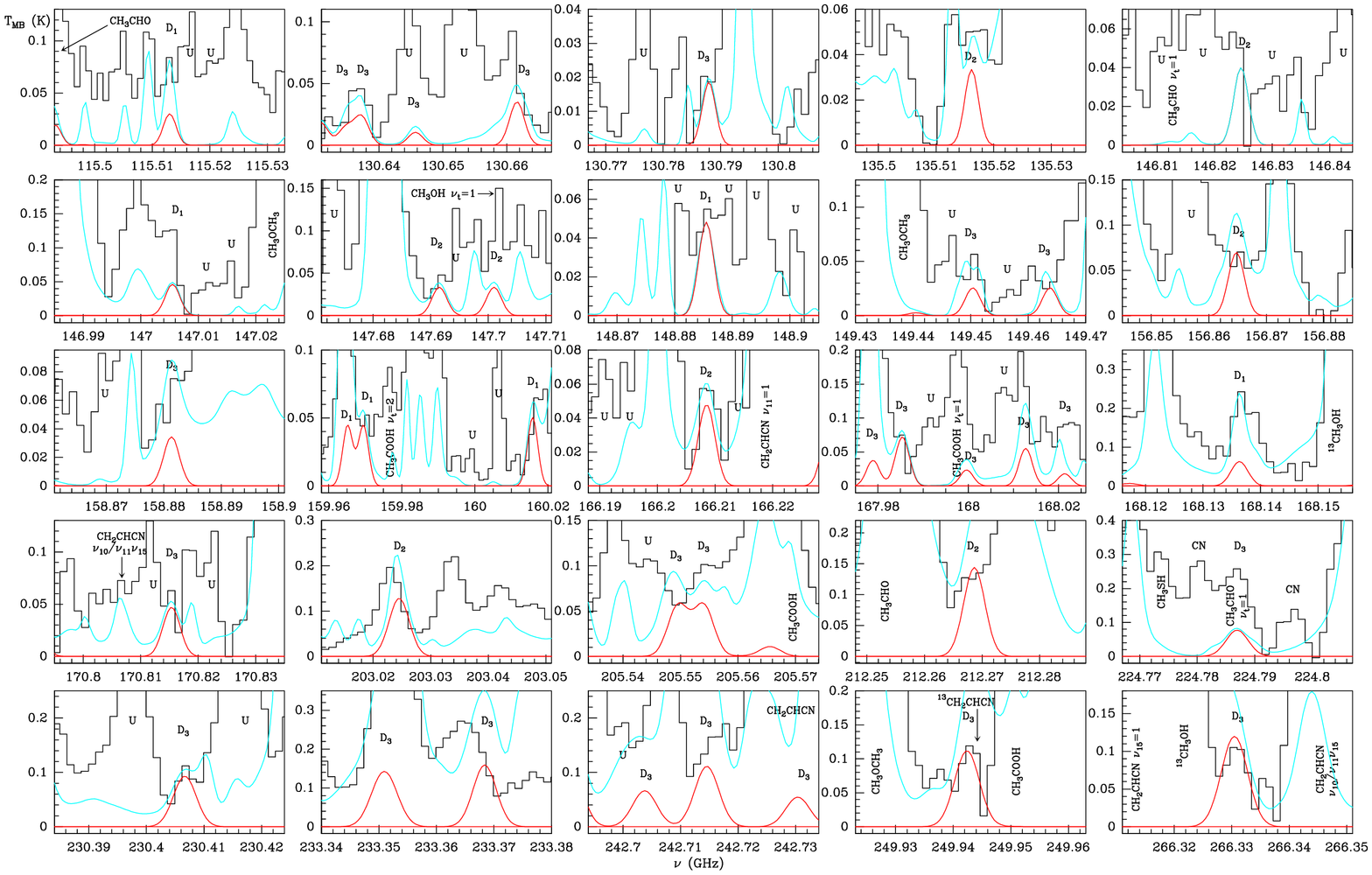}
\caption{Observed lines from Orion-KL (histogram spectra) and model (thin red
curves) of deuterated isotopes for CH$_2$CHCN in the ground state. The subindex
of D$_i$ (i=1, 2, 3) are corresponded with the position of the isotope in the
molecule (D$_{1}$CD$_{2}$CD$_{3}$CN). The cyan line corresponds to the model of
the molecules we have already studied in this survey (see text Sect. \ref{sec_cd})
including the CH$_2$CHCN species.
A v$_{\rm LSR}$ of 5\,km s$^{-1}$ is assumed.}
\label{fig_deu}
\vspace{2.cm} 
\end{figure*}

\begin{figure*}[ht]
\centering
\includegraphics[angle=270, width=0.5\textwidth]{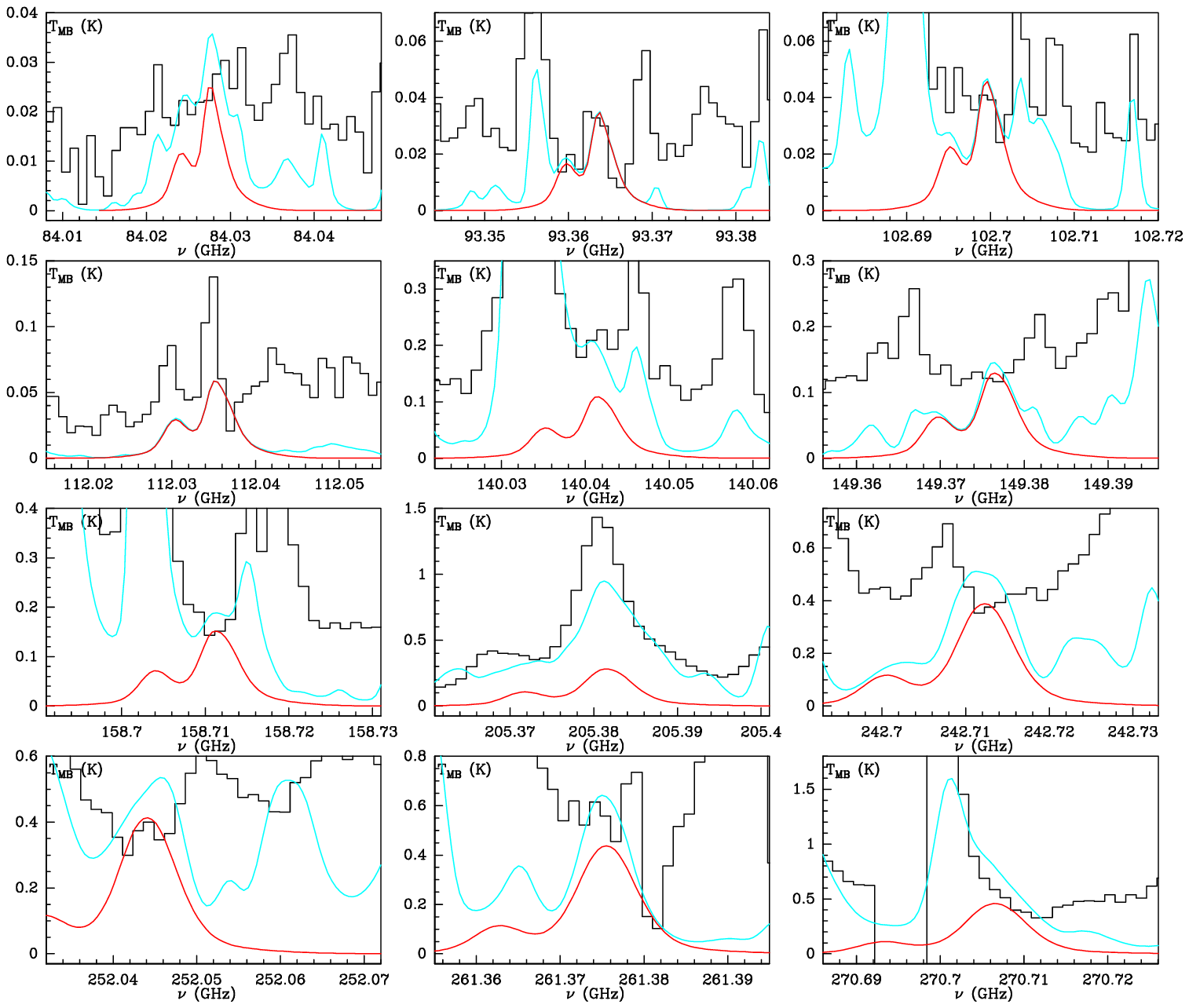}
\caption{Observed lines from Orion-KL (histogram spectra) and model (thin red 
curves) of 3-imino-1,2-propadienyllidene in its ground state. The cyan line corresponds to the model of
the molecules we have already studied in this survey (see text Sect. \ref{sec_cd})
including the CH$_2$CHCN species. A v$_{\rm LSR}$ of 5\,km\,s$^{-1}$ is assumed.}
\label{fig_HNCCC}
\end{figure*}

\begin{figure*}[ht]
\centering
\includegraphics[angle=270,width=0.8\textwidth]{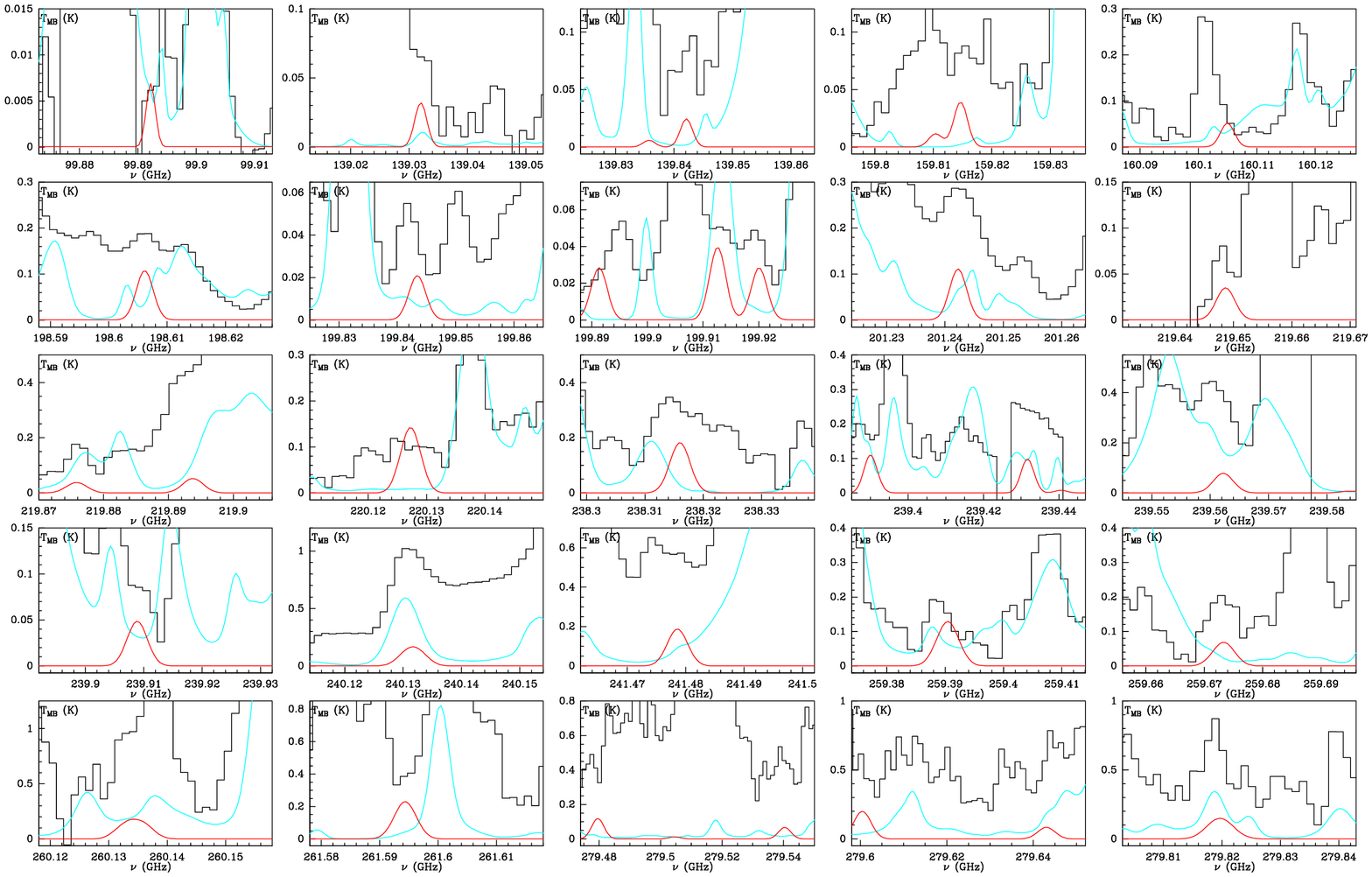}
\caption{Observed lines from Orion-KL (histogram spectra) and model (thin red
curves) of cyanamide in its ground state. The cyan line corresponds to the model of
the molecules we have already studied in this survey (see text Sect. \ref{sec_cd})
including the CH$_2$CHCN species. A v$_{\rm LSR}$ of 5\,km\,s$^{-1}$ is assumed.}
\label{fig_NH2CN}
\vspace{2.cm} 
\end{figure*}

\begin{figure*}[ht]
\centering
\includegraphics[angle=270, width=0.9\textwidth]{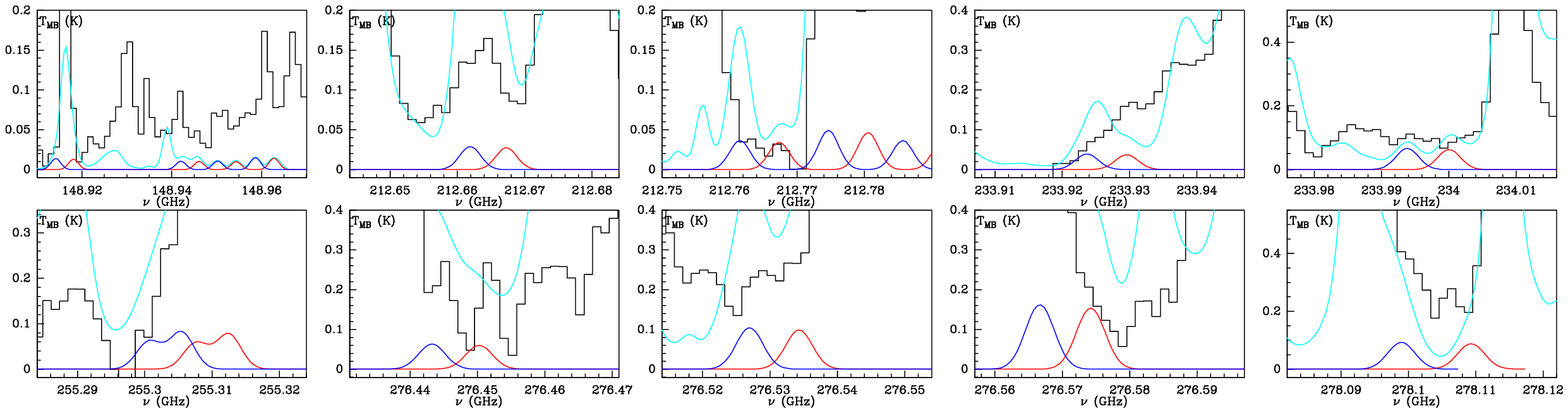}
\caption{Observed lines from Orion-KL (histogram spectra) and model
 (thin red and blue curves) of isocyanamide in its ground state.
The cyan line corresponds to the model of
the molecules we have already studied in this survey (see text Sect. \ref{sec_cd})
including the CH$_2$CHCN species.
A v$_{\rm LSR}$ of 5\,km\,s$^{-1}$ is assumed.}
\label{fig_NH2NC}
\end{figure*}

\clearpage

%\centering \tiny
%\centering 
%{

\longtab{6}{
{\setlength{\extrarowheight}{5.0pt}
\small
\begin{longtable}{clrllclc}
\caption{Detected lines of CH$_2$CHCN g.s. \label{tab_gs}.}\\
\hline\hline                      
Transition		     &  Predicted          &  $S_{ij}$ & $E_u$    & $v_{LSR}$  & $\Delta$$v$ & $T_{MB}$         & $\int T_{MB} dv$\\
$J_{K_a,K_c}-J'_{K'_a,K'_c}$ & frequency  (MHz)    &           &  (K)     & km s$^{-1}$& km s$^{-1}$ &   (K)            & (K km s$^{-1}$)  \\
\hline
\endfirsthead
\caption{continued.}\\
\hline\hline
Transition		     &  Predicted          &  $S_{ij}$ & $E_u$    & $v_{LSR}$  & $\Delta$$v$ & $T_{MB}$         & $\int T_{MB} dv$\\
$J_{K_a,K_c}-J'_{K'_a,K'_c}$ & frequency  (MHz)    &           &  (K)     & km s$^{-1}$& km s$^{-1}$ &   (K)            & (K km s$^{-1}$)  \\
\hline
\endhead
\hline
\endfoot
 9$_{1,9}$-8$_{1,8}$         &    83207.507          &  8.89           &   22.1    &    5.0$^{(1)}$       &               & 0.24$^{(2)}$    &                  \\
                             &                       &Wide comp.       &           &    3.5$\pm$0.4       &  14$\pm$1     & 0.12            & 1.8$\pm$0.2      \\
                             &                       &Narrow comp.     &           &    5.2$\pm$0.3       &   0.8$\pm$0.4 & 0.12            & 0.78$\pm$0.11    \\
 9$_{0,9}$-8$_{0,8}$         &    84946.003          &  9.00           &   20.4    &    4.4$^{(1)}$       &               & 0.24$^{(2)}$    &                  \\
                             &                       &                 &           &    4.4$\pm$0.4       &  11$\pm$1     & 0.23            & 2.7$\pm$0.2      \\
 9$_{2,8}$-8$_{2,7}$         &    85302.649          &  8.56           &   29.1    &    5.2$^{(1)}$       &               & 0.24$^{(2)}$    &                  \\
                             &                       &                 &           &    5.2$\pm$0.3       &   9.5$\pm$0.7 & 0.22            & 2.26$\pm$0.14    \\

      \vdots                 &      \vdots           &     \vdots      &           &      \vdots          &               &    \vdots       &                  \\

14$_{0,14}$-13$_{0,13}$      &   131267.475          & 14.00           &   47.5    &    2.5$^{(1)}$       &               & 2.08$^{(2)}$    &                  \\
14$_{2,13}$-13$_{2,12}$      &   132524.586          & 13.70           &   56.4    &    $^{(14)}$         &               & ...             &                  \\
14$_{5,10}$-13$_{5,9}$       &   132900.001          & 12.20           &  101.9    &    5.7$^{(1)}$       &               & 1.18$^{(2)}$    &                  \\
14$_{5,9}$-13$_{5,8}$        &   132900.012$\dagger$ & 12.20           &  101.9    &    5.8$^{(1)}$       &               &  ''             &                  \\
                             &                       &                 &           &    6.1$\pm$0.5       &   7.8$\pm$0.9 & 1.02            & 8.4$\pm$0.9      \\
14$_{6,9}$-13$_{6,8}$        &   132905.283          & 11.40           &  125.6    &    5.1$^{(1)}$       &               & 1.02$^{(2)}$    &                  \\
14$_{6,8}$-13$_{6,7}$        &   132905.283$\dagger$ & 11.40           &  125.6    &    5.1$^{(1)}$       &               &  ''             &                  \\

      \vdots                 &      \vdots           &     \vdots      &           &      \vdots          &               &    \vdots       &                  \\

21$_{2,20}$-20$_{2,19}$      &   198258.130          & 20.80           &  113.6    &    2.7$^{(1,12)}$    &               & 6.30$^{(2)}$    &                  \\
21$_{6,16}$-20$_{6,15}$      &   199396.979          & 19.30           &  183.0    &    4.7$^{(1)}$       &               & 2.25$^{(2)}$    &                  \\
21$_{6,15}$-20$_{6,14}$      &   199396.985$\dagger$ & 19.30           &  183.0    &    4.7$^{(1)}$       &               &  ''             &                  \\
21$_{7,15}$-20$_{7,14}$      &   199401.677          & 18.70           &  211.0    &    5.1$^{(1)}$       &               & 2.45$^{(2)}$    &                  \\
21$_{7,14}$-20$_{7,13}$      &   199401.677$\dagger$ & 18.70           &  211.0    &    5.1$^{(1)}$       &               &  ''             &                  \\

      \vdots                 &      \vdots           &     \vdots      &           &      \vdots          &               &    \vdots       &                  \\

\end{longtable}
}
%\onecolumn
%\flushleft
\renewcommand{\baselinestretch}{1.5}
\tablefoot{
%{\footnotesize{\sf{{\bf{Notes.}}}}} \small Emission lines of CH$_2$CHCN ground state 
\small Emission lines of CH$_2$CHCN ground state 
present in the spectral scan of the Orion-KL from the 
radio-telescope of IRAM 30-m. Column 1 indicates the line transition, Col. 2 
gives the predicted frequency in the laboratory, Col. 3 the line strength,
Col. 4 upper level energy, Col. 5 observed radial velocities relative to the local system rest (v$_{LSR}$), 
Col. 6 the line width, Col. 7 main beam temperature, and Col. 8 shows the area of the line. $\dagger$ blended with the previous line. $\ast$ noise level. $\ast$$\ast$ hole in the observed spectrum. \\
(1) peak channel line observed velocity.
(2) peak channel line intensity.
(3) blended with CH$_3$CCH.
(4) blended with $^3$$^4$SO$_2$.
(5) blended with t-CH$_3$CH$_2$OH.
(6) blended with OC$^3$$^4$S.
(7) blended with H$^1$$^3$CCCN.
(8) blended with CH$_3$OH.
(9) blended with CH$_3$COOH $\varv_{t}$=0.
(10) blended with $^1$$^3$CH$_3$OH.
(11) blended with U-line.
(12) blended with CH$_3$CH$_2$CN.
(13) blended with CH$_2$CHCN $\varv_{15}$=1.
(14) blended with CH$_3$OCH$_3$.
(15) blended with HCOOCH$_3$.
(16) blended with H$_2$C$^3$$^4$S.
(17) blended with CH$_3$CH$_2$$^1$$^3$CN.
(18) blended with $^1$$^3$CH$_3$CH$_2$CN.
(19) blended with CH$_3$CH$_2$CN $\varv_{13}$/$\varv_{21}$.
(20) blended with HCOOCH$_3$ $\varv_{t}$=1.
(21) blended with $^1$$^3$CH$_3$CN.
(22) blended with CH$_3$CH$_2$CN $\varv_{20}$=1.
(23) blended with $^3$$^3$SO$_2$.
(24) blended with CH$_2$CHCN $\varv_{11}$=1.     
(25) blended with CH$_3$CHO.     
(26) blended with H$_{33}$$\alpha$.
(27) blended with (CH$_3$)$_2$CO. %%%%%%%%%
(28) blended with SiS.
(28) blended with HCCCN.
(30) blended with SO$_2$.
(31) blended with CH$_3$CN.
(32) blended with HCCCN $\varv_{6}$=1. %%%%%%%
(33) blended with CH$_2$CHCN $\varv_{10}$/$\varv_{11}$$\varv_{15}$.
(34) blended with H$_2$CO.
(35) blended with HCCCN $\varv_{7}$=1.
(36) blended with HCOO$^1$$^3$CH$_3$.
(37) blended with SO$_2$ $\varv_{2}$=1.
(38) blended with CO.
(39) blended with C$_3$H$_2$.
(40) blended with CH$_3$$^1$$^3$CN.
(41) blended with CH$_2$CHCN $\varv_{11}$=2.
(42) blended with HDCO.
(43) blended with HDCS.
(44) blended with H$^1$$^3$COOCH$_3$.
(45) blended with $^3$$^4$SO.
(46) blended with CH$_3$SH.             
(47) blended with $^2$$^9$SiO.
(48) blended with CH$_3$CN $\varv_{8}$=1.
(49) blended with SO.
(50) blended with HCN.
(51) blended with HDO.
(52) blended with NO.
(53) blended with HCO$^+$.
(54) blended with NH$_2$CHO.
(55) blended with CH$_3$$^1$$^3$CH$_2$CN.
(56) blended with CH$_3$OD.\\
(This table is available in its entirety at CDS via http://cdsweb.u-strasbg.fr. A portion is shown here for guidance regarding its form and content.)\\
}
}

%\clearpage

%\centering \scriptsize
\longtab{7}{
%{\setlength{\extrarowheight}{4.0pt}
\small
\begin{longtable}{clccllll}
\caption{Detected b-type lines of CH$_2$CHCN g.s. \label{tab_gs_b}.}\\
\hline\hline                      
Transition		           & Predicted           &  $S_{ij}$ & $E_u$    & $v_{LSR}$$^{(1)}$   & Observed       & Observed             & Model        \\
$J_{K_a,K_c}-J'_{K'_a,K'_c}$       & frequency (MHz)     &           &  (K)     & km s$^{-1}$         & frequency (MHz)& $T_{MB}$ (K)$^{(2)}$ & $T_{MB}$ (K) \\
\hline
\endfirsthead
\caption{continued.}\\
\hline\hline
Transition		           & Predicted           &  $S_{ij}$ & $E_u$    & $v_{LSR}$$^{(1)}$   & Observed       & Observed             & Model        \\
$J_{K_a,K_c}-J'_{K'_a,K'_c}$       & frequency (MHz)     &           &  (K)     & km s$^{-1}$         & frequency (MHz)& $T_{MB}$ (K)$^{(2)}$ & $T_{MB}$ (K) \\
\hline
\endhead
\hline
\endfoot
18$_{1,17}$-17$_{0,18}$      &  95212.208         & 11.40     & 81.7    &  4.95$^{(3)}$       &  95212.2   &  0.03               &  0.01        \\
20$_{1,19}$-20$_{0,20}$      & 108813.600         & 11.40     & 99.7    &  3.74               & 108814.0   &  0.02               &  0.01        \\
18$_{2,16}$-18$_{1,17}$      & 113831.149         & 14.00     & 87.1    &  5.48               & 113831.0   &  0.02               &  0.02        \\

      \vdots                 &      \vdots           &     \vdots      &           &      \vdots          &               &    \vdots       &                  \\

29$_{2,27}$-29$_{1,28}$      & 131168.734         & 22.70     &209.5    &  3.88$^{(5)}$       & 131169.2   &  0.04               &  0.01        \\
17$_{0,17}$-16$_{1,16}$      & 136855.602         & 11.00     & 69.0    &  4.75$^{(4)}$       & 136855.7   &  0.05               &  0.02        \\

      \vdots                 &      \vdots           &     \vdots      &           &      \vdots          &               &    \vdots       &                  \\

39$_{2,37}$-39$_{1,38}$      & 199913.795         & 22.40     &369.5    &  5.41               & 199913.5 &  0.02               &  0.02        \\
20$_{1,20}$-19$_{0,19}$      & 200364.538         & 14.20     & 95.2    &  6.62               & 200363.4 &  0.08               &  0.06        \\
24$_{0,24}$-23$_{1,23}$      & 211519.057         & 18.10     &134.5    &  6.62$^{(4)}$       & 211518.4 &  0.12               &  0.07        \\
      \vdots                 &      \vdots           &     \vdots      &           &      \vdots          &               &    \vdots       &                  \\

\end{longtable}
%}
%\onecolumn
%\flushleft
%\renewcommand{\baselinestretch}{1.5}
{\footnotesize{\sf{{\bf{Notes.}}}}} \small Emission b-type lines of CH$_2$CHCN ground state 
present in the spectral scan of the Orion-KL from the 
radio-telescope of IRAM 30-m. Column 1 indicates the line transition, Col. 2 
gives the predicted frequency in the laboratory, Col. 3 the line strength,

Col. 4 upper level energy, Col. 5 observed radial velocities relatives (v$_{LSR}$),
Col. 6 observed centroid frequencies assuming a $v$$_{LSR}$ of 5\,km\,s$^{-1}$, Col. 7 observed mean beam temperature, 
and Col. 8 mean beam temperature obtained with the model. $\dagger$ blended with the last one. \\
(1) peak line observed velocity.
(2) peak line intensity.
(3) blended with $\varv_{11}$=1.
(4) blended with U-line.
(5) blended with DCOOCH$_3$.\\
(This table is available in its entirety at CDS via http://cdsweb.u-strasbg.frl. A portion is shown here for guidance regarding its form and content.)\\
}

%\clearpage

%\centering \scriptsize
\longtab{8}{
%{\setlength{\extrarowheight}{4.0pt}
\small
\begin{longtable}{clrclclc}
\caption{Detected lines of CH$_2$CHCN $\varv_{11}$=1 \label{tab_v11}.}\\
\hline\hline                      
Transition		     &  Predicted          &  $S_{ij}$ & $E_u$    & $v_{LSR}$  & $\Delta$$v$ & $T_{MB}$         & $\int T_{MB} dv$\\
$J_{K_a,K_c}-J'_{K'_a,K'_c}$ & frequency  (MHz)    &           &  (K)     & km s$^{-1}$& km s$^{-1}$ &   (K)            & (K km s$^{-1}$)  \\
\hline
\endfirsthead
\caption{continued.}\\
\hline\hline
Transition		     &  Predicted          &  $S_{ij}$ & $E_u$    & $v_{LSR}$  & $\Delta$$v$ & $T_{MB}$         & $\int T_{MB} dv$\\
$J_{K_a,K_c}-J'_{K'_a,K'_c}$ & frequency  (MHz)    &           &  (K)     & km s$^{-1}$& km s$^{-1}$ &   (K)            & (K km s$^{-1}$)  \\
\hline
\endhead
\hline
\endfoot
 9$_{1,9}$-8$_{1,8}$         &    83398.992          &  8.89           &  350.6    &    5.4$^{(1,3)}$     &               & 0.08$^{(2)}$    &                  \\
 9$_{0,9}$-8$_{0,8}$         &    85167.948          &  9.00           &  349.0    &    5.3$^{(1)}$       &               & 0.03$^{(2)}$    &                  \\
                             &                       &                 &           &    5.3$\pm$0.5       &   6$\pm$1     & 0.04            & 0.24$\pm$0.04    \\
 9$_{2,8}$-8$_{2,7}$         &    85547.124          &  8.56           &  357.5    &    6.4$^{(1)}$       &               & 0.04$^{(2)}$    &                  \\
                             &                       &                 &           &    6.6$\pm$0.5       &   8$\pm$1     & 0.03            & 0.29$\pm$0.03    \\

      \vdots                 &      \vdots           &     \vdots      &           &      \vdots          &               &    \vdots       &                  \\

14$_{0,14}$-13$_{0,13}$      &   131561.145          & 14.00           &  376.1    &    4.9$^{(1,10)}$&               & 0.37$^{(2)}$    &                  \\
14$_{2,13}$-13$_{2,12}$      &   132894.527          & 13.70           &  384.9    &    $^{(11)}$         &               & ...             &                  \\

      \vdots                 &      \vdots           &     \vdots      &           &      \vdots          &               &    \vdots       &                  \\

21$_{2,20}$-20$_{2,19}$      &   198781.116          & 20.80           &  442.2    &    5.0$^{(1)}$       &               & 0.21$^{(2)}$    &                  \\
                             &                       &                 &           &    5.00$\pm$0.11     &   7.5$\pm$0.3 & 0.21            & 1.70$\pm$0.07    \\
21$_{7,15}$-20$_{7,14}$      &   199974.194          & 18.70           &  538.2    &    2.9$^{(1)}$       &               & 0.48$^{(2)}$    &                  \\
21$_{7,14}$-20$_{7,13}$      &   199974.194$\dagger$ & 18.70           &  538.2    &    2.9$^{(1)}$       &               &  ''             &                  \\
21$_{6,16}$-20$_{6,15}$      &   199976.624$\dagger$ & 19.30           &  510.6    &    6.6$^{(1)}$       &               &  ''             &                  \\
21$_{6,15}$-20$_{6,14}$      &   199976.631$\dagger$ & 19.30           &  510.6    &    6.6$^{(1)}$       &               &  ''             &                  \\
                             &                       &                 &           &    4.8$\pm$0.6       &  12$\pm$1     & 0.41            & 5.1$\pm$0.5      \\

      \vdots                 &      \vdots           &     \vdots      &           &      \vdots          &               &    \vdots       &                  \\

\end{longtable}
%}
%\onecolumn
%\flushleft
\renewcommand{\baselinestretch}{1.5}
\tablefoot{
%{\footnotesize{\sf{{\bf{Notes.}}}}} \small Emission lines of CH$_2$CHCN $\varv_{11}$=1 
\small Emission lines of CH$_2$CHCN $\varv_{11}$=1 
present in the spectral scan of the Orion-KL from the 
radio-telescope of IRAM 30-m. Column 1 indicates the line transition, Col. 2 
gives the predicted frequency in the laboratory, Col. 3 the line strength,
Col. 4 upper level energy, Col. 5 observed radial velocities relative to the local system rest (v$_{LSR}$), 
Col. 6 the line width, Col. 7 main beam temperature, and Col. 8 shows the area of the line. $\dagger$ blended with the previous line. $\ast$$\ast$ hole in the observed spectrum. \\
(1)  peak channel line observed velocity.
(2)  peak channel line intensity.
(3)  blended with U-line.
(4)  blended with H$^+$ (H 42$\alpha$).
(5)  blended with $^1$$^3$CH$_3$OH. 
(6)  blended with OCS.
(7)  blended with (CH$_3$)$_2$CO.
(8) blended with H 49$\beta$.
(9) blended with CH$_3$CH$_2$CN.
(10) blended with $^3$$^3$SO$_2$.
(11) blended with CH$_3$OH.
(12) blended with CH$_3$OCH$_3$.
(13) blended with O$^1$$^3$CS.
(14) blended with CH$_3$CH$_2$CN $\varv_{20}$=1.
(15) blended with HCOOCH$_3$.
(16) blended with CH$_3$$^1$$^3$CH$_2$CN.
(17) blended with $^1$$^3$CH$_3$CN.
(18) blended with CH$_3$CH$_2$CN $\varv_{13}$/$\varv_{21}$.
(19) blended with H$_2$CO.
(20)  blended with CH$_2$CHCN.
(21) blended with $^3$$^4$SO$_2$.
(22) blended with CH$_2$CHCN $\varv_{10}$/$\varv_{11}$$\varv_{15}$.
(23) blended with HCOO$^1$$^3$CH$_3$.
(24) blended with H$^1$$^3$COOCH$_3$.
(25) blended with CH$_2$CHCN $\varv_{15}$=1.
(26) blended with $^2$$^9$SiO.
(27) blended with HCCCN $\nu$=0.
(28) blended with H$_2$CCO.
(29) blended with SO$_2$.
(30) blended with CH$_2$CHCN $\varv_{11}$=2.
(31) blended with CH$_3$CN $\varv_{8}$=1.
(32) blended with CH$_2$CH$^1$$^3$CN.
(33) blended with HCCCN $\varv_{6}$=1.
(34) blended with H$_2$CCC.
(35) blended with NH$_2$CHO.
(36) blended with $\mid$g$_+$-g$_-$$\mid$-CH$_3$CH$_2$OH.
(37) blended with SO$^1$$^7$O.
(38) blended with CCCS.
(39) blended with HNCO.
(40) blended with CH$_3$CCH.
(41) blended with HCC$^1$$^3$CN $\varv_{6}$=1. 
(42) blended with HCCCN $\varv_{7}$=2.
(43) blended with HDCO.
(44) blended with CH$_3$OD.
(45) blended with H$_2$$^1$$^3$CS.
(46) blended with CH$_3$CH$_2$$^1$$^3$CN. 
(47) blended with $^3$$^4$S$^1$$^8$O.     
(48) blended with CH$_3$COOH $\varv_{t}$=0. 
(49) blended with HCCCN $\varv_{7}$=3.      
(50) blended with CH$_2$DCCH.            
(51) blended with SO$^1$$^8$O.           
(52) blended with O$^1$$^3$C$^3$$^4$S.    
(53) blended with HNC$^1$$^8$O.           
(54) blended with CH$_3$C$^1$$^5$N.       
(55) blended with CH$_3$CN.               
(56) blended with CH$_3$$^1$$^3$CN.      
(57) blended with SO.   
(58) blended with $^1$$^3$CH$_2$CHCN.                    
(59) blended with HDCS.                  
(60) blended with H$_2$C$^3$$^3$S.        
(61) blended with H$_2$CS.\\
(This table is available in its entirety at CDS via http://cdsweb.u-strasbg.frl. A portion is shown here for guidance regarding its form and content.)\\
}
}

\clearpage

%\centering \scriptsize
\longtab{9}{
%{\setlength{\extrarowheight}{4.0pt}
\small
\begin{longtable}{clrclclc}
\caption{Detected lines of CH$_2$CHCN $\varv_{11}$=2 \label{tab_2v11}.}\\
\hline\hline                      
Transition		     &  Predicted          &  $S_{ij}$ & $E_u$    & $v_{LSR}$  & $\Delta$$v$ & $T_{MB}$         & $\int T_{MB} dv$\\
$J_{K_a,K_c}-J'_{K'_a,K'_c}$ & frequency  (MHz)    &           &  (K)     & km s$^{-1}$& km s$^{-1}$ &   (K)            & (K km s$^{-1}$)  \\
\hline
\endfirsthead
\caption{continued.}\\
\hline\hline
Transition		     &  Predicted          &  $S_{ij}$ & $E_u$    & $v_{LSR}$  & $\Delta$$v$ & $T_{MB}$         & $\int T_{MB} dv$\\
$J_{K_a,K_c}-J'_{K'_a,K'_c}$ & frequency  (MHz)    &           &  (K)     & km s$^{-1}$& km s$^{-1}$ &   (K)            & (K km s$^{-1}$)  \\
\hline
\endhead
\hline
\endfoot
 9$_{1,9}$-8$_{1,8}$         &    83586.209          &  8.89           &  680.0    &    $^{(3,4)}$        &               & ...             &                  \\
 9$_{0,9}$-8$_{0,8}$         &    85384.841          &  8.99           &  678.3    &    6.8$^{(1)}$       &               & 0.02$^{(2)}$    &                  \\
                             &                       &                 &           &    6.8$\pm$1.8       &   4$\pm$2     & 0.02            & 0.06$\pm$0.04    \\
 9$_{2,8}$-8$_{2,7}$         &    85787.036          &  8.56           &  686.8    &    $^{(5)}$          &               & ...             &                  \\

      \vdots                 &      \vdots           &     \vdots      &           &      \vdots          &               &    \vdots       &                  \\

14$_{0,14}$-13$_{0,13}$      &   131846.587          & 14.00           &  705.5    &    $^{(7)}$          &               & ...             &                  \\
14$_{2,13}$-13$_{2,12}$      &   133257.141          & 13.70           &  714.2    &    8.0$^{(1)}$       &               & 0.07$^{(2)}$    &                  \\
14$_{6,9}$-13$_{6,8}$        &   133664.633          & 11.40           &  781.4    &    5.9$^{(1)}$       &               & 0.04$^{(2)}$    &                  \\
14$_{6,8}$-13$_{6,7}$        &   133664.633$\dagger$ & 11.40           &  781.4    &    5.9$^{(1)}$       &               &  ''             &                  \\
14$_{5,10}$-13$_{5,9}$       &   133666.986$\dagger$ & 12.20           &  758.4    &    6.8$^{(1)}$       &               &  ''             &                  \\
14$_{5,9}$-13$_{5,8}$        &   133667.002$\dagger$ & 12.20           &  758.4    &    6.8$^{(1)}$       &               &  ''             &                  \\

      \vdots                 &      \vdots           &     \vdots      &           &      \vdots          &               &    \vdots       &                  \\

21$_{2,20}$-20$_{2,19}$      &   199292.525          & 20.80           &  771.7    &    $^{(29)}$         &               & ...             &                  \\
21$_{7,15}$-20$_{7,14}$      &   200537.645          & 18.70           &  866.4    &    3.8$^{(1,6)}$     &               & 0.30$^{(2)}$    &                  \\
21$_{7,14}$-20$_{7,13}$      &   200537.645$\dagger$ & 18.70           &  866.4    &    3.8$^{(1,6)}$     &               &  ''             &                  \\
21$_{6,16}$-20$_{6,15}$      &   200546.938          & 19.30           &  839.2    &    0.9$^{(1,30)}$    &               & 0.18$^{(2)}$    &                  \\
21$_{6,15}$-20$_{6,14}$      &   200546.947$\dagger$ & 19.30           &  839.2    &    0.9$^{(1,30)}$    &               &  ''             &                  \\
21$_{8,14}$-20$_{8,13}$      &   200551.088$\dagger$ & 18.00           &  897.1    &    7.1$^{(1,30)}$    &               &  ''             &                  \\

21$_{8,13}$-20$_{8,12}$      &   200551.088$\dagger$ & 18.00           &  897.1    &    7.1$^{(1,30)}$    &               &  ''             &                  \\

      \vdots                 &      \vdots           &     \vdots      &           &      \vdots          &               &    \vdots       &                  \\

\end{longtable}
%}
%\onecolumn
%\flushleft
\renewcommand{\baselinestretch}{1.5}
{\footnotesize{\sf{{\bf{Notes.}}}}} \small Emission lines of CH$_2$CHCN $\varv_{11}$=2 
present in the spectral scan of the Orion-KL from the 
radio-telescope of IRAM 30-m. Column 1 indicates the line transition, Col. 2 
gives the predicted frequency in the laboratory, Col. 3 the line strength,
Col. 4 upper level energy, Col. 5 observed radial velocities relative to the local system rest (v$_{LSR}$), 
Col. 6 the line width, Col. 7 main beam temperature, and Col. 8 shows the area of the line. $\dagger$ blended with the previous line. $\ast$$\ast$ hole in the observed spectrum. \\
(1) peak channel line observed velocity.
(2) peak channel line intensity.
(3) blended with H 53$\beta$.
(4) blended with Si$^1$$^7$O.
(5) blended with HCOOCH$_3$.
(6) blended with U-line.
(7) blended with HNCO.
(8) blended with CH$_3$CH$_2$CN.
(9) blended with CH$_2$CHCN $\varv_{15}$=1.
(10) blended with CH$_2$$^1$$^3$CHCN.
(11) blended with SO$_2$.
(12) blended with $^3$$^4$SO$_2$.
(13) blended with $^1$$^3$CH$_3$OH.
(14) blended with (CH$_3$)$_2$CO.
(15) blended with CH$_2$CHCN.
(16) blended with CH$_3$OH.
(17) blended with CH$_3$CHO.
(18) blended with CH$_3$CH$_2$CN $\varv_{13}$/$\varv_{21}$.
(19) blended with CH$_3$CH$_2$CN $\varv_{20}$=1.
(20) blended with H 54$\delta$.
(21) blended with NO.
(22) blended with H$^1$$^3$COOCH$_3$.
(23) blended with CH$_2$CHCN $\varv_{10}$/$\varv_{11}$$\varv_{15}$.
(24) blended with CH$_3$OCH$_3$.
(25) blended with CH$_3$CH$_2$$^1$$^3$CN.
(26) blended with H$_2$CCO.
(27) blended with c-C$_2$H$_4$O.
(28) blended with $\mid$g$_+$-g$_-$$\mid$-CH$_3$CH$_2$OH.
(29) blended with HC$^1$$^3$CCN.
(30) blended with HCC$^1$$^3$CN.
(31) blended with HCCCN $\varv_{7}$=1.
(32) blended with HCCCN $\varv_{4}$+$\varv_{7}$.
(33) blended with CH$_3$CH$_2$C$^1$$^5$N.
(34) blended with CH$_2$CHCN $\varv_{11}$=1.
(35) blended with CH$_2$$^1$$^3$CHCN.
(36) blended with NH$_2$CHO.
(37) blended with $^1$$^3$CH$_3$CH$_2$CN.
(38) blended with HCCCN $\varv_{7}$=2.
(39) blended with SO.
(40) blended with H$_2$CCO.
(41) blended with SO$^1$$^8$O.
(42) blended with HCCCN.
(43) blended with H$^1$$^3$CCCN.
(44) blended with CH$_2$CH$^1$$^3$CN.
(45) blended with SO$_2$ $\varv_{2}$=1.
(46) blended with CH$_3$$^1$$^3$CN.
(47) blended with DCOOCH$_3$.
(48) blended with CH$_3$CN.
(49) blended with H$_2$CS.
(50) blended with CCCS.
(51) blended with CH$_3$COOH $\varv_{t}$=0.
(52) blended with HNC$^1$$^8$O.
(53) blended with CH$_3$$^1$$^3$CH$_2$CN.
(54) blended with OCS.
(55) blended with H$_2$CCC.
(56) blended with CH$_3$CN $\varv_{8}$=1.
(57) blended with HDCO.
(58) blended with CH$_3$C$^1$$^5$N.
(59) blended with HCO$^+$.
(60) blended with $^1$$^3$CH$_3$CN.
(61) blended with HCOO$^1$$^3$CH$_3$.
(62) blended with t-CH$_3$CH$_2$OH
(63) blended with H$^1$$^5$NCO.
(64) blended with H$_2$C$^1$$^8$O.\\
(This table is available in its entirety at CDS via http://cdsweb.u-strasbg.frl. A portion is shown here for guidance regarding its form and content.)\\
}

\clearpage

%\centering \scriptsize
\longtab{10}{
%{\setlength{\extrarowheight}{4.0pt}
\small
\begin{longtable}{clrclclc}
\caption{Detected lines of CH$_2$CHCN $\varv_{11}$=3 \label{tab_3v11}.}\\
\hline\hline                      
Transition		     &  Predicted          &  $S_{ij}$ & $E_u$    & $v_{LSR}$  & $\Delta$$v$ & $T_{MB}$         & $\int T_{MB} dv$\\
$J_{K_a,K_c}-J'_{K'_a,K'_c}$ & frequency  (MHz)    &           &  (K)     & km s$^{-1}$& km s$^{-1}$ &   (K)            & (K km s$^{-1}$)  \\
\hline
\endfirsthead
\caption{continued.}\\
\hline\hline
Transition		     &  Predicted          &  $S_{ij}$ & $E_u$    & $v_{LSR}$  & $\Delta$$v$ & $T_{MB}$         & $\int T_{MB} dv$\\
$J_{K_a,K_c}-J'_{K'_a,K'_c}$ & frequency  (MHz)    &           &  (K)     & km s$^{-1}$& km s$^{-1}$ &   (K)            & (K km s$^{-1}$)  \\
\hline
\endhead
\hline
\endfoot
 9$_{1,9}$-8$_{1,8}$         &    83769.181          &  8.89           & 1007.6    &    $^{(3)}$          &               & ...             &                  \\
 9$_{0,9}$-8$_{0,8}$         &    85596.639          &  8.99           & 1006.1    &    $^{(3)}$          &               & ...             &                  \\
 9$_{2,8}$-8$_{2,7}$         &    86022.191          &  8.56           & 1014.4    &    $^{(4)}$          &               & ...             &                  \\
 9$_{5,5}$-8$_{5,4}$         &    86148.048          &  6.22           & 1057.9    &    4.5$^{(1)}$       &               & 0.01$^{(2)}$    &                  \\
 9$_{5,4}$-8$_{5,3}$         &    86148.048$\dagger$ &  6.22           & 1057.9    &    4.5$^{(1)}$       &               &  ''             &                  \\

      \vdots                 &      \vdots           &     \vdots      &           &      \vdots          &               &    \vdots       &                  \\

14$_{1,14}$-13$_{1,13}$      &   130055.036          & 13.90           & 1034.4    &    $^{(11,12)}$      &               & ...             &                  \\
14$_{0,14}$-13$_{0,13}$      &   132123.977          & 14.00           & 1033.3    &    $^{(13)}$         &               & ...             &                  \\
14$_{2,13}$-13$_{2,12}$      &   133612.205          & 13.70           & 1041.9    &    $^{(4)}$          &               & ...             &                  \\
14$_{6,9}$-13$_{6,8}$        &   134034.186          & 11.40           & 1108.3    &    $^{(6)}$          &               & ...             &                  \\
14$_{6,8}$-13$_{6,7}$        &   134034.187$\dagger$ & 11.40           & 1108.3    &    $^{(6)}$          &               & ...             &                  \\

      \vdots                 &      \vdots           &     \vdots      &           &      \vdots          &               &    \vdots       &                  \\

21$_{2,20}$-20$_{2,19}$      &   199792.621          & 20.80           & 1099.5    &    $^{(19,27,28)}$   &               & ...             &                  \\
21$_{7,15}$-20$_{7,14}$      &   201090.968          & 18.70           & 1193.1    &    $^{(10)}$         &               & ...             &                  \\
21$_{7,15}$-20$_{7,14}$      &   201090.968$\dagger$ & 18.70           & 1193.1    &    $^{(10)}$         &               & ...             &                  \\
21$_{8,14}$-20$_{8,13}$      &   201098.099$\dagger$ & 18.00           & 1224.1    &    $^{(10)}$         &               & ...             &                  \\
21$_{8,13}$-20$_{8,12}$      &   201098.099$\dagger$ & 18.00           & 1224.1    &    $^{(10)}$         &               & ...             &                  \\

      \vdots                 &      \vdots           &     \vdots      &           &      \vdots          &               &    \vdots       &                  \\

\end{longtable}
%}
%\onecolumn
%\flushleft
\renewcommand{\baselinestretch}{1.5}
{\footnotesize{\sf{{\bf{Notes.}}}}} \small Emission lines of CH$_2$CHCN $\varv_{11}$=3 
present in the spectral scan of the Orion-KL from the 
radio-telescope of IRAM 30-m. Column 1 indicates the line transition, Col. 2 
gives the predicted frequency in the laboratory, Col. 3 the line strength,
Col. 4 upper level energy, Col. 5 observed radial velocities relative to the local system rest (v$_{LSR}$), 
Col. 6 the line width, Col. 7 main beam temperature, and Col. 8 shows the area of the line. $\dagger$ blended with the previous line. $\ast$ noise level. $\ast$$\ast$ hole in the observed spectrum. \\
(1) peak channel line observed velocity.
(2) peak channel line intensity.
(3) blended with U-line.
(4) blended with HCOOCH$_3$.
(5) blended with CH$_3$OCH$_3$.
(6) blended with CH$_3$CH$_2$CN.
(7) blended with CH$_3$CH$_2$C$^1$$^5$N.
(8) blended with H 49$\beta$.
(9) blended with CH$_2$CHCN $\varv_{11}$=1.
(10) blended with CH$_3$OH.
(11) blended with O$^1$$^3$C$^3$$^4$S.
(12) blended with CH$_3$$^1$$^3$CH$_2$CN.
(13) blended with $^3$$^4$SO$_2$.
(14) blended with H$^1$$^3$CS.
(15) blended with H$^1$$^3$COOCH$_3$.
(16) blended with CH$_2$CHOH.
(17) blended with (CH$_3$)$_2$CO.
(18) blended with SO$_2$ $\varv_{2}$=1.
(19) blended with CH$_3$CH$_2$CN $\varv_{13}$/$\varv_{21}$.
(20) blended with HDO.
(21) blended with CH$_3$OD.
(22) blended with SO$^1$$^7$O.
(23) blended with HCCCN.
(24) blended with $^2$$^9$SiO.
(25) blended with CH$_3$CH$_2$$^1$$^3$CN.
(26) blended with $^1$$^3$CH$_2$CHCN.
(27) blended with CH$_2$CHCN $\varv_{15}$=1.
(28) blended with HCC$^1$$^3$CN $\varv_{7}$=1.
(29) blended with CHDCHCN.
(30) blended with CH$_2$$^1$$^3$CHCN.
(31) blended with H$_2$CCO.
(32) blended with SO$_2$.
(33) blended with CH$_3$CH$_2$CN $\varv_{20}$=1.
(34) blended with H$_2$CS.
(35) blended with $^3$$^3$SO$_2$.
(36) blended with DCOOCH$_3$.
(37) blended with c-C$_3$H$_2$.
(38) blended with CH$_2$CHCN.
(39) blended with $^1$$^3$CH$_3$OH.
(40) blended with HN$^1$$^3$CO.
(41) blended with $^1$$^3$CH$_3$CH$_2$CN.
(42) blended with HNC$^1$$^8$O.
(43) blended with SO$^1$$^8$O.
(44) blended with H$^1$$^3$CCCN $\varv_{7}$=1.
(45) blended with HCOO$^1$$^3$CH$_3$.
(46) blended with $\mid$g$_+$-g$_-$$\mid$-CH$_3$CH$_2$OH.
(47) blended with t-CH$_3$CH$_2$OH.
(48) blended with D$_2$CO.
(49) blended with CH$_3$C$^1$$^5$N.
(50) blended with NH$_2$CHO.
(51) blended with CH$_3$CN $\varv_{8}$=1.
(52) blended with $^3$$^4$SO.
(53) blended with $^3$$^3$SO.
(54) blended with CH$_3$COOH $\varv_{t}$=0.
(55) blended with CH$_3$OD.
(56) blended with H$^1$$^3$CN.
(57) blended with H$_2$CCS.
(58) blended with HDCO.
(59) blended with CH$_2$CHCN $\varv_{11}$=2.
(60) blended with CH$_2$DCN.
(61) blended with DCCCN.\\
(This table is available in its entirety at CDS via http://cdsweb.u-strasbg.frl. A portion is shown here for guidance regarding its form and content.)\\
}

\clearpage

%\centering \scriptsize
\longtab{11}{
%{\setlength{\extrarowheight}{4.0pt}
\small
\begin{longtable}{clrclclc}
\caption{Detected lines of CH$_2$CHCN $\varv_{15}$=1 \label{tab_v15}.}\\
\hline\hline                      
Transition		     &  Predicted          &  $S_{ij}$ & $E_u$    & $v_{LSR}$  & $\Delta$$v$ & $T_{MB}$         & $\int T_{MB} dv$\\
$J_{K_a,K_c}-J'_{K'_a,K'_c}$ & frequency  (MHz)    &           &  (K)     & km s$^{-1}$& km s$^{-1}$ &   (K)            & (K km s$^{-1}$)  \\
\hline
\endfirsthead
\caption{continued.}\\
\hline\hline
Transition		     &  Predicted          &  $S_{ij}$ & $E_u$    & $v_{LSR}$  & $\Delta$$v$ & $T_{MB}$         & $\int T_{MB} dv$\\
$J_{K_a,K_c}-J'_{K'_a,K'_c}$ & frequency  (MHz)    &           &  (K)     & km s$^{-1}$& km s$^{-1}$ &   (K)            & (K km s$^{-1}$)  \\
\hline
\endhead
\hline
\endfoot
 9$_{1,9}$-8$_{1,8}$         &    83349.865          &  8.89           &  500.9    &    $^{(3)}$          &               & ...             &                  \\
 9$_{0,9}$-8$_{0,8}$         &    85075.547          &  9.00           &  499.1    &    5.6$^{(1)}$       &               & 0.03$^{(2)}$    &                  \\
 9$_{2,8}$-8$_{2,7}$         &    85416.767          &  8.56           &  508.0    &    $^{(4)}$          &               & ...             &                  \\
 9$_{4,6}$-8$_{4,5}$         &    85528.061          &  7.22           &  534.5    &    4.0$^{(1)}$       &               & 0.02$^{(2)}$    &                  \\
 9$_{4,5}$-8$_{4,4}$         &    85528.107$\dagger$ &  7.22           &  534.5    &    4.2$^{(1)}$       &               &  ''             &                  \\
 9$_{5,5}$-8$_{5,4}$         &    85532.951          &  6.22           &  554.4    &    4.1$^{(1)}$       &               & 0.02$^{(2)}$    &                  \\
 9$_{5,4}$-8$_{5,3}$         &    85532.951$\dagger$ &  6.22           &  554.4    &    4.1$^{(1)}$       &               &  ''             &                  \\
                             &                       &                 &           &    4.3$\pm$0.2       &  7.2$\pm$0.7  & 0.02            & 0.12$\pm$0.01    \\

      \vdots                 &      \vdots           &     \vdots      &           &      \vdots          &               &    \vdots       &                  \\

14$_{0,14}$-13$_{0,13}$      &   131504.447          & 14.00           &  526.2    &    6.6$^{(1,13)}$    &               & 0.11$^{(2)}$    &                  \\
14$_{2,13}$-13$_{2,12}$      &   132709.220          & 13.70           &  535.3    &    6.1$^{(1)}$       &               & 0.08$^{(2)}$    &                  \\
14$_{5,10}$-13$_{5,9}$       &   133073.970          & 12.20           &  581.7    &    $^{(4)}$          &               & ...             &                  \\
14$_{5,9}$-13$_{5,8}$        &   133073.979$\dagger$ & 12.20           &  581.7    &    $^{(4)}$          &               & ...             &                  \\

      \vdots                 &      \vdots           &     \vdots      &           &      \vdots          &               &    \vdots       &                  \\

21$_{2,20}$-20$_{2,19}$      &   198556.992          & 20.80           &  592.6    &    5.5$^{(1)}$       &               & 0.16$^{(2)}$    &                  \\
                             &                       &                 &           &    5.60$\pm$0.11     &  5.2$\pm$0.4  & 0.17            & 0.91$\pm$0.08    \\
21$_{6,16}$-20$_{6,15}$      &   199658.717          & 19.30           &  663.4    &    $^{(4,32)}$       &               & ...             &                  \\
21$_{6,15}$-20$_{6,14}$      &   199658.722$\dagger$ & 19.30           &  663.4    &    $^{(4,32)}$       &               & ...             &                  \\
21$_{7,15}$-20$_{7,14}$      &   199669.152          & 18.70           &  692.0    &    $^{(4,32)}$       &               & ...             &                  \\
21$_{7,14}$-20$_{7,13}$      &   199669.152$\dagger$ & 18.70           &  692.0    &    $^{(4,32)}$       &               & ...             &                  \\
21$_{5,17}$-20$_{5,16}$      &   199683.011          & 19.80           &  639.2    &    $^{(6)}$          &               & ...             &                  \\
21$_{5,16}$-20$_{5,15}$      &   199683.411$\dagger$ & 19.80           &  639.2    &    $^{(6)}$          &               & ...             &                  \\

      \vdots                 &      \vdots           &     \vdots      &           &      \vdots          &               &    \vdots       &                  \\

\end{longtable}
%}
%\onecolumn
\flushleft
\renewcommand{\baselinestretch}{1.5}
{\footnotesize{\sf{{\bf{Notes.}}}}} \small Emission lines of CH$_2$CHCN $\varv_{15}$ =1
present in the spectral scan of the Orion-KL from the 
radio-telescope of IRAM 30-m. Column 1 indicates the line transition, Col. 2 
gives the predicted frequency in the laboratory, Col. 3 the line strength,
Col. 4 upper level energy, Col. 5 observed radial velocities relative to the local system rest (v$_{LSR}$), 
Col. 6 the line width, Col. 7 main beam temperature, and Col. 8 shows the area of the line. $\dagger$ blended with the previous line. $\ast$$\ast$ hole in the observed spectrum. \\
(1) peak channel line observed velocity.
(2) peak channel line intensity.
(3) blended with HCOOCH$_3$.
(4) blended with CH$_2$CHCN.
(5) blended with CH$_2$CHCN $\varv_{11}$=1.
(6) blended with U-line.
(7) blended with CCH.
(8) blended with $^1$$^3$CH$_3$OH.
(9) blended with CH$_2$$^1$$^3$CHCN.
(10) blended with CH$_3$OH.
(11) blended with $^3$$^4$SO.
(12) blended with CH$_3$COOH $\varv_{t}$=0.
(13) blended with t-CH$_3$CH$_2$OH.
(14) blended with (CH$_3$)$_2$CO.
(15) blended with CH$_3$CH$_2$CN.
(16) blended with SO$_2$.
(17) blended with $^1$$^3$CH$_3$CH$_2$CN.
(18) blended with CH$_3$C$^1$$^5$N.
(19) blended with DNCO.
(20) blended with CH$_3$CH$_2$CN $\varv_{13}$/$\varv_{21}$.
(21) blended with HCC$^1$$^3$CN $\varv_{7}$=1.
(22) blended with H$^1$$^3$COOCH$_3$.
(23) blended with HCOOCH$_3$ $\varv_{t}$=1.
(24) blended with CH$_3$OCH$_3$.
(25) blended with CH$_3$OCOD.
(26) blended with HNCO.
(27) blended with H$_2$CCO.
(28) blended with NS.
(29) blended with CH$_2$DCCH.
(30) blended with CH$_3$CH$_2$CN $\varv_{20}$=1.
(31) blended with $^1$$^8$OCS.
(32) blended with SiS.
(33) blended with HCCCN.
(34) blended with CH$_3$CH$_2$C$^1$$^5$N.
(35) blended with H$_2$C$_3$.
(36) blended with $\mid$g$_+$-g$_-$$\mid$-CH$_3$CH$_2$OH.
(37) blended with SO$^1$$^8$O.
(38) blended with CH$_3$CHO.
(39) blended with HCCCN $\varv_{6}$=1.
(40) blended with H$_2$CO.
(41) blended with HN$^1$$^3$CO.
(42) blended with HCCCN $\varv_{7}$=1.
(43) blended with $^3$$^3$SO$_2$.
(44) blended with OCS. 
(45) blended with DNCS.
(46) blended with CH$_3$CN $\varv_{8}$=1.
%(47) blended with H$_2$C$_3$.
(47) blended with HCCCN $\varv_{7}$=2.
(48) blended with H$^1$$^3$CCCN $\varv_{7}$=1.
(49) blended with $^1$$^3$CH$_2$CHCN.
(50) blended with HCCCN $\varv_{6}$+$\varv_{7}$.
(51) blended with H$^1$$^3$CCCN $\nu$=0.
(52) blended with CH$_2$CHCN $\varv_{10}$/$\varv_{11}$$\varv_{15}$.
(53) blended with CH$_3$CN. 
(54) blended with CH$_3$$^1$$^3$CN.
(55) blended with HCOO$^1$$^3$CH$_3$.
(56) blended with CH$_3$$^1$$^3$CH$_2$CN.
(57) blended with $^3$$^4$SO$_2$.
(58) blended with HDCS.
(59) blended with O$^1$$^3$CS.
(60) blended with CH$_2$CHCN $\varv_{11}$=3.
(61) blended with HCCCN $\varv_{7}$=3.
(62) blended with DCOOCH$_3$.
(63) blended with H$^1$$^3$COOCH$_3$.
(64) blended with NH$_2$CHO.
(65) blended with CH$_3$OD.
(66) blended with g$_-$-CH$_3$CH$_2$OH.\\
(This table is available in its entirety at CDS via http://cdsweb.u-strasbg.frl. A portion is shown here for guidance regarding its form and content.)\\
}

\clearpage

%\centering \scriptsize
\longtab{12}{
%{\setlength{\extrarowheight}{4.0pt}
\small
\begin{longtable}{clccllll}
\caption{Detected lines of CH$_2$CHCN $\varv_{10}$=1$\Leftrightarrow$($\varv_{11}$=1,$\varv_{15}$=1) \label{tab_v10v11v15}.}\\
\hline\hline                      
Transition		           & Predicted           &  $S_{ij}$ & $E_u$    & $v_{LSR}$$^{(1)}$   & Observed       & Observed             & Model        \\
$J_{K_a,K_c,v}-J'_{K'_a,K'_c,v'}$  & frequency (MHz)     &           &  (K)     & km s$^{-1}$         & frequency (MHz)& $T_{MB}$ (K)$^{(2)}$ & $T_{MB}$ (K) \\
\hline
\endfirsthead
\caption{continued.}\\
\hline\hline
Transition		           & Predicted           &  $S_{ij}$ & $E_u$    & $v_{LSR}$$^{(1)}$   & Observed       & Observed             & Model        \\
$J_{K_a,K_c,v}-J'_{K'_a,K'_c,v'}$  & frequency (MHz)     &           &  (K)     & km s$^{-1}$         & frequency (MHz)& $T_{MB}$ (K)$^{(2)}$ & $T_{MB}$ (K) \\
\hline
\endhead
\hline
\endfoot
 9$_{1,9,0}$-8$_{1,8,0}$           &   83116.219         &  8.89     &  830.7    &  4.4                &83116.4         &  0.02               &  0.01        \\
 9$_{1,9,1}$-8$_{1,8,1}$           &   83527.527         &  8.89     &  834.3    &  1.9                &83528.4         &  0.03               &  0.01        \\
 9$_{0,9,0}$-8$_{0,8,0}$           &   84834.239         &  8.99     &  829.0    &  7.4                &84833.6         &  0.01               &  0.01        \\
 9$_{2,9,0}$-8$_{2,8,0}$           &   85222.373         &  8.55     &  837.7    &  $^{(3)}$           &   ...          &  ...                &  ...         \\

      \vdots                 &      \vdots           &     \vdots      &           &      \vdots          &               &    \vdots       &                  \\

14$_{0,14,0}$-13$_{0,13,0}$        &  131110.736         & 14.00     &  856.1    &  $^{(22)}$          &  ...           &  ...                &  ...         \\
14$_{2,13,0}$-13$_{2,12,0}$        &  132389.964         & 13.70     &  864.9    &  3.0                &132390.8        &  0.04               &  0.04        \\
14$_{5,10,0}$-13$_{5,9,0}$         &  132761.289         & 12.20     &  910.1    &  $^{(5)}$           &  ...           &  ...                &  ...         \\
14$_{5,9,0}$-13$_{5,8,0}$          &  132761.289$\dagger$& 12.20     &  910.1    &  $^{(5)}$           &  ...           &  ...                &  ...         \\
14$_{6,9,0}$-13$_{6,8,0}$          &  132767.225         & 11.40     &  933.6    &  $^{(5)}$           &  ...           &  ...                &  ...         \\
14$_{6,8,0}$-13$_{6,7,0}$          &  132767.225$\dagger$& 11.40     &  933.6    &  $^{(5)}$           &  ...           &  ...                &  ...         \\

      \vdots                 &      \vdots           &     \vdots      &           &      \vdots          &               &    \vdots       &                  \\

21$_{2,20,0}$-20$_{2,19,0}$        &  198041.306         & 20.80     &  922.0    &  $^{(16)}$          &  ...           &  ...                &  ...         \\
21$_{2,20,1}$-20$_{2,19,1}$        &  199062.857         & 20.80     &  926.1    &  6.0                &199062.2        &  0.08               &  0.10        \\
21$_{6,16,0}$-20$_{6,15,0}$        &  199190.101         & 19.30     &  991.0    &  7.5$^{(5,16)}$     &199188.4        &  0.15               &  0.15        \\
21$_{6,15,0}$-20$_{6,14,0}$        &  199190.107$\dagger$& 19.30     &  991.0    &  7.5$^{(5,16)}$     &199188.4        &  ''                 &  0.15        \\
21$_{7,14,0}$-20$_{7,13,0}$        &  199196.532         & 18.70     & 1018.8    &  4.0                &199197.1        &  0.14               &  0.13        \\
21$_{7,15,0}$-20$_{7,14,0}$        &  199196.532$\dagger$& 18.70     & 1018.8    &  4.0                &199197.1        &  ''                 &  0.13        \\

      \vdots                 &      \vdots           &     \vdots      &           &      \vdots          &               &    \vdots       &                  \\

\end{longtable}
%}
%\onecolumn
\flushleft
\renewcommand{\baselinestretch}{1.5}
{\footnotesize{\sf{{\bf{Notes.}}}}} \small Emission lines of CH$_2$CHCN $\varv_{10}$=1$\Leftrightarrow$($\varv_{11}$=1,$\varv_{15}$=1)
present in the spectral scan of the Orion-KL from the 
radio-telescope of IRAM 30-m. The quantum number $\nu$ is corresponded with the vibrational level, 
and take the value $\nu$=0 and $\nu$=1 whether the state is $\varv_{10}$=1 or ($\varv_{11}$=1,$\varv_{15}$=1), respectively.
Column 1 indicates the line transition, Col. 2 
gives the predicted frequency in the laboratory, Col. 3 the line strength,
Col. 4 upper level energy, Col. 5 observed radial velocities relative to the local system rest (v$_{LSR}$), 
Col. 6 observed centroid frequencies assuming a $v$$_{LSR}$ of 5\,km\,s$^{-1}$, Col. 7 observed main beam temperature, 
y Col. 8 mean beam temperature obtained with the model. $\dagger$ blended with the last one. $\ast$$\ast$ hole in the observed spectrum. \\
(1) peak channel line observed velocity.
(2) peak channel line intensity.
(3) blended with $^3$$^3$SO.
(4) blended with HCS$^+$.
(5) blended with U-line.
(6) blended with HCOOCH$_3$.
(7) blended with SO.
(8) blended with CH$_3$CH$_2$C$^1$$^5$N.
(9) blended with (CH$_3$)$_2$CO.
(10) blended with CH$_3$OCH$_3$.
(11) blended with CH$_3$OH.
(12) blended with HDCS.
(13) blended with $^1$$^3$CH$_3$OH. 
(14) blended with CH$_2$$^1$$^3$CHCN.
(15) blended with CH$_2$CHCN.
(16) blended with CH$_3$CH$_2$CN.
(17) blended with $^3$$^4$SO$_2$.
(18) blended with CH$_3$C$^1$$^5$N.
(19) blended with CH$_3$CN $\varv_{8}$=1.
(20) blended with c-C$_2$H$_4$O.              
(21) blended with CH$_3$CH$_2$$^1$$^3$CN.
(22) blended with CH$_2$CHCN $\varv_{15}$=1.
(23) blended with SO$_2$.                 
(24) blended with CH$_3$CH$_2$CN $\varv_{13}$/$\varv_{21}$.
(25) blended with $\mid$g$_+$-g$_-$$\mid$-CH$_3$CH$_2$OH.
(26) blended with CH$_3$OD.
(27) blended with CH$_3$$^1$$^3$CH$_2$CN.
(28) blended with CH$_3$CH$_2$CN $\varv_{20}$=1.
(29) blended with CH$_2$CHCN $\varv_{11}$=2.
(30) blended with CH$_3$CHO.
(31) blended with CH$_2$CHCN $\varv_{11}$=1.
(32) blended with $^1$$^3$CH$_2$CHCN.
(33) blended with H$^1$$^3$COOCH$_3$.
(34) blended with SO$^1$$^7$O.
(35) blended with OC$^3$$^6$S.
(36) blended with SO$_2$ $\varv_{2}$=1.
(37) blended with HCCCN $\varv_{7}$=2.
(38) blended with CH$_2$CH$^1$$^3$CN.
(39) blended with H$_2$$^1$$^3$CS.
(40) blended with NH$_2$CHO.
(41) blended with H$_2$CS.
(42) blended with $^3$$^3$SO$_2$.
(43) blended with HCOO$^1$$^3$CH$_3$.
(44) blended with $^1$$^3$CH$_3$CH$_2$CN.
(45) blended with HCC$^1$$^3$CN $\varv_{7}$=2.
(46) blended with CH$_3$CN.
(47) blended with DCOOCH$_3$.
(48) blended with HC$^1$$^3$CCN $\nu$=0.
(49) blended with SiS.
(50) blended with O$^1$$^3$CS.
(51) blended with H$_2$CO.
(52) blended with HCCCN.
(53) blended with t-CH$_3$CH$_2$OH.
(54) blended with CH$_3$$^1$$^8$OH.
(55) blended with C$^1$$^8$O.
(56) blended with O$^3$$^4$S$^1$$^8$O.
(57) blended with HDCO.
(58) blended with HCCCN $\varv_{6}$=1.
(59) blended with HNC$^1$$^8$O.
(60) blended with SO$^1$$^8$O.
(61) blended with CO.
(62) blended with HCCCN $\varv_{6}$=2.
(63) blended with HCCCN $\varv_{7}$=3.
(64) blended with CH$_3$COOH $\varv_{t}$=0.
(65) blended with CH$_2$CHCN $\varv_{11}$=3.
(66) blended with CH$_2$CDCN.
(67) blended with CH$_3$CN $\varv_{8}$=1.
(68) blended with CCCS.
(69) blended with HCC$^1$$^3$CN.
(70) blended with $^3$$^4$SO.
(71) blended with H$^1$$^3$CCCN.
(72) blended with HDO.
(73) blended with $^1$$^3$CH$_3$CN.
(74) blended with CH$_3$CCH.
(75) blended with HDCS.
(76) blended with CH$_3$$^1$$^3$CN.
(77) blended with HCOOH.
(78) blended with SiO.
(79) blended with SiO $\nu$=1.
(80) blended with DCOOH.
(81) blended with HCN.
(82) blended with H$_2$C$^3$$^4$S.
(83) blended with NH$_2$CHO $\varv_{12}$=1.
(84) blended with CH$_2$CHC$^1$$^5$N.
(85) blended with CH$_3$OD.
(86) blended with DCHCHCN.\\
(This table is available in its entirety at CDS via http://cdsweb.u-strasbg.frl. A portion is shown here for guidance regarding its form and content.)\\
}

\clearpage

%\centering \scriptsize
\longtab{13}{
%{\setlength{\extrarowheight}{4.0pt}
\small
\begin{longtable}{clccllll}
\caption{Detected lines of $^{13}$C$_1$,  $^{13}$C$_2$, and $^{13}$C$_3$ isotopologues of CH$_2$CHCN. \label{tab_13C}.}\\
\hline\hline                      
Transition		           & Predicted           &  $S_{ij}$ & $E_u$    & $v_{LSR}$$^{(1)}$   & Observed       & Observed             & Model        \\
$J_{K_a,K_c}-J'_{K'_a,K'_c}$       & frequency (MHz)     &           &  (K)     & km s$^{-1}$         & frequency (MHz)& $T_{MB}$ (K)$^{(3)}$ & $T_{MB}$ (K) \\
\hline
\endfirsthead
\caption{continued.}\\
\hline\hline
Transition		           & Predicted           &  $S_{ij}$ & $E_u$    & $v_{LSR}$$^{(1)}$   & Observed       & Observed             & Model        \\
$J_{K_a,K_c}-J'_{K'_a,K'_c}$       & frequency (MHz)     &           &  (K)     & km s$^{-1}$         & frequency (MHz)& $T_{MB}$ (K)$^{(3)}$ & $T_{MB}$ (K) \\
\hline
\endhead
\hline
\endfoot

\hline
\multicolumn{8}{c}{\bfseries{Detected lines of $^{13}$CH$_2$CHCN}}           \\
\hline
 9$_{1,9}$-8$_{1,8}$          &    81051.736         &  8.89    &  21.6    &    5.54             &81051.6       &  0.01               &  0.01         \\
 9$_{2,8}$-8$_{2,7}$          &    83064.074         &  8.56    &  28.5    &    5.66$^{(2)}$     &83063.9       &  0.01               &  0.01         \\
 9$_{4,6}$-8$_{4,5}$          &    83172.144         &  7.22    &  54.2    &    2.66             &83172.8       &  0.01               &  0.01         \\
 9$_{4,5}$-8$_{4,4}$          &    83172.189$\dagger$&  7.22    &  54.2    &    2.82             &      "       &   "                 &  "            \\
 9$_{3,6}$-8$_{3,5}$          &    83187.779         &  8.00    &  39.2    &    6.34$^{(2)}$     &83187.4       &  0.02               &  0.01         \\
 9$_{6,4}$-8$_{6,3}$          &    83188.121$\dagger$&  5.00    &  96.8    &    7.57$^{(2)}$     &      "       &   "                 &  "            \\
 9$_{6,3}$-8$_{6,2}$          &    83188.121$\dagger$&  5.00    &  96.8    &    7.57$^{(2)}$     &      "       &   "                 &  "            \\

      \vdots                 &      \vdots           &     \vdots      &           &      \vdots          &               &    \vdots       &                  \\

14$_{1,13}$-13$_{1,12}$       &   131955.171         & 13.90    &  49.7    &    8.06$^{(3)}$     &131953.8      &  0.13               &  0.03         \\
15$_{2,14}$-14$_{2,13}$       &   138228.563         & 14.70    &  61.7    &    5.78             &138228.3      &  0.08               &  0.03         \\
15$_{5,11}$-14$_{5,10}$       &   138657.101         & 13.30    & 106.6    &    1.60$^{(2)}$     &138658.6      &  0.08               &  0.05         \\
15$_{5,10}$-14$_{5,9}$        &   138657.119$\dagger$& 13.30    & 106.6    &    1.64$^{(2)}$     &      "       &   "                 &  "            \\
15$_{4,12}$-14$_{4,11}$       &   138678.606         & 13.90    &  87.4    &    3.79             &138679.1      &  0.08               &  0.07         \\
15$_{4,11}$-14$_{4,10}$       &   138678.418$\dagger$& 13.90    &  87.4    &    7.71             &      "       &   "                 &  "            \\
15$_{7,9}$-14$_{7,8}$         &   138680.928$\dagger$& 11.70    & 157.8    &    8.81             &      "       &   "                 &  "            \\
15$_{7,8}$-14$_{7,7}$         &   138680.928$\dagger$& 11.70    & 157.8    &    8.81             &      "       &   "                 &  "            \\

      \vdots                 &      \vdots           &     \vdots      &           &      \vdots          &               &    \vdots       &                  \\

22$_{0,22}$-21$_{0,21}$       &   198591.415         & 22.00    & 110.7    &    3.79$^{(7)}$     &198592.3      &  0.18               &  0.08         \\
22$_{4,19}$-21$_{4,18}$       &   203539.866         & 21.30    & 110.7    &    0.49$^{(2)}$     &203542.9      &  0.21               &  0.08         \\
22$_{10,12}$-21$_{10,11}$     &   203542.989$\dagger$& 17.50    & 325.0    &    5.09$^{(2)}$     &      "       &   "                 &  "            \\
22$_{10,13}$-21$_{10,12}$     &   203542.989$\dagger$& 17.50    & 325.0    &    5.09$^{(2)}$     &      "       &   "                 &  "            \\

      \vdots                 &      \vdots           &     \vdots      &           &      \vdots          &               &    \vdots       &                  \\

%\end{longtable}
%}}
%}

%\centering \scriptsize
%\begin{longtable}{ccccllcc}
%\caption{Detected lines of CH$_2$$^{13}$CHCN ground state.}\label{tab_13C2} \\
%\hline\hline
%Transition		           & Predicted           &  $S_{ij}$ & $E_u$    & $v_{LSR}$$^{(1)}$   & Observed       & Observed             & Model        \\
%$J_{K_a,K_c,v}-J'_{K'_a,K'_c,v'}$  & frequency (MHz)     &           &  (K)     & km s$^{-1}$         & frequency (MHz)& $T_{MB}$ (K)$^{(3)}$ & $T_{MB}$ (K) \\
%\hline
%\endfirsthead
%\caption{continued.}\\
%\hline\hline
%Transition		           & Predicted           &  $S_{ij}$ & $E_u$    & $v_{LSR}$$^{(1)}$   & Observed       & Observed             & Model        \\
%$J_{K_a,K_c,v}-J'_{K'_a,K'_c,v'}$  & frequency (MHz)     &           &  (K)     & km s$^{-1}$         & frequency (MHz)& $T_{MB}$ (K)$^{(3)}$ & $T_{MB}$ (K) \\
%\hline
%\endhead
%\hline
%\endfoot

%###########
%CH2-13CHCN:
%###########

\hline
\multicolumn{8}{c}{\bfseries{Detected lines of CH$_2$$^{13}$CHCN}}           \\ 
\hline
 9$_{4,6}$-8$_{4,5}$          &    84961.208         &  7.22    &  54.1    &    0.86             &84962.9       &  0.01               &  0.01         \\
 9$_{4,5}$-8$_{4,4}$          &    84961.208$\dagger$&  7.22    &  54.1    &    0.65             &      "       &   "                 &  "            \\
 9$_{5,5}$-8$_{5,4}$          &    84963.011$\dagger$&  6.22    &  73.0    &    5.50             &      "       &   "                 &  "            \\
 9$_{5,4}$-8$_{5,3}$          &    84963.011$\dagger$&  6.22    &  73.0    &    5.50             &      "       &   "                 &  "            \\
 9$_{2,7}$-8$_{2,6}$          &    85278.270         &  8.56    &  28.9    &    8.17$^{(2)}$     &85277.4       &  0.02               &  0.01         \\

      \vdots                 &      \vdots           &     \vdots      &           &      \vdots          &               &    \vdots       &                  \\

14$_{0,14}$-13$_{0,13}$       &   130481.709         & 14.00    &  47.2    &    5.48             &130481.5      &  0.09               &  0.03         \\
14$_{5,10}$-13$_{5,9}$        &   132191.757         & 12.20    & 100.2    &    4.38$^{(2)}$     &132192.0      &  0.13               &  0.05         \\
14$_{5,9}$-13$_{5,8}$         &   132191.770$\dagger$& 12.20    & 100.2    &    4.41$^{(2)}$     &      "       &   "                 &  "            \\
14$_{6,9}$-13$_{6,8}$         &   132194.320$\dagger$& 11.40    & 123.3    &    7.81$^{(2)}$     &      "       &   "                 &  "            \\
14$_{6,8}$-13$_{6,7}$         &   132194.320$\dagger$& 11.40    & 123.3    &    7.81$^{(2)}$     &      "       &   "                 &  "            \\

      \vdots                 &      \vdots           &     \vdots      &           &      \vdots          &               &    \vdots       &                  \\

21$_{7,15}$-20$_{7,14}$       &   198336.172         & 18.70    & 207.7    &    4.26             &198336.7      &  0.18               &  0.18         \\
21$_{7,14}$-20$_{7,13}$       &   198336.172$\dagger$& 18.70    & 207.7    &    4.26             &      "       &   "                 &  "            \\
21$_{6,16}$-20$_{6,15}$       &   198336.544$\dagger$& 19.30    & 180.4    &    4.83             &      "       &   "                 &  "            \\
21$_{6,15}$-20$_{6,14}$       &   198336.552$\dagger$& 19.30    & 180.4    &    4.84             &      "       &   "                 &  "            \\
21$_{8,14}$-20$_{8,13}$       &   198358.019         & 18.00    & 239.1    &    1.45$^{(2)}$     &198360.4      &  0.12               &  0.07         \\
21$_{8,13}$-20$_{8,12}$       &   198358.019$\dagger$& 18.00    & 239.1    &    1.45$^{(2)}$     &      "       &   "                 &  "            \\

      \vdots                 &      \vdots           &     \vdots      &           &      \vdots          &               &    \vdots       &                  \\

%\end{longtable}
%}}
%}

%\centering \scriptsize
%\begin{longtable}{ccccllcc}
%\caption{Detected lines of CH$_2$CH$^{13}$CN.}\label{tab_13C3} \\
%\hline\hline
%Transition		           & Predicted           &  $S_{ij}$ & $E_u$    & $v_{LSR}$$^{(1)}$   & Observed       & Observed             & Model        \\
%$J_{K_a,K_c,v}-J'_{K'_a,K'_c,v'}$  & frequency (MHz)     &           &  (K)     & km s$^{-1}$         & frequency (MHz)& $T_{MB}$ (K)$^{(3)}$ & $T_{MB}$ (K) \\
%\hline
%\endfirsthead
%\caption{continued.}\\
%\hline\hline
%Transition		           & Predicted           &  $S_{ij}$ & $E_u$    & $v_{LSR}$$^{(1)}$   & Observed       & Observed             & Model        \\
%$J_{K_a,K_c,v}-J'_{K'_a,K'_c,v'}$  & frequency (MHz)     &           &  (K)     & km s$^{-1}$         & frequency (MHz)& $T_{MB}$ (K)$^{(3)}$ & $T_{MB}$ (K) \\
%\hline
%\endhead
%\hline
%\endfoot

%###########
%CH2CH-13CN:
%###########

\hline
\multicolumn{8}{c}{\bfseries{Detected lines of CH$_2$CH$^{13}$CN}}           \\
  \hline
 9$_{5,5}$-8$_{5,4}$          &    85041.672         &  6.22    &  74.4    &    2.61$^{(2)}$     &85042.4       &  0.02               &  0.01         \\
 9$_{5,4}$-8$_{5,3}$          &    85041.672$\dagger$&  6.22    &  74.4    &    2.61$^{(2)}$     &      "       &   "                 &  "            \\
 9$_{6,4}$-8$_{6,3}$          &    85053.074         &  5.00    &  98.1    &    3.16             &85055.4       &  0.01               &  0.01         \\
 9$_{6,3}$-8$_{6,2}$          &    85053.074$\dagger$&  5.00    &  98.1    &    3.16             &      "       &   "                 &  "            \\
 9$_{3,6}$-8$_{3,5}$          &    85055.519$\dagger$&  8.00    &  39.9    &    5.47             &      "       &   "                 &  "            \\
10$_{0,10}$-9$_{0,9}$         &    93864.835         &  9.99    &  24.8    &    3.50$^{(2)}$     &93865.2       &  0.02               &  0.01         \\

      \vdots                 &      \vdots           &     \vdots      &           &      \vdots          &               &    \vdots       &                  \\

21$_{9,12}$-20$_{9,11}$       &   198588.933         & 17.10    & 279.3    &    7.61$^{(2,26)}$  &198587.2      &  0.31               &  0.11         \\
21$_{9,13}$-20$_{9,12}$       &   198588.933$\dagger$& 17.10    & 279.3    &    7.61$^{(2,26)}$  &      "       &   "                 &  "            \\
21$_{4,17}$-20$_{4,16}$       &   198654.522         & 20.20    & 139.4    &    4.75$^{(27)}$    &198654.6      &  0.20               &  0.07         \\
21$_{13,9}$-20$_{13,8}$       &   198864.217         & 13.00    & 467.0    &    6.22$^{(2)}$     &198863.4      &  0.10               &  0.04         \\
21$_{13,8}$-20$_{13,7}$       &   198864.217$\dagger$& 13.00    & 467.0    &    6.22$^{(2)}$     &      "       &   "                 &  "            \\
22$_{1,22}$-21$_{1,21}$       &   201496.193         & 21.95    & 113.7    &    3.60             &201497.1      &  0.10               &  0.13 \\

      \vdots                 &      \vdots           &     \vdots      &           &      \vdots          &               &    \vdots       &                  \\

\end{longtable}
%}
%\onecolumn
%\flushleft
\renewcommand{\baselinestretch}{1.5}
{\footnotesize{\sf{{\bf{Notes.}}}}} \small Emission lines of  of $^1$$^3$CH$_2$CHCN, CH$_2$$^1$$^3$CHCN and CH$_2$CH$^1$$^3$CN 
isotopologues in its ground state present in the spectral scan of the Orion-KL from the 
radio-telescope of IRAM 30-m. Column 1 indicates the line transition, Col. 2 
gives the predicted frequency in the laboratory, Col. 3 the line strength,
Col. 4 upper level energy, Col. 5 observed radial velocities relatives (v$_{LSR}$),
Col. 6 observed centroid frequencies assuming a $v$$_{LSR}$ of 5\,km\,s$^{-1}$, Col. 7 observed mean beam temperature, 
and Col. 8 mean beam temperature obtained with the model. $\dagger$ blended with the last one. \\
(1)  peak line intensity.
(2)  blended with U-line.
(3)  blended with HCOOCH$_3$.
(4)  blended with CH$_2$CHC$^1$$^5$N.
(5)  blended with DCOOCH$_3$.
(6)  blended with HCOO$^1$$^3$CH$_3$.
(7)  blended with CH$_2$CH$^1$$^3$CN.
(8)  blended with g$_+$-CH$_3$CH$_2$OH.
(9)  blended with H$_2$CCS.
(10)  blended with CH$_3$CH$_2$CN $\varv_{13}$/$\varv_{21}$
(11)  blended with CH$_3$CCD.
(12)  blended with t-CH$_3$CH$_2$OH.
(13)  blended with CH$_3$CH$_2$CN.
(14)  blended with CH$_2$CHCN $\varv_{11}$=1.
(15)  blended with CH$_3$COOH $\varv_{t}$=0.
(16)  blended with (CH$_3$)$_2$CO.
(17)  blended with CH$_2$CHCN $\varv_{10}$/$\varv_{11}$$\varv_{15}$.
(18)  blended with H$^1$$^3$COOCH$_3$.
(19)  blended with SO$^1$$^8$O.
(20)  blended with CH$_3$CHO.
(21)  blended with SO$^1$$^7$O.
(22)  blended with $^1$$^3$CN.
(23)  blended with CH$_2$CH$^1$$^3$CN.
(24)  blended with CH$_2$$^1$$^3$CHCN.
(25)  blended with HCOOCH$_3$ $\varv_{t}$=1.
(26)  blended with $^1$$^3$CH$_2$CHCN.
(27)  blended with NH$_2$D.
(28)  blended with $^1$$^3$CH$_3$OH.
(29)  blended with $^3$$^3$SO$_2$.
(30)  blended with SO$_2$ $\varv_{2}$=2.\\
(This table is available in its entirety at CDS via http://cdsweb.u-strasbg.frl. A portion is shown here for guidance regarding its form and content.)\\
}

\clearpage

%\centering \scriptsize
\longtab{14}{
%{\setlength{\extrarowheight}{4.0pt}
\small
\begin{longtable}{clccllll}
\caption{Detected lines of vinyl isocyanide (CH$_2$CHNC) \label{tab_VyISOCN}.}\\
\hline\hline                      
Transition		      & Predicted           &  $S_{ij}$ & $E_u$    & $v_{LSR}$$^{(1)}$   & Observed       & Observed             & Model        \\
$J_{K_a,K_c}-J'_{K'_a,K'_c}$  & frequency (MHz)     &           &  (K)     & km s$^{-1}$         & frequency (MHz)& $T_{MB}$ (K)$^{(2)}$ & $T_{MB}$ (K) \\
\hline
\endfirsthead
\caption{continued.}\\
\hline\hline
Transition		      & Predicted           &  $S_{ij}$ & $E_u$    & $v_{LSR}$$^{(1)}$   & Observed       & Observed             & Model        \\
$J_{K_a,K_c}-J'_{K'_a,K'_c}$  & frequency (MHz)     &           &  (K)     & km s$^{-1}$         & frequency (MHz)& $T_{MB}$ (K)$^{(2)}$ & $T_{MB}$ (K) \\
\hline
\endhead
\hline
\endfoot
 9$_{2,8}$-8$_{2,7}$               &   92222.557         &  8.56     &   31.0    &2.74             &92223.3       &  0.02               &  0.01        \\
 9$_{3,7}$-8$_{3,6}$               &   92376.457         &  8.00     &   42.2    &8.82$^{(3)}$     &92375.3       &  0.03               &  0.01        \\
 9$_{6,4}$-8$_{6,3}$               &   92379.404         &  5.00     &  102.0    &5.48$^{(7)}$     &92379.3       &  0.01               &  0.01        \\
 9$_{6,3}$-8$_{6,2}$               &   92379.404$\dagger$&  5.00     &  102.0    &5.48$^{(7)}$     &92379.3       &  0.01               &  0.01        \\
 9$_{3,6}$-8$_{3,5}$               &   92386.902         &  8.00     &   42.2    &6.95             &92386.3       &  0.02               &  0.01        \\

      \vdots                 &      \vdots           &     \vdots      &           &      \vdots          &               &    \vdots       &                  \\

13$_{2,12}$-12$_{2,11}$            &  133062.524         & 12.70     &   53.6    &3.41$^{(6)}$     &133063.2      &  0.04               &  0.03        \\
13$_{3,10}$-12$_{3,9}$             &  133566.376         & 12.30     &   64.8    &6.48             &133565.7      &  0.04               &  0.03        \\
13$_{2,11}$-12$_{2,10}$            &  134563.575         & 12.70     &   53.9    &4.75$^{(7)}$     &134563.7      &  0.08               &  0.04        \\
14$_{0,14}$-13$_{0,13}$            &  141702.008         & 14.00     &   51.3    &5.82             &141701.6      &  0.04               &  0.04        \\

      \vdots                 &      \vdots           &     \vdots      &           &      \vdots          &               &    \vdots       &                  \\

19$_{2,17}$-18$_{2,16}$            &  198094.397         & 18.80     &  103.3    &4.50$^{(10)}$    &198094.8      &  0.09               &  0.10        \\
19$_{1,18}$-18$_{1,17}$            &  198245.681         & 18.90     &   97.7    &8.79             &198242.9      &  0.10               &  0.10        \\
20$_{3,18}$-19$_{3,17}$            &  205532.002         & 19.50     &  123.5    &8.41             &205529.7      &  0.10               &  0.10        \\
21$_{1,21}$-20$_{1,20}$            &  208561.673         & 20.90     &  112.7    &2.99             &208563.0      &  0.15               &  0.12        \\

      \vdots                 &      \vdots           &     \vdots      &           &      \vdots          &               &    \vdots       &                  \\

\end{longtable}
%}
%\onecolumn
\flushleft
\renewcommand{\baselinestretch}{1.5}
{\footnotesize{\sf{{\bf{Notes.}}}}} \small Emission lines of vinyl isocyanide (CH$_2$CHNC) 
present in the spectral scan of the Orion-KL from the 
radio-telescope of IRAM 30-m. Column 1 indicates the line transition, Col. 2 
gives the predicted frequency in the laboratory, Col. 3 the line strength,
Col. 4 upper level energy, Col. 5 observed radial velocities relatives (v$_{LSR}$),
Col. 6 observed centroid frequencies assuming a $v$$_{LSR}$ of 5\,km\,s$^{-1}$, Col. 7 observed mean beam temperature, 
and Col. 8 mean beam temperature obtained with the model. $\dagger$ blended with the last one. \\
(1) peak line observed velocity.
(2) peak line intensity.
(3) blended with HCOOCH3 $\varv_t$=1.
(4) blended with HCOOCH$_3$.
(5) blended with H$^{13}$COOCH$_3$.
(6) blended with CH$_2$CHCN.
(7) blended with U-line.
(8) blended with SO$^{18}$O.
(9) blended with DCOOCH$_3$.
(10) blended with O-H$_2$CS.
(11) blended with HCDCHCN.
(12) blended with CH$_3$CH$_2$CN.
(13) blended with (CH$_3$)$_2$CO.
(14) blended with CH$_3$CH$_2$C$^{15}$N.
(15) blended with HCOO$^{13}$CH$_3$.\\
(This table is available in its entirety at CDS via http://cdsweb.u-strasbg.frl. A portion is shown here for guidance regarding its form and content.)\\
}

\end{appendix}{}
\clearpage

%%%%%%%%%%%%%%%%%%%%%%%%%%%%%%%%%%%%%%%%%%%%%%%%%%%%%%%%%%%%%%%%%%%%%%%%%%%%%%%%
%%%%%%%%%%%%%%%%%%%%%%%%%%%%%%%%%%%%%%%%%%%%%%%%%%%%%%%%%%%%%%%%%%%%%%%%%%%%%%%%
%  ONLINE APPENDIX B: typographical error
%%%%%%%%%%%%%%%%%%%%%%%%%%%%%%%%%%%%%%%%%%%%%%%%%%%%%%%%%%%%%%%%%%%%%%%%%%%%%%%%

\begin{appendix}{}
%\label{AppendixI}
%\section{\small{Appendix I}}

\begin{table*}
\section{\small{Typographical error in \citet{adm13}}}
%\scriptsize
\caption[Physico-chemical conditions of Orion-KL from CH$_3$CH$_2$CN.]
{Physico-chemical conditions of Orion-KL from CH$_3$CH$_2$CN} \label{tab_prop_ethyl}
\centering
%\tablewidth{0pt}
%\tabletypesize{\tiny}
\resizebox{0.9\textwidth}{!}{%
\begin{tabular}{lcccc}
\hline\hline
                                                 &\bfseries Hot core 1   &\bfseries Hot core 2   &\bfseries Hot core 3 &                                    \\
\hline
$d$$_{\rm sou}$   ('')                           &        4              &       10              &      25              &                                   \\
\textit{offset} ('')                             &        5              &        5              &       5              &                                   \\
$v$$_{\rm exp}$  (km s$^{-1}$)                   &        5              &       13              &      22              &                                   \\
$v$$_{\rm LSR}$  (km s$^{-1}$)                   &        5              &        3              &       3              &                                   \\
\hline
$T$$_{ETL}$ (K)                                  &      275              &      110              &      65               &$N$$_{TOTAL}$ (cm$^{-2}$)         \\
\hline\hline
$N$(CH$_3$CH$_2$CN) (cm$^{-2}$)                  &\bfseries(6$\pm$2)$\times$10$^{16}$&   (8$\pm$2)$\times$10$^{15}$      &  (3.0$\pm$0.9)$\times$10$^{15}$   & \bfseries(7$\pm$2)$\times$10$^{16}$\\
$N$(CH$_3$CH$_2$CN $\varv_{13}=1/\varv_{21}=1$) (cm$^{-2}$) &\bfseries (8$\pm$2)$\times$10$^{15}$ &  (1.1$\pm$0.3)$\times$10$^{15}$ &  (4$\pm$1)$\times$10$^{14}$      &\bfseries(1.0$\pm$0.3)$\times$10$^{16}$\\
$N$(CH$_3$CH$_2$CN $\varv_{20}$) (cm$^{-2}$)       &\bfseries (3$\pm$1)$\times$10$^{15}$      &   (4$\pm$1)$\times$10$^{14}$      &  (1.7$\pm$0.5)$\times$10$^{14}$  &\bfseries(4$\pm$1)$\times$10$^{15}$\\
$N$(CH$_3$CH$_2$CN $\varv_{12}$) (cm$^{-2}$)       & \bfseries (1.2$\pm$0.6)$\times$10$^{15}$ &   (1.6$\pm$0.5)$\times$10$^{14}$  &  (6$\pm$3)$\times$10$^{13}$      &\bfseries(1.4$\pm$0.7)$\times$10$^{15}$\\
$N$($^{13}$CH$_3$CH$_2$CN) (cm$^{-2}$)           &   (7$\pm$2)$\times$10$^{14}$      &   (1.9$\pm$0.6)$\times$10$^{14}$  &  (7$\pm$2)$\times$10$^{13}$       &\\
$N$(CH$_3$$^{13}$CH$_2$CN) (cm$^{-2}$)           &   (7$\pm$2)$\times$10$^{14}$      &   (1.9$\pm$0.6)$\times$10$^{14}$  &  (7$\pm$2)$\times$10$^{13}$       &\\
$N$(CH$_3$CH$_2$$^{13}$CN) (cm$^{-2}$)           &   (7$\pm$2)$\times$10$^{14}$      &   (1.9$\pm$0.6)$\times$10$^{14}$  &  (7$\pm$2)$\times$10$^{13}$       &\\
$N$(CH$_3$CH$_2$C$^{15}$N) (cm$^{-2}$)           &   (2$\pm$1)$\times$10$^{14}$      &   (5$\pm$3)$\times$10$^{13}$      &  (1.7$\pm$0.8)$\times$10$^{13}$   &\\
$N$(A-CH$_2$DCH$_2$CN) (cm$^{-2}$)               &$\leq$ 6$\times$10$^{14}$          &$\leq$ 2$\times$10$^{14}$          &$\leq$ 6$\times$10$^{13}$          &\\
$N$(S-CH$_2$DCH$_2$CN) (cm$^{-2}$)               &$\leq$ 7$\times$10$^{14}$          &$\leq$ 1$\times$10$^{14}$          &$\leq$ 6$\times$10$^{13}$          &\\
$N$(CH$_3$CHDCN) (cm$^{-2}$)                     &$\leq$ 6$\times$10$^{14}$          &$\leq$ 2$\times$10$^{14}$          &$\leq$ 6$\times$10$^{13}$          &\\
\hline
\end{tabular}
}
\flushleft
Note. \small Physico-chemical conditions of Orion-KL from the analysis of ethyl
cyanide emission lines in the range of 80-280 GHz. In bold type, we display the
revised values by a factor 2. The other values are the same.
The revised values corresponded to the hot narrow component (1) for the ground and
excited states. Vibrational temperatures are not affected while isotopic ratios have to
be changed by the same correction factor.\\
\vspace{2cm}
\end{table*}

\begin{table*}
%\scriptsize
\caption[Isotopic abundances for CH$_3$CH$_2$CN in the Orion-KL region.]
{Isotopic abundances for CH$_3$CH$_2$CN in the Orion-KL region.} \label{isot_abund_ethyl}
\centering
%\tablewidth{0pt}
%\tabletypesize{\tiny}
\resizebox{0.6\textwidth}{!}{%
\begin{tabular}{lccc}
\hline\hline
                                                 & $^{12}$C/$^{13}$C                   & $^{14}$N/$^{15}$N                     & H/D                       \\
\hline
Isotopic abundance (ratio X)                     &   \bfseries 73$\pm$22             &    \bfseries  256$\pm$128           &    \bfseries 0.012$\pm$0.005  \\
\hline\hline
\end{tabular}
}
\flushleft
Note. \small Owing to the error in the column densities of hot narrow component for the
ground and excited states, isotopic abundances are increased by a factor less
than 2 for $^{12}$C/$^{13}$C and $^{14}$N/$^{15}$N ratios of ethyl cyanide. On
the other hand, the H/D ratio is decreased by a factor 2.
\\
\end{table*}

\end{appendix}{}

\end{document}